\documentclass[%
 reprint,
 amsmath,amssymb,
 aps,
endfloat
]{revtex4-2}
\usepackage{hyperref}
\usepackage{graphicx}
\usepackage{dcolumn}
\usepackage{comment}
\usepackage{bm}
\usepackage[labelformat=empty,font=small, justification=raggedright,format=plain]{caption}
\usepackage{setspace}

\begin{document}

\preprint{Agility from instability}

\title{Ciliary flocking and emergent instabilities enable collective agility in a non-neuromuscular animal\\  
\small{Part 2 (of 3): Tissue scale }}
\thanks{Part 2 of 3 for a series reporting the stable yet sensitive collective dynamics of ciliary flocking across scales.}%

\author{Matthew S. Bull}
\email{bullm@stanford.edu}
\affiliation{Department of Applied Physics\\
 Stanford University, Stanford, CA 94305, USA}

\author{Vivek N. Prakash}
\affiliation{%
Department of Bioengineering\\
 Stanford University, Stanford, CA 94305, USA}
\affiliation{Department of Physics\\ University of Miami, Coral Gables, FL 33146, USA
}%

\author{Manu Prakash}
 \email{manup@stanford.edu}
\affiliation{
 Department of Bioengineering\\
 Stanford University, Stanford, CA 94305, USA
}%

\date{\today}
             
\singlespacing

\begin{abstract}
Effective organismal behavior is characterized by the ability to respond appropriately to changes in the surrounding environment. Attaining this delicate balance of sensitivity and stability is a hallmark of the animal kingdom. By studying the locomotory behavior of a simple animal (\textit{Trichoplax adhaerens}) without muscles or neurons, here we demonstrate how monociliated epithelial cells work collectively to give rise to an agile non-neuromuscular organism. Via direct visualization of large ciliary arrays, we report the discovery of sub-second ciliary reorientations under a rotational torque that is mediated by collective tissue mechanics and the adhesion of cilia to the underlying substrate. In an illuminating toy model, we show a mapping of this system onto an "active-elastic resonator". This framework explains how perturbations propagate information in this array as linear speed traveling waves in response to mechanical stimulus. Next, we explore the implications of periodic and noisy parametric driving in this active-elastic resonator and show that such driving can excite mechanical 'spikes'. These spikes in collective mode amplitudes are consistent with a system driven by parametric amplification and a saturating nonlinearity. We conduct extensive numerical experiments to corroborate these findings within a polarized active-elastic sheet. These results indicate that periodic and stochastic forcing are critical for increasing the sensitivity of collective ciliary flocking. We support these theoretical predictions via direct experimental observation of linear speed traveling waves which are made possible through the hybridization of spin and overdamped density waves. We map how these ciliary flocking dynamics result in agile motility via coupling between an amplified resonator and a tuning (Goldstone-like) mode of the system. This sets the stage for how activity and elasticity can self-organize into behavior which benefits the organism as a whole.\\


\textsl{Significance:}\\

Agility -- the ability to convert sensory stimulation to action rapidly -- is a cornerstone of the success of the animal kingdom. Despite growing to millions, billions, and even trillions of cells (humans $\sim10^{13}$), animals maintain their agility through a cascade of evolutionary innovations in development, division of labor, and cellular communication -- a prime example being the neuro-muscular system. Here, we study a simple, motile animal which coordinates its behavior across millions of cells without neurons or muscles. We discover sub-second reorientation dynamics of walking cilia akin to "ciliary flocking". We utilize tools from the physics of active matter and flocking to develop a simple yet effective description of "ciliary flocking". Through parallel experimental and numerical studies, we identify a suite of phenomena including linear speed traveling waves, and a long length-scale instability. We show that these phenomena are consistent with: 1) an active-elastic resonator which generates a phonon-like wave equation through the delicate dance of a reorientation and displacement and 2) an activity fluctuation-driven parametric instability. In this work, we reconcile the conundrum of how linear speed traveling waves can arise in the absence of behavioral inertia and show that agility can arise from harnessing this instability -- conceptually analogous to the roll/pitch instabilities of a modern fighter jet plane. We suggest that the presence of qualitatively similar phenomena arise from very different micro-physics which affirms the importance of excitability as a organism/super-organism behavioral motif. We expect that both the introduction of this model organism and our results will be of broad intellectual interest to communities ranging from the physics of collective motion/proto-nervous systems, ciliary biophysics, physical origins of ethology to swarm robotics, embodied intelligence and the self-organization of distributed actuation.
\begin{description}
\item[Usage]
Secondary publications and information retrieval purposes.
\item[Keywords]
placozoa, nonequilibrium collective motion, active matter, flocking, parametric resonance

\item[A note to readers]
This work is the second of three complementary stories we have posted together describing multiple scales of ciliary flocking. While we encourage readers to read all three for a more complete picture, each of these works are written to be as self-contained as possible. In the rare case where we evoke a result from another manuscript, we make effort to motivate it as a falsifiable assumption on the grounds of existing literature and direct readers to the relevant manuscript for a more in-depth understanding.

\item [Part 1 ] \textsl{Excitable mechanics embodied in a walking cilium} (posted on arXiv concurrently)\cite{bullpart1} describes the emergent mechanics of ciliary walking as a transition between ciliary swimming and ciliary stalling with increasing adhesion. 

\item [Part 2 ] \textsl{Non-neuromuscular collective agility by harnessing instability in ciliary flocking} (this work) reports the implications of an effective rotational degree-of-freedom governing the direction of ciliary walking for agile locomotive behavior in an animal without a brain. 

\item [Part 3 ] \textsl{Mobile defects born from an energy cascade shape the locomotive behavior of a headless animal} (posted on arXiv concurrently)\cite{bullpart3} describes the emergent locomotive behavior of the animal in terms of a low dimensional manifold which is identified by both top-down and bottom-up approaches to find agreement.
\end{description}
\end{abstract}

\maketitle


\section{Introduction}
Agility is critical to the survival of multi-cellular collectives competing in the evolutionary arms race for ever-increasing sensitivity and fast response \cite{ keijzer_moving_2015,moroz_independent_2009, jekely_chemical_2021, jekely2010origin, wan_origins_2021, sponberg_emergent_2017}. Agility -- at its core -- is not about inherent speed, but the ability to pass information from sensing to appropriate actions quickly and efficiently. Intuition serves us well here: large organisms will require longer distance communication between sensing and actuation to coordinate this time sensitive response \cite{more_scaling_nodate, pratt_neural_2017}. Thus, we might expect that single cells will be more agile than animals with their highly local information transmission \cite{wan_origins_2021}. However, the distinct advantages of multi-cellularity (e.g. size, division of labor) combined with innovations such as an animal's neuromusuclar system that enables non-local information transfer based on neuronal wiring meant that early animals could also be agile and hence evolutionarily competitive to their single cell counterparts \cite{grosberg_evolution_2007, pentz_ecological_2020, sponberg_emergent_2017}. In this work, we explore the limits of an alternative, non-neuromuscular paradigm for agile multi-cellular behavior: distributed sensory-actuation pairs self-organized into organism-scale consensus by the physics of active matter. How agile can this non-neuromuscular solution for multicellularity be?

\subsection{Agility and information transfer in active matter}
Active matter is a rapidly maturing field of research aiming to understand, shape and harness how energy injection at the shortest length scale can be coordinated into collective action at longer length scales of the problem \cite{marchetti_hydrodynamics_2013, gompper_2020_2020, shaebani_computational_2020,alert_active_2021}. This field works across many scales of living matter from cells \cite{giavazzi_flocking_2018,szabo_phase_2006, ferrante_elasticity-based_2013, henkes_dense_2020} to insects \cite{van_der_vaart_environmental_2020} to migratory animals \cite{attanasi_information_2014, tunstrom_collective_2013, katz_inferring_2011}. Critically, the collective phenomena in active matter map directly onto locomotive behavioral motifs which have important implications for the coherence, agility and survival of the collective\cite{attanasi_information_2014, van_der_vaart_environmental_2020, blanch-mercader_hydrodynamic_2017}.  Thus, the study of active matter is important for our understanding of the transition from many individuals to a single collective. Despite its diverse successes there is still much to learn about how the animal kingdom exploits cellular and sub-cellular activity and forces in the service of the animal as a whole\cite{gilpin_multiscale_2020, jekely_chemical_2021, Mongera2018AElongation, Gilpin2017}. 

Propagation of a wave in active matter and agility are intimately linked\cite{attanasi_information_2014, tunstrom_collective_2013} via the rate of mechanical information transfer in a system. With only local coupling amongst components, information about a time and location of stimulus must be shared through the collective via a communication channel \cite{attanasi_information_2014, van_der_vaart_environmental_2020, geyer_sounds_2018}. One common and simple form of communication is mediated through traveling waves \cite{tu_sound_1998, attanasi_information_2014, pajic-lijakovic_mechanical_2020}. Traveling waves can come in many forms from nonlinear trigger waves \cite{mathijssen_collective_2019, armon_epithelial_2020, boocock_theory_2021} to emergent mechanics \cite{banerjee_propagating_2015, peyret_sustained_2019, blanch-mercader_hydrodynamic_2017}, and play a fundamental role in coordinating the behavior of a collective at an incredible speed. There is a rich history coupling active mechanical work with elastic materials. This work ranges from 'active crystals'\cite{ferrante_elasticity-based_2013}, vertex models\cite{giavazzi_flocking_2018} and voronoi models \cite{Bi2015a}. When these active systems are embedded within a strongly dissipative media (commonly at short enough length scales to be in low Reynolds number), momentum change is effectively instantaneous relative to the important timescales of the problem. This maps the governing elastic equations onto gradient descent which is then modified by non-conservative, active contribution governed by its own kinetics. This overdamped solution precludes the oscillations that one would commonly associate with the harmonic response of an elastic structure. And yet, oscillation is abound in active elastic structures ranging from tissue response to expansion\cite{serra-picamal_mechanical_2012, banerjee_propagating_2015, blanch-mercader_hydrodynamic_2017, pajic-lijakovic_mechanical_2020} to ciliary beat oscillations \cite{camalet2000generic, ma2014active} and even cell growth\cite{rojas_response_2014}. 

Here, we present the discovery of flocking behavior in ciliary arrays in two dimensions, including propagation of spin-waves that globally coordinate the locomotive behavior of a living animal \textit{Trichoplax adhaerens}. Next, we show a simple mapping of polarized activity onto a momentum-like response which approaches a simple harmonic oscillator in a specific parameter regime. We further provide support that our experimental model-- \textit{Trichoplax adhaerens} -- lives in this region of parameter space which supports linear speed traveling waves. We show that this minimal set of ingredients permits both rapid information propagation and when coupled with ciliary driving can amplify small external inputs as a mechanically excitable parametric amplifier.

\subsection{Placozoa - a headless animal as experimental model}

Comparative experimental work between evolutionarily distant organisms is a powerful technique for highlighting the governing principles underlying a plethora of biological phenomena\cite{russell_non-model_2017}. Motivated by the question of "how cells collectively build organisms" - going from millions of collaborating units, we sought out an experimental system which was both simple enough for us to understand but complicated enough teach us new lessons. 

\textit{Trichoplax adhaerens} is an understudied early divergent marine animal known predominately for its simplicity in cell types \cite{Smith2014}, radial body plan \cite{dubuc_dorsalventral_2019}, and its non-neuromuscular coordination\cite{smith2015coordinated, Smith2016AdherensAdhaerens., Senatore2017NeuropeptidergicSynapses., Ueda1999DynamicAdhaerence, romanova_sodium_2020} and locomotion\cite{smith_coherent_2019}.The organism has shown early promise as an experimentally tractable model organism for a number of studies ranging from mechanobiology to resilience in epithelial tissue to highly dynamic loading. The top tissue layer exhibits ultra-fast cellular contractions\cite{armon2018ultrafast, armon_epithelial_2020} and the bottom tissue layer behaves as an epithelial alloy tuned to balance the tissue's ductility when pushed to the edge of rupture\cite{prakash_motility-induced_2021}.  Here, via experimental and theoretical tools - we directly observe the dynamics of thousands of cilia in a freely moving animal and for the first time discover - through direct sub-cellular tracking - sub-second ciliary reorientation dynamics in a living metazoan. We further develop tools to measure this re-orientation dynamics at the scale of the whole animal and demonstrate how real-time self-organization of this active matter directly enables locomotive dynamics of this headless animal.

\subsection{Roadmap}
The experimental study of \textit{T. adhaerens} presents an opportunity to map collective mechanics at multiple scales to decipher how real-time activity of a group of monociliated cells transforms into real-time behavior of an individual organism - a dream for a physical ethologist. In this work, we focus on the locomotive mechanics emergent from the fast time scale mechanical interaction of cilia to an underlying substrate (modelled as cellularized chemo-mechanical oscillators - aka walking cilia [cite Part 1]). Here, we directly measure the mechanical communication between the tissue and the ciliary forcing direction. We report sub-second ciliary reorientation of both the bending axis and stepping direction. These dynamics couple the tissue forces to the direction of motion in a mechanically interacting system of two spatio-temporal fields - ciliary dynamics and tissue dynamics. To lowest order, these dynamics represent the physical embodiment of a polarized active elastic material.

In order to develop a model for this collective locomotive behavior, we map the active forcing from a walking cilia to an oscillatory force consistent with observations of ciliary dynamics. A minimal model (based on models of walking cilia developed in Part 1\cite{bullpart1}) couples the ciliary phase (oscillator phase) to the forces within the epithelial tissue that is locomoting. This framework gives rise to a flocking-oscillator model with a single parameter which controls the system's deviation from steady-active-forcing. In this minimal model, we find two distinct boundaries between an polarized-unpolarized and synchronized-unsynchronized regime giving rise to four possible qualitative behaviors of phase space tuneable by two parameters. 

Equipped with a minimal model, we study the displacement dynamics of the tissue experimentally in a freely moving animal. By tagging and tracking individual cells in a whole animal, we identify three key observations: 1) anisotropic traveling waves (broken by the spontaneous symmetry breaking of the direction of travel), 2) an underdamped-like response in an overdamped environment (at $\sim70$ mHz), and 3) an effective mode coupling the activity direction to the displacement from the mean tissue position orthogonal to the direction of tissue polarization. 

We corroborate these findings numerically in our theoretical framework (focusing initially on the low active periodicity limit) and find that fast ciliary reorientation relative to the environmental damping is important for supporting traveling waves, which represent the hybridization between the spin mode and overdamped elastic displacement mode. We suggest a new non-dimensional parameter connecting these two quantities and explain its utility in understanding the underlying dynamics. 

This set of experimental and numerical approaches finally motivates the derivation of a toy model which we term as the 'active-elastic resonator'. Beginning with the equation of motion for a single active cell (walking cilia embedded in the epithelial tissue), we project the dynamics of force generation onto the axis orthogonal to the axis of symmetry breaking. This gives rise to a two dimensional dynamical system linking the orthogonal displacement to the orthogonal projection of the active force. We note that the linearization of this dynamical system maps conveniently onto the simple harmonic oscillator with new interpretations for the effective damping, the effective mass, and the effective stiffness. This new formulation of ciliary activity onto an elastic tissue (epithelium) opens the door to a rich base of intuition associated with effective modes of this active-elastic resonator. We describe results regarding the Q-factor and response to external driving of a characteristic frequency. 

We next look experimentally at the activity fluctuations in the direction of tissue symmetry breaking. To accomplish this we designed two experiments, where we monitor the displacement of the bottom layer of epithelial cells as the animal moves freely via a single cell staining technique in live animals. Secondly, we also study the displacement dynamics of bright fluorescent beads embedded in a soft agar medium as reporters of ciliary walking activity above. In both experiments, we found evidence of higher frequency forcing driving the tissue forward. 

We incorporate these dynamics of periodic forcing into our toy model to better understand the implications of this form of parametric driving on an active-elastic resonator. Thus we are able to map a very well-known phenomena in parametric amplification onto the dynamics of periodically driven active matter. We show that the amplitude versus frequency of the driving has the classic tongue profile of a damped-Mathieu equation combined with a nonlinear saturation. Furthermore, the dynamics of the fixed point undergoes a bifurcation transforming the steady state dynamics into a limit cycle when the driving frequency is positioned on-resonance. We further show that the response of this system exhibits a tuneable excitation threshold in response to mechanical stimuli of different strength. This tuneable excitability has implications for not only the steady-state behaviors but also the agile response the animal in response to environmental coupling. 

To lend further support to the relevance of these parametric amplification phenomena to an active elastic sheet, we relaxed the one-dimensional projection used to highlight the active-elastic resonator mode and study a two-dimensional (2D) active element attached to a tissue characterized solely by its behavioral inertia. [Here we define behavioral inertia in the sense introduced in flocking models for birds \cite{attanasi_information_2014} which refers to relationship between ciliary orientation and direction of tissue displacement]. For low behavioral inertia (large coupling parameter), the tissue responds rapidly to changes in the motion of the individual cell. For high behavioral inertia (low coupling parameter), the tissue continues on its original path independent of the individual cell state. We find that in the limit of high tissue behavioral inertia, oscillations are preserved and by extension so is parametric amplification around the resonance. However, at low behavioral inertia the power of this displacement mode couples strongly into a zero-stiffness mode associated with the spontaneous symmetry breaking of a continuous degree of freedom (a Goldstone-like mode). We present our understanding in the form of two modes: 1) a resonator mode (which under the right conditions can amplify), and 2) a turning mode. Power leaks from the amplified mode to the turning mode at a rate which is controlled by the inverse of the behavioral inertia of the tissue. If power leaks too fast, the resonator mode cannot grow in magnitude through amplification and the entire system is overdamped. However, at intermediate values of the coupling parameter, the amplified mode leaks out to the turning mode resulting in an excitable turning response. 

Having built a suite of toy models to understand our experimentally observed agility, we next present a unified framework of how active-elastic sheets can be viewed as a collection of active-elastic resonator modes coupled weakly by nonlinear up-conversion; parametric coupling between modes and activity mediated parametric pumping. We show that the steady state power distribution is heavily modified by the presence of parametric driving which can result in the growth of intermediate wavelength modes without evoking an emergent negative viscosity. We generate deeper support for the role of activity fluctuations in driving multi-cellular agility by studying the response of a simulated tissue of flocking oscillators in response to cell-localized mechanical stimulus and show that parametric driving can increase the turning sensitivity by a factor of $\sim 4.5$ while maintaining collective locomotive stability of the longest length scale mode which controls the organism's speed of travel. Accompanying this experiment, we show that the position of stimulus carries important information which shapes the amplitude and direction of the response to stimulus in a fashion consistent with embodied computation. 

\section{Ciliary flocking}
Without a known developmental head-tail encoded symmetry breaking, placozoa have traditionally been described as gliding along surfaces \cite{Senatore2017NeuropeptidergicSynapses., smith_coherent_2019}. In first part of our work\cite{bullpart1}, we demonstrate that cilia interact strongly with the substrate in a fashion similar to a 'walking gait' without specialized gait control. Via imaging at multiple length and time scales, to the best of our knoweldge, here we show for the first time that the resulting self-organizing dynamics of cilia via fast ciliary reorientation can generate organism-scale coherent structures of ciliary spin waves which we term 'ciliary flocking'.
\subsection{Large scale coherent structures in ciliary fields}
Single-organelle resolved imaging across full organisms is a far-reaching challenge \cite{prakash_motility-induced_2021} especially when compounded with fast temporal dynamics. Most attempts to this problem are achieved with engineering complex microscopy experiments merging genetic engineering, efficient 3D optical sectioning and computational tracking to keep the organism in frame\cite{nguyen_whole-brain_2016}. In our work, we leverage the experimental advantages of studying an effectively two dimensional behaving animal to reduce the complexity of the required equipment while still achieving high temporal and spatial resolution measurements across entire organisms. 

These techniques are characterized by a comprehensive suite of experiments ranging from high-speed, single organelle resolve trans-illumination microscopy to course grained tissue flow using fluorescent markers and chemically adhered micro-spheres. 

To achieve single organelle resolved imaging at 100 Hz, we employed an inverted microscope (60x TIRF objective, NA = 1.43, Nikon TE200U and an ASI RAMM setup) with a high NA condenser (NA=0.85 and NA=0.5) with an absorption filter passing $\lambda = 550$nm to reduce chromatic aberration. Importantly, these experiments are not photon-limited like high-speed fluorescence microscopy and we could resolve single cilium steps in the plane parallel to the substrate. This gives us clean access to both ciliary bending and steps as the monociliated bottom tissue locomotes across the substrate. 

\begin{figure}
  \includegraphics[width=1\textwidth]{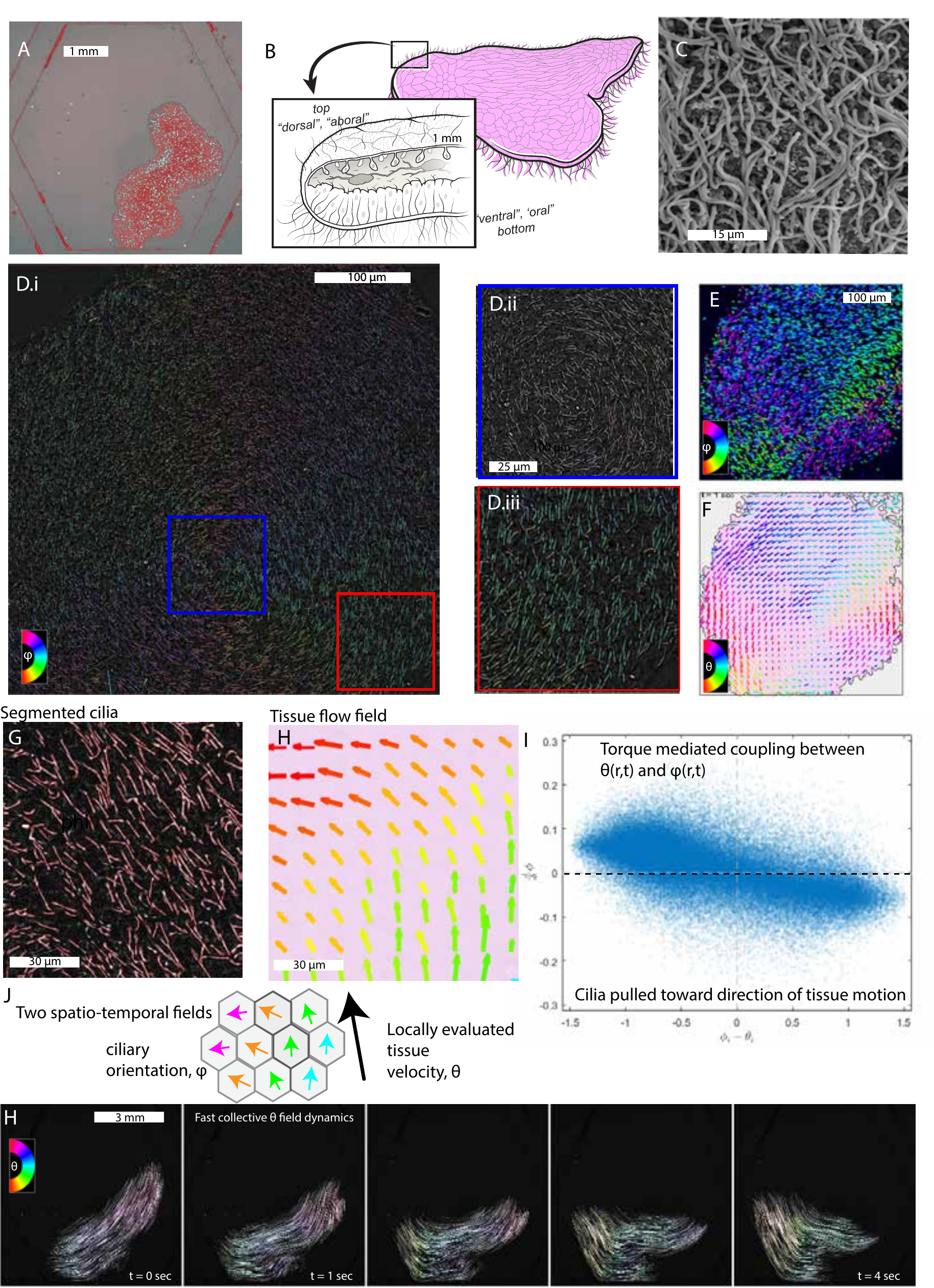}
  \caption{FIG 1. (Caption next page.)}
  \label{fig:fig1}
\end{figure}

After the removal of the long-wavelength noise using a numerical bandpass-pass filter (MATLAB, 1-10 pixels), the dynamics of the in-focal-plane component of the cilia become more visible and computationally measurable. By coloring the pixels by the orientation of of their minimum eigenvalue of the image Hessian, we can begin identifying the bulk orientation of these ciliary fields and observing deviations from polarized order in a single snapshot of this highly dynamic field. Some of the most striking patterns are vortex-like whirls in which the ciliary field appears to circulate around a single point (see figure \ref{fig:fig1}Dii). 

These trans-illumination fields are directly corroborated by parallel experiments conducted with epi-illumination using the membrane dye, Cell Mask Orange [Thermo Fisher, 1/1000 ratio of stock solution into Artificial Sea Water]. The fluorescent marker is active only when incorporated to the membrane allowing for long time period imaging without insurmountable photobleaching at continuous excitation illumination [Lambda XL 80\% through excitation filter]. These dynamics are consistent with that seen in the above experiment up to the long-exposure-induced stretching of the underlying ciliary dynamics. [See Supplementary figures].

\subsection{Subsecond ciliary reorientation}
Despite many advances in microscopy, the high temporal resolution imaging a large highly dynamic tissue ($um/s$ displacements) at single organelle resolution remains an experimental challenge. Here, we use a combination of fixed plane and tracking imaging sampled at high frequency to directly observe ciliary dynamics in a single plane. We are fortunate that the characteristic frequency of placozoan ciliary beating is about 5 fold lower than the more commonly observed 30 Hz among model ciliates\cite{gilpin_multiscale_2020} allowing us to directly image the dynamics of this field cilia as the animal crawls along the substrate. In part 1 of this series of manuscripts, we show that watching these dynamics revealed a delicate yet essential interplay between height and locomotive forcing with implications on collective locomotive control and sensing of environmental conditions. 

In this work, we adopt numerous imaging perspectives to develop our model by construction of the dynamics of ciliary reorientation which plays a complementary role to the dynamics described in part 1.

In one guiding series of experiments, we used both high temporal resolution bright field and fluorescent imaging of cell membrane markers to image the ciliary dynamics from the bottom of the organism. By choosing a single axial plane, we watch the dynamics of the organism as it moves through a stationary lab frame of reference. 

We can slow down the reorientation dynamics of the organism by squeezing it between two substrates using a 15 $\mu m$ spacer. In this imaging configuration, we can stretch normally sub-second events into much longer timescale reorientations. This technique enables us to measure with higher sampling the reorientation of the cilia under tissue applied torques. In these movies, we observe many examples of cilia 'taking a step' in one direction and then another in a new orientation after torque mediated alignment (see SI fig SIfig 1a). We use this as the first of two supporting data for the hypothesis that ciliary orientation in plane can serve as a reliable proxy for ciliary beating direction.

\subsection{Mapping ciliary orientation structure and locomotive behavior}
The tissue moves in response to the ciliary forces, but how do the ciliary forces evolve in time? We sought out an experimental link between the evolution of ciliary orientation and the motion of the tissue. To design an experiment to close our understanding, we first had to recognize that our data is rich in information about both of these interacting fields. In the high spatial frequency components, we carry measurable information about the local ciliary orientation. Whereas- the lower-spatial frequency fields (> 10 pixel wavelengths) carries slightly blurred information about how the the tissue is displacing. By extracting these parallel channels of information from the same image, we can monitor the simultaneous evolution of both the ciliary orientation field $\vec \phi (\vec r, t)$ and the tissue displacement field $\vec \theta (\vec r, t)$.

First, we separate the high spatial frequencies and the low spatial frequencies of the image using parallel applications of the bandpass filter (MATLAB). The high spatial frequency image is processed to segment cilia using a line detector which studies the local phase symmetry of the image (MATLAB, phasesym.m from \cite{kovesi_symmetry_1997}). This routine transforms the asymmetric structure into a local intensity which can be subsequently segmented using a locally adjusted image threshold. Using morphological operations, we can extract the orientation, center and intensity of these cilia in the optical plane. Orientation is corroborated by the direction of the local image gradient which can be calculated only within the masked regions of the successfully segmented cilia. When the two measurements disagree, the corresponding datum can be marked with higher uncertainty. 

Once, the cilia are segmented and the orientations are extracted, we bin the orientations over a 32 by 32 pixel square and report the local average of the orientation field $\vec\phi(\vec r, t)$ for each bin in the frame. 

With the low spatial frequency images, we apply a digital image correlation to find the local displacement field for each portion of the tissue using an iterative algorithm which ends with 32 x32 pixel bins summarized by a locally calculated displacement angle and magnitude\cite{Thielicke2014}. To extract how the two fields interact, we plotted a comparison of the local disagreement between the two fields with respect to how quickly the ciliary orientation field evolved at each point in time. 

\subsection{Torque mediated reorientation self-organizes these coupled fields}

The resulting curve in Figure \ref{fig:fig1}I shows a negative correlation between the velocity of the orientation and the difference in the angles. This direct measurement suggests that the dynamics of the ciliary orientation are driven predominantly by the changes in the direction of the tissue motion above. In this way, these dynamics can be compared analogously to a castor wheel, where motion of the base of the wheel induces reorientation of the wheel itself as it pulled into alignment. The caveat here is that these castor wheels are injecting their own power and thus it may be better captured by a "motorized" castor wheel. We use this analogy of a reorientable "castor wheel" throughout our work only as an illustrative example - since this is the simplest framework with sufficient degrees of freedom to exhibit ciliary flocking. The molecular mechanisms of ciliary re-orientation; such as either at the basal body or in the buckling of the cilia itself - are currently under investigation. Previously, basal body reorientation has been seen in a ciliate Opalina \cite{tamm_relation_1970} - but here we describe the very first example of ciliary re-orientation in a metazoan. 

\subsection{Ciliary flocking governs the locomotive behavior of the animal}

The combination of these results leads us to interpret the highly dynamic tissue motion of \textit{T. Adhaerens} as a direct consequence of the collective action of the underlying subsecond ciliary reorientation. In relation to similar manifestations in schools of fish \cite{tunstrom_collective_2013}, birds \cite{attanasi_information_2014} and tissues \cite{giavazzi_flocking_2018}, we term this phenomena 'ciliary flocking'. 

In the remainder of the paper, we set out to understand how this flocking behavior can give rise to such coherent and rapid locomotive behavior. Does it borrow the same innovations such as behavioral inertia \cite{attanasi_information_2014}, a negative viscosity \cite{slomka_geometry-dependent_2017} or does \textit{T. Adhaerens} have something new to teach us?

\section{A minimal model by construction}
\subsection{Ciliary walking meets torque mediated reorientation}
Experimentally, we have observed the interplay of two fields: the ciliary field $\vec \phi(\vec r, t)$ and the tissue displacement field $\vec \theta(\vec r, t)$. Our experimental allows us to measure the essential coupling between these two. The evolution of the cell-displacement field is precisely the manifestation of locomotive behavior in \textit{T. Adhaerens}. Unlike many other models systems for studying animal behavior where the internal dynamics is mapped via complex encoding schemes to motor neurons and hence animal activity; at short time scales, the ciliary dynamics of \textit{T. Adhaerens} map directly onto the manifestation of observable behavior including locomotion. This remarkable insight and natural seperation of time scale - allows us to build detailed models of mechanics of ciliary dynamics in large two-dimensional sheet and map it to mechanical information propagation in an entire animal. 

There are numerous types of governing dynamics that one can consider for how each of these fields evolve. As a start, we have shown experimentally that the orientation of the ciliary field is mediated by the tissue elasticity (cells that anchor the cilia into a confluent tissue) which allows us to map the simplest architecture of torque mediated reorientation of a cilia interacting with a substrate:
\[
\partial_t \phi_i(\vec r,t) \sim -\ell\hat\phi_i(\vec r,t) \times F_i
\]
where $\ell$ defines the length scale between point of contact and cell body of a mono-ciliated epithelial cell, $\hat\phi_i$ is the direction unit vector associated with the ciliary orientation, and $F_i$ is the force arising from the other cells on the tissue at the $i$th cell. This formulation is consistent with ideas in other domains such a bird flocks \cite{szabo_phase_2006, ferrante_elasticity-based_2013, giavazzi_flocking_2018}; although our system provides an explicit mechanical parameters to this reorientation dynamics. In a more general form the tissue force can be written as the gradient of an elastic configuration energy: $F_i = -\vec \nabla_r E$. Note that in the overdamped limit (with zero momentum transients) the net force can be transformed onto the local cell speed through a linear relationship between the terminal speed and the instantaneously applied force. 

Also note that these dynamics differ from many of the classic flocking models where there is a direct coupling between locations of the orientation field $\phi_i$ and $\phi_j$ \cite{tu_sound_1998, attanasi_information_2014, cavagna_silent_2015}. In some formulations it has been suggested that, keeping track of the medium through which the communication occurs (the tissue forces) is superfluous \cite{tu_sound_1998}, but we will show the important qualitatively distinct role the elasticity plays later on in generating a richer array of transients. Next, we can discuss how to account for unique aspects of "cilia" as a fluctuating force generator. 

\subsection{Flocking ciliary oscillators (aka Swarmalators)}
In part 1 of this series\cite{bullpart1}, we demonstrated that the organism walks on surface using cilia. This ciliary walking is punctuated by a slipping-stalling transition mediated in part by the local tissue height. With the goal of making this paper sufficiently self-contained, we adopt an idea from this work as an axiom which we encourage the reader can interpret on its own ground in the context of part 1. 

Here, we take the axiom: "walking" cilia (as described in part 1 \cite{bullpart1}) generate a time varying active force centered around a dominant beat frequency. These time varying forces communicate via a classic phase response curve mediated through tissue forces. This starting point is born out of independent work presented in part 1 \cite{bullpart1}. The approximate functional form of these phase response curves of perturbed oscillators often fall into two camps 1) always positive, or 2) a crossover from negative to positive \cite{cestnik_inferring_2018}. In this work, we present our efforts to understand the collective dynamics of force mediated phase oscillators coupled by phase response curves of the second type. 

The next essential ingredient to understand ciliary flocking is the experimentally measured reorientation of ciliary step generation. We can minimally encode this with a very simple degree of freedom, $\phi_i$ the active force direction. The combination of these two ingredients above gives rise to a single 'cell' model which can be classified by its direction of travel $\phi$ and its amplitude of forcing, $a(t)$. Further, we define the phase angle of ciliary oscillator by an oscillator phase field $\psi(r,t)$ and elastic forces acting on individual cells $|F(r,t)|$


These dynamics then take the form of a phase angle $a(t) = a_o(1+\epsilon cos(\psi(t)))$ where the evolution of the phase angle is governed by a simple stochastic differential equation:
\[
\partial_t \psi_i = \Omega + Z(\psi)\vec F_i\cdot \hat \phi_i + \xi_{\psi}(t)
\]
where we can choose a minimal model for the phase response curve $Z(\psi) \sim sin(\psi)$ \cite{dayan_theoretical_2001}. We also define $\Omega$ as related to the native frequency by the division of a $2\pi$, $F_i \cdot \hat \phi_i$ is the force on the cell in the direction of the ciliary orientation and $\xi_{\psi}(t)$ denotes noise in the system. 

The simultaneous evolution of the dynamics takes the form of:
\[
\partial_t r_i = (1+\epsilon cos(\psi(t)))\cdot \hat\phi_i + F_i
\]
\[
\partial_t \phi_i = -\Gamma \hat\phi_i \times F_i
\]
\[
\partial_t \psi_i = \Omega + J sin(\psi) \vec F_i \cdot \hat \phi_i + \xi_{\psi}(t)
\]

where we define three non-dimensional quantities $\Gamma$ which is the ratio between timescales of reorientation versus displacement, $J$ which is the ratio between timescales of phase change and displacement, and $\epsilon$ which controls the normalized amplitude of driving. Recall that $F_i$ is the gradient of the elastic energy and can applied to more diverse tissue elasticity functions. For simplicity, we apply it to a grid of springs governed by the energy function: $E = \sum_{\text{neighbors } i, j} \frac{k}{2} (|r_i - r_j| - l_o)^2$.

This model represents a more general system of flocking oscillators with conceptual links to the emerging literature on 'swarmalators' \cite{okeeffe_oscillators_2017}. We caution, that our flocking oscillators model do not abide by the same coupling choices and thus fall in a distinct class. To try to avoid confusion, we will call them 'flocking  oscillators'. 

\subsection{Self-organizing collective dynamics of the tissue}
The first question we set out to answer: are there any non-random steady state solutions to these many-body non-equilibrium dynamics in a system of flocking cilia? To study this numerically, we first initialize our in-silico model to random initial conditions (in orientation, phase and tissue force) and allow the dynamics to evolve forward via an Euler-Maruyama integration with a sufficiently small step size to maintain stability ($dt = 1e-3$). In a numerical experiment with the choice of the parameters: $\epsilon = 0.5, J = -10, \Gamma = 10$, and $\eta = 1e-1$, we observe fast self-organization of the random initial state into a symmetry broken system with cilia oriented and aligned to facilitate movement of the entire tissue copherently in one direction, although with no activity synchronization. 

By visualizing the time evolution of these two fields ($\phi$ and $\psi$), we can see that their progression follows a route characterized by the coarsening of defects akin to spin waves and supports the conceptual connection to the relaxation of spins in magnetic systems\cite{cavagna_silent_2015}. In these numerical experiments, we can further visualize the tissue force field (stress and strain in the elastic network) as a measure for the time integrated disagreement which puts emphasis on the locations of these local defects and shows a significant reduction in amplitude as the tissue comes into greater agreement. 

\subsection{Two distinct crossovers: polarization and synchronization}
Conducting this experiment for many choices of parameters sweeps out a rich phase space characterized by the three summary statistics: i) the mean polarization, ii) the mean synchronization and iii) the mean force amplitude acting on the cells across the tissue (Figure \ref{fig:fig2}G,H). These choices of diagnostics in the qualitative behavior of the model capture two distinct crossovers which have well-defined (yet coupled) relationships to the underlying parameters. We find numerical evidence for all four qualitative behaviors: polarized and synchronized, polarized only, synchronized only, and unpolarized and unsychronized. 
As a simplification, when $\epsilon \rightarrow 0$ and phase does not play a role, we recover a classic model for an active-elastic sheet \cite{ferrante_elasticity-based_2013} with fixed forcing directly analogous to the castor wheel model. This model's behavior is characterized primarily by the dynamics of non-dimensional parameter $\Gamma$; defined as reorientation-displacement time scale.

When $J, \Omega << \xi_{\psi}$, the dynamics of the activity become the integral of a stochastic process. These dynamics take the form of an active elastic sheet driven by activity noise. These dynamics undergo a order-to-disorder transition with increasing activity noise. 

When $J$ is positive (and $\epsilon \neq 0$), the oscillatory forces will have a tendency to synchronize in phase $\psi_i \rightarrow \psi_j$, while when $J$ is negative, the oscillator forces will synchronize anti-phase $\psi_i - \psi_j \rightarrow \pi$. This anti-phase synchronization on a long-range lattice is likely to exhibit the hallmarks of frustration\cite{gilpin_multiscale_2020}. In the intermediate regimes, the noise will dominate and the system will remain largely unsynchronized.

We also note that there are deep mathematical connections between the reorientation and coupling models established by a transformation in to a rotating frame of reference. Thus some of the observed similarities between the time evolution of these fields may be attributed to this deeper connection. 

\section{Active elastic oscillations and waves}
"{\textsl{The nerve-force is transmitted from molecule to molecule by some sort of vibritory action, as sound is transmitted through a stretched wire.}} - {Henry Bowditch \cite{cobb_idea_2020}}" - Mislead as he was, Bowditch's 1886 devotion to century old ideas, reflects the perceived power of information transmission via vibrations. However, early detractors, such as Alexander Monroe in 1749, had pointed out that "the nerves are unfit for vibrations because their extremities ... are quite soft and pappy"\cite{cobb_idea_2020}. In this section, we will show how even soft and overdamped tissues can transmit vibratory information through the combination of polarized active injection of locomotive forcing and overdamped elasticity. 

While the animal kingdom's nervous system may have foregone 'vibratory action' to pass information over long distances, the mechanism for information passing might be ripe for the collaborative dynamics of animals, like placozoa, which have no neuro-muscular system. In this section, we report on the discovery of tissue scale waves in the tissue without an obvious behavioral inertia\cite{attanasi_information_2014}.

The dynamics of cellular displacement in a large-to-intermediate-size placozoan is punctuated by dramatic changes in direction, the creation and collision of quasi-particles and the constant battle for the tissue to come to locomotive consensus. In this work, we focus our attention on two phenomena: (1) the effective underdamped response of the tissue to perturbation and (2) the emergence of organism scale traveling waves arising from the competition between activity and elasticity. Another parallel publication (part 3) studies the emergence of low-order models from the study of quasi-particles.

Our study of these dynamics begins by imaging a collection of 5$\times 10^5$ cells walking together as a single animal without a neuromuscular system. We confine this organism to a 10mm scale microfluidic chamber (Hexagonal) to keep the organism in the field of view and reduce tissue buckling out of plane. These PDMS chambers are fabricated curing PDMS [DOW] over a prefabricated positive tone photoresist (SU-8 [Kayaku Advanced Materials]) mold on 100mm Silicon wafers. 

This data is collected at 10Hz sampling rate on a upright microscope [Nikon AZ100, 5x objective with 3x optical zoom] in a epi-illumination configuration through a Texas Red filter set [Thorlabs, MDF-TXRED].
\subsection{Tracking chemically bonded fluorescent micro-spheres}
Each organism placed into the flowcell is coated with a coating of WGA-coated-beads\cite{prakash_motility-induced_2021} designed to firmly bind to the upper epithelium. The controllable density of these bound beads creates an dynamic array of randomly placed fiducial markers which can be tracked with high spatio-temporal precision over long periods of time at organism scale fields of view with reduced photo-toxicity \cite{prakash_motility-induced_2021}. 

We use Flowtrace\cite{gilpin2017flowtrace} to project the displacement dynamics into a sequence of simulated 'long exposure' images where the motion of the bright bead streaks out a visible trajectory. These trajectories carry more information than the instantaneous dynamics reconstructed by methods such as particle image velocimetry as their shape represents the time integral of the position over the past 2 seconds.  This enables us to directly visualize patterns such as 'squiggles' or 'loops' present in the displacement time dynamics. These higher order patterns (beyond the position and first derivative of the displacement) encode the evolution of otherwise hidden dynamical signatures. We use these properties of the Flowtrace method \cite{gilpin2017flowtrace} to visualize signatures of an underdamped oscillation in the tissue frame of reference and organism-scale traveling waves with strongly direction dependent speeds. 

\subsection{Measuring perpendicular acceleration of the tissue to identify traveling disturbances}

We also employ conservative particle tracking [MATLAB, Nearest neighbor] to extract long time series tracks of individual beads over the course of 100's of seconds in particular windows of these hour long movies. We use these tracks in three primary functions during this analysis. 

\subsection{Active elastic oscillations of an overdamped system}

The first, we project the displacement in the direction perpendicular to local tissue velocity and extract the spectral content of the the perpendicular velocity field via a power spectral density [Python]. These power spectral densities reveal a notable peak (which is preserved across experimental data-sets of similar size organisms) around 70 mHz. This frequency peak corresponds to the order 10 s period tissue oscillations in response to running into the walls of the confinement.

\subsection{Traveling waves in experimental trajectories}
Second, we employed the tracks to extract the spatial traveling speed of changes in direction. Treating each bead as an individual trajectory, we calculated the local acceleration of the bead projected along the direction orthogonal to the last displacement direction. This acceleration metric was then smoothed using a Gaussian filter to extract the persistent turning dynamics. Since this metric keeps track of a CW (clockwise) or CCW type of acceleration, we can disentangle direction changes which contribute to and oppose an organism-scale turns. 
\subsection{Anisotropic wave speeds}

By comparing the time of the extrema in these acceleration traces to their spatial location relative to a origin of the turn, we can extract the wave speed traveling across the tissue \cite{attanasi_information_2014}. Consistent with the anisotropic dispersion relationships of active matter \cite{tu_sound_1998, cavagna_silent_2015, yang_hydrodynamics_2015, geyer_sounds_2018}, we find that the speed of the disturbance arrival appears to be strongly dependent upon the direction of travel. We report the linear speed fits of each of these for discrete bins of the collected data to find that the maximum speed of propagation occurs orthogonal to the direction of tissue motion. [See figure \ref{fig:fig3} d]. 
\subsection{Projecting the bead dynamics into a low dimensional phase space with oscillations}
In a third use of the bead tracks, we can project both the displacement and the velocity onto the direction perpendicular to tissue travel. When we plot the average flow in this two dimensional space for all observed points, we find that the trajectories on average spiral inward toward the point (0,0) in this 2D space. The lack of a exactly radial response suggests that these degrees of freedom behave similar to an underdamped mass spring system, where a perturbation causes a circular flow around the fixed point before reaching it. This flow agrees with the observations that the projected ciliary forcing field follows the displacement of the tissue orthogonal to mean tissue motion (as we will demonstrate through derivation in the following section). 

\subsection{The history of wave phenomena in tissues}

The complex spatio-temporal dynamics of these multicellular collective locomotion (propelled by ciliary walking, see part 1), shows a significant behavioral repertoire of the tissue.

To understand these dynamics at a deeper level and as an effort to look past the complication of these rich dynamics to the underlying phenomena, we will first focus our inquiry on the response to turning. Large parts of the tissue exhibit a oscillatory response as a result of contact with the chamber walls. While the tissue is strongly overdamped, the response appears like an underdamped response to a poke and has connections to instabilities observed in numerical filament models. These dynamical signatures show up as $S-shaped$ profiles of the local tissue trajectories (translation plus oscillatory response).

There is a rich history of understanding the dynamics of oscillations within healing tissues \cite{banerjee_propagating_2015} and we have previously identified a mechanism for second scale linear speed waves propogate through the top layer of this tissue \cite{armon2018ultrafast, armon_epithelial_2020}. Yet, how these map onto the locomotive response of the organism is yet unknown.

We can experimentally view the manifestations of these oscillations as traveling waves capable of steering the tissue as a whole. We can visualize the manifestation of these traveling waves in the system by looking for spin wave like excitations that propagate over large distances in the tissue. It has been shown that equations governing a type of Laplcian of local disagreement result in waves that travel diffusively, and dynamics which are supplemented by a type of behavioral inertia can gives rise of dynamics which propagate information in a ballistic like manner (linear speed traveling waves). In this system do we share similar signatures in our dynamics as has been observed in flocks of birds (very different physical and biological systems)? 

\section{Wave phenomena in response to mechanical stimulus in active-elastic flocking with $\epsilon = 0$}

We can study this phenomena in a cleaner manner by conducting a number of \textit{in silico} experiments on the model we constructed in the form of a polarized active elastic sheet. To accomplish this, we take our flocking oscillator model in the limit of $\epsilon <<1$ to reconstruct the simplest case. In this limit, our model is related to models from Ferrante et al\cite{ferrante_elasticity-based_2013}), Szabo et al 2006\cite{szabo_phase_2006}, and Copenhagen et al\cite{copenhagen_frustration-induced_2018}. Our goal is to study the response to mechanical stimulus by applying a pulling force in direction orthogonal to motion of a single cell of the tissue initialized from a polarized state (as $\epsilon \rightarrow 0$, the $\psi$ field decouples from the rest of the equation).

\subsection{Traveling waves in an active elastic sheet}

The response of the \textit{in silico} flocking tissue is strongly dependent upon the non-dimensional parameter for the reorientation time of the active forcing, $\Gamma$. When $\Gamma >1$, the active forcing vector reorients quickly compared to the displacement of the tissue. In the numerical experiments, at small $\Gamma$ and $\Gamma$ approaching 1, the tissue responds with a heavily overdamped response consistent with pulling on a single degree of freedom in an network submerged in honey: The displacement from the equilibrium position grows slowly (exponential with a timescale determined by the $\sqrt{k/\gamma_{honey}}$) with no finite time extrema in the acceleration. When $\Gamma$ grows sufficiently large (shown for $\Gamma = 10$), the character of the response changes dramatically to support linear speed traveling waves with an anisotropic wave speed. [See figure \ref{fig:fig3}K]. 

Polarized active elastic systems can control their wave propagation by modifying the reorientation time relative to the displacement time under the same applied force. Less wave damping with higher relative reorientation speed. In placozoa, we suggest that this non-dimensional parameter, $\Gamma = L \gamma_{r}/\gamma_{\phi}$ can be controlled by the static ciliary curvature \cite{geyer_independent_2016} and the ciliary length, which taken together give rise to the tune-able length between the basal body and the ciliary point of surface contact. 

\section{A toy model: the active elastic resonator}
\subsection{Deriving the active elastic resonator by projecting the dynamics of a polarized active elastic sheet}
As shown above, despite overdamping, the system can display an oscillatory response with linear speed traveling waves. We derive a simple understanding of the model by entering the frame of reference of the moving organism. We can then look at a single cell in a reduce dimensional description (the coupling of activity and elasticity along a single axis - say x-axis). 

We can use our linear elasticity of the chain to define the overdamped dynamics of this system with a simple, in-line activity.
$$
\frac{\partial}{\partial t} \vec x = -\mathcal{K}\vec x + \vec\alpha_{\vec x}
$$
Here, $x$ is the list of displacements from the equilibrium position, $\mathcal{K}$ is the stiffness matrix, and $\alpha$ is the activity defined for each degree of freedom.

Let's assume the dynamics of the activity are governed by a simple set of equations relating a continuous degree of freedom (the orientation of the locomotive force): $\alpha_i = acos(\phi_i)$, where $a$ is a constant and $\phi_i$ is the angular representation of that activity vector. In our 1D problem forces which act perpendicular to the chain of cells are exactly canceled by the holonomic constraints (a constraint that allows one to reduce the degrees of freedom of the problem, e.g. fixed length of a pendulum rod).

The equation that governs this time evolution takes the form of:
$$
\dot \alpha_i = \frac{\partial \alpha_i}{\partial \phi} \dot \phi_i
$$ 
with
$$
\dot \phi = -\Gamma \ell \hat \alpha_i \times \nabla_x E
$$
where we have $\Gamma$ representing both the characterstic length and orientational damping parameter.

As a first pass, let's consider a single cell centered by a compliant spring. We can later expand to the full chain.

In this simple case, the equations of motion for the single cell takes the form of:
$$
\gamma \dot{x_i} = -kx + \alpha_i
$$
$$
\dot \alpha_i = -a \Gamma k x_i\left(1-\left(\frac{\alpha_i}{a}\right)^2 \right)
$$

This form generates a two dimensional phase space with $x$ position and $\alpha$ activity. The cooresponding null-clines occur at:
\[
\dot x = 0 \rightarrow \alpha_i = kx
\]
\[
\dot \alpha = 0 \rightarrow 
\begin{cases}
    \alpha_i = \pm a \\ x=0 
\end{cases}
\]
There are two noteable observations from Figure 3L:
i) The dynamics are quite reminiscent of a damped harmonic oscillator or damped pendulum.
ii) The decay to the stable $\{0,0\}$ fixed point are strongly dependent upon your choice of the linear damping to the angular damping, captured in this system with the choice of $\Gamma$.

\subsection{Exploring the phase space of the active elastic resonator}

We can study the qualitative character of this 2D dynamical system to learn more about the active elastic resonators response to stimuli. At low $\Gamma$, the system is reminiscent of an overdamped system taking an exponential time to reach the equilibrium position. This limit corresponds to slow reorientation of the active force relative to the response timescale of a cell in its viscous medium. Said another way, the slower the response of the spin, the more 'overdamped' the dynamics appear.

At high $\Gamma$ the response takes the form of a underdamped system oscillating around the equilibrium before reaching the fixed point. This spiral nature is strongly reminiscent of the 2D phase space, reconstructed from data in Figure \ref{fig:fig3}I-J. This suggests that the implications of having a fast reorientation observed in the ciliary dynamics may be critical for establishing an effectively underdamped response.

The simple interpretation of the fixed point at $(0,0)$ is that the energy will eventually couple into the system such that the active vector is pointing perpendicular to the axis of the degree of freedom thus no longer costing elastic energy. We expect that the system will return to this point under no stimulus or driving consisent both with the observations of the full polarized active elastic sheet model and that seen experimentally.

These result suggests that active matter systems in overdamped environments can indeed support complex oscillitory modes combining elasticity and behavioral inertia (this has been elegantly shown by Cavange et al \cite{attanasi_information_2014, cavagna_silent_2015} in the context of flocking starlings). 

\subsection{Linearization and mapping onto a harmonic oscillator}
If we linearize the dynamics of the Active-Elastic resonator, the surface level similarity to a simple harmonic oscillator can be mapped onto an analogous equation for the active elastic resonator. We begin by dropping the second order terms in the limit that $\alpha_i / a << 1$. By taking the derivative of $\dot x$ we find that we can write:
\[
\gamma \ddot x + k \dot x + a \Gamma k x = 0
\]
This is precisely the form of the simple harmonic oscillator if we redefine the terms. The effective mass of the active elastic resonator becomes : $m \rightarrow \gamma$, the effective damping becomes $\beta \rightarrow k$, and the effective stiffness of the mode becomes: $K \rightarrow a\gamma k$.

The $Q$-factor for a simple harmonic oscillator takes the form of: $
Q = \frac{\sqrt{Mk}}{D}
$
which can be mapped simply onto the $Q$-factor for the active elastic resonator.
\[
Q = \sqrt{\frac{\gamma a \Gamma}{k}}
\]

In the limit where the restoring force goes to zero (e.g. the stiffness of the excited active-elastic resonator goes to zero), the Q factor, as written diverges. Since the period ($T = 2\pi \sqrt{\gamma/ a \Gamma k}$) is also diverging, the system in this limit will not oscillate and will purely couple into the massless mode with zero restoring force. 

Massless modes can arise in either details of the tissue physics imposed by topological invariants\cite{huber_topological_2016, shankar_topological_2020} or rigidity transitions (both density dependent and independent)\cite{Bi2015a, kim_embryonic_2021} or can be associated with Goldstone like modes of the orientational degree of freedom of the activity vector\cite{attanasi_information_2014}. 

\subsection{Driving an active elastic resonator results in harmonic up-conversion}

A natural way to study a resonator is by looking toward its response to driving. While there is energy injection at the shortest length scale driving this system out of equilibrium (and causing it to translate), when we move into the frame of reference of the moving cell, this transformation maps this out-of-equilibrium system onto one where the balance of the energy injection and dissipation is equal and the dynamics take the form of an expansion around a fixed point. 

We can study this fixed point by looking toward the system response to periodic driving. In our derivation of this simple active-elastic resonator, we keep the nonlinearities associated with the saturation of the activity up the maximum speed of the active element. This nonlinearity manifests itself in a saturation of the 'momentum-like' term as the amplitude of the activity approaches the speed-limit of the active element. 

$$
\gamma \dot{x_i} = -kx + \alpha_i
$$
$$
\dot \alpha_i = -a \Gamma k x_i\left(1-\left(\frac{\alpha_i}{a}\right)^2 \right)
$$

The square dependence on the curve suggests that this nonlinearity behaves something like a Second Harmonic generation where the excitation at a given frequency results in harmonic overtones at twice the drive frequency. At sufficiently high amplitudes of driving, and $\Gamma$ reaching sufficiently high, the response will cascade power to higher frequencies. We can study this response numerically, by driving our simple active elastic resonator at a given amplitude and frequency for different values of $\Gamma$. We expect that as $\Gamma >> 1$, the system will respond by passing energy into higher harmonic generation. 

This harmonic generation shows an example of up-conversion of the input driving via small, amplitude dependent nonlinearities in the simple active-elastic resonator. 

However, in nature, there are few external stimuli that take such a predictable response. A series of future experiments could probe this type of resonant response of these active elastic resonators by driving the tissue at a known amplitude and frequency (orthogonal to the direction of motion) and measuring the orthongonal response.

\subsection{'Active-elastic phonons'}
Next, let's consider the dynamics of a chain of these active cells. Our equations are:
$$
\frac{\partial}{\partial t} \vec x = -\mathcal{K}\vec x + \vec\alpha_{\vec x}
$$
$$
\frac{\partial}{\partial t}\vec\alpha = -a \Gamma \mathcal{K} \vec x \left(1-\left(\frac{\vec\alpha}{a}\right)^2 \right)
$$

We can again toss away the nonlinear terms if $\alpha_i/a << 1$, which allows us to rewrite our dynamical system as in its analogy to the simple harmonic oscillator.
$$
\frac{\partial^2}{\partial t^2} \vec x = -\mathcal{K}\frac{\partial}{\partial t}  \vec x - a\Gamma\mathcal{K}\vec x
$$

This approximation gives us access to the full suite of quantities. The resonance frequency is of the form: $\omega_o = \sqrt{a \Gamma \mathcal{K}}$ and the Q factor takes the form of: $Q = \frac{2\sqrt{a\Gamma}}{\sqrt{\mathcal{K}}}$. Critical damping occurs at $Q_{critical} = 1/2$ which corresponds to $\sqrt{a\Gamma} = \sqrt{\mathcal{K}}$.

The implications of this linearization is that the transverse modes (perpendicular to the direction of the net polarization) support the idea that the lowest stiffness modes of the structure are the most 'underdamped' for fixed system parameters. This has interesting implications for the response of the tissue as it scales with system size.
$$
a(\ell) \sim \ell
$$

Since this system is translationally invariant under the peridic transfermation, we can apply Bloch's theorem to study its dynamics.

The simplest way to do so is to ansatz a solution for the form of a traveling wave with a characterstic spatial frequency, $k$. For this system of equations, we need both:
$$
\vec x = Xe^{\kappa x - \omega t}
$$
$$
\vec\alpha = Ae^{\kappa x - \omega t}
$$

In the limit of small $\alpha/a << 1$, our dynamics are governed by the standard second order dynamics of the damped Harmonic oscillator.

$$
\frac{\partial^2}{\partial t^2} \vec x = -\mathcal{K}\frac{\partial}{\partial t}  \vec x - a\Gamma\mathcal{K}\vec x
$$

The $\mathcal{K}$-matrix can be written in terms of a second order central difference in the finite difference approximation of the spatial gradient. This tells us that our dynamics takes on the continuum limit:
$$
\frac{\partial^2}{\partial t^2} \vec x = -k\frac{\partial^2}{\partial x^2}\frac{\partial}{\partial t}  \vec x - a\Gamma k \frac{\partial^2}{\partial x^2}\vec x
$$

In the limit that $a \Gamma >> 1$, we may obtain approximate solutions by neglecting the 'damping-like' term to find the classic wave equation.
$$
\frac{\partial^2}{\partial t^2} \vec x \approx - a\Gamma k \frac{\partial^2}{\partial x^2}\vec x
$$

In this limit, the wave speed is controlled by:
$$
c \sim \sqrt{a\Gamma k}
$$

The entire equation is then governed by the D'Almbertian and represents the effective linearized composite of the spin-waves hybridizing with the overdamped elastic modes. These active-elastic phonons propagate as a linear speed wave orthogonal to the direction of motion.

\subsection{The dispersion relation of 'active-elastic phonons'}
Using the above D'Almbertian, we can ansatz the solution of traveling waves of the form:
\[
x(\kappa,t) = A e^{i(\kappa x - \omega t)} + CC
\]
Plugging this into the linearized equations of motion gives us:
\[
\omega^2 \vec x = -a\Gamma k ( 2 - e^{i \kappa} + e^{-i \kappa} )\vec x
\]
Using the trigonometric identity we find that the dispersion relation on the 1D chain takes for the form of:
\[
\omega(\kappa) = \sqrt{4a\Gamma k} sin \left( \frac{|\kappa|}{2}\right)
\]
For small values of $\kappa$ which correspond to long wavelengths, the relationship is approximately linear. This means that low spatial frequency waves have an approximately linear or non-dispersive dispersion relationship.

One of the implications of this is that provided that $\kappa$ remains sufficiently small the wave speed is constant for all frequencies which gives us phase matching for second harmonic generation as second harmonic generation efficiency is maximized when the phase velocity of both waves is matched\cite{boyd_nonlinear_2020}.

\subsection{A difference between ciliary flocking and bird flocking}
Ciliary flocking is distinct from bird flocking. First, birds have been reported to leverage behavioral inertia in their turn curvature to generate linear speed, effectively undamped waves across the flock. This behavioral inertia has no analog in a ciliary system where the tissue force acts directly on the rotation of the active force vector and when removed the rotation speed drops instantaneously (on the relevant timescales of the problem) down to zero. Without the explicit inclusion of the tissue forces accumulating through displacement, we would naively expect the spin waves to propagate diffusely in accordance with a Fick's law type of equation. 

The ballistic active-elastic phonons observed in the above chain are consistent with ballistic phonon transport where the lengthscales of interest are shorter than the mean free path length of the phonons themselves. It is noteworthy to comment that the mean free path length can be reduced by phonon-phonon scattering, or quenched disorder in the medium which can initiate a crossover between effectively ballistic and effectively diffusive transport of these phonon type waves. 

\section{Cilia generate periodic forcing}
A second major difference between ciliary flocking and bird flocking is the amplitude of the speed fluctuations relative to the mean speed of travel. While birds must maintain a sufficiently high speed to remain in flight, a cilia is marked by its periodic driving defined by the power and recovery stroke respectively. The time-averaged force is in the direction of the orientation, but the periodic fluctuations can be substantial\cite{hill2010force,ma2014active, klindt2016load, gilpin_multiscale_2020}. Coupling this periodic force generation with an overdamped environment means that the resulting velocity is prone to more substantial fluctuations. 

In this section, we make measurements of ciliary force fluctuations and study the potential impact of periodic forcing on the collective dynamics of the tissue by using our toy model to link activity driving to a parametric amplification which can be employed by the animal to increase its agility. 

\subsection{Lyso tracker data allows for more direct measurement of changes to ciliary forcing}
Briefly, the bottom tissue of \textit{T. Adhaerens} is comprised of an epithelial mixture of cell types with about 10\% comprised of large lipid filled cells which are well stained by a lipophillic fluorescent dye [LysoTracker Deep Red, ThermoFisher] \cite{prakash_motility-induced_2021, Smith2014a}. Staining the tissue, results in a uniformly distributed sparse sampling of this highly dynamic tissue. By filtering, segmenting and tracking these cell locations as a function of time [MATLAB] we can extract the periodic fluctuations in the direction of tissue motion using our previously described power spectral analysis [python, scipy, N = 1183]. 

While these experiments are limited by a high noise floor and shorter tracks relative to beads experiments outlined in figure \ref{fig:fig3}, we can extract two critical phenomena from them. The first is that the 70 mHz peak is consistent with the data and the second is that there is a new peak at $\sim 0.3 Hz$. It is noteworthy that this new peak presents close to a factor of 4 times the original peak identified in the beads experiments. 

\subsection{Beads in soft support 1Hz periodicity in forcing}
To complement these experiments on tissue response and to push the noise floor down at higher frequencies, we conducted a complementary set of experiments measuring the active force induced fluctuation of beads suspended in soft substrates. Traction force microscopy is an established technique to measure nN scale forces arises from cells crawling on substrates \cite{Grell1974ElektronenmikroskopischePlacozoa}. By suspending fluorescent beads [1um, Flurospheres Red, Thermofisher] in low concentration, agar [0.35 \% wt] films, the cells have induce elastic deformation which can be precisely measured by tracking the displacement of the fiducial marker beads.

By allowing the organism to walk over the top of these embedded markers, we note a strong increase in the visible fluctuations of the particle position aligned in the direction of organism travel (imaged at 20Hz). By extracting the position of these beads as a function of time, we can calculate their displacement from the unforced position. Even without a direct calibration of the elastic modulus (which we expect to be in $<30$ kPa range) the position can be mapped as a scaled proxy for force \cite{toyjanova_3d_2014}. 

Extracting the power spectrum from these traces shows a defined peak in the neighborhood of 1 Hz. This quantitatively agrees with the extracted step frequencies found in part 1 of this series \cite{bullpart1}, and corresponds to a peak in the distribution of frequencies for tissues which are locomoting in a highly polarized state (flocking).

\subsection{Power in three peaks}

The frequency content of the underlying dynamics of the tissues and the response of the underlying substrate shows up in three main peaks: 70 mHz, 300 mHz and 1Hz. In the following sections, we explore if this approximate $16x$ to $4x$ to $1x$ relationship between these frequencies can be accounted for via energy passing between the frequencies through a simple form of parametric down-conversion \cite{boyd_nonlinear_2020}. 

\section{Parametric driving of an active elastic resonator}
External periodic driving, while interesting, is not an essential form of out-of-equilibrium dynamics to study for this form of active elastic-resonator. These simplified systems (motivated by the dynamics we see in a placozoan) are instead driven out of equilibrium by velocity and length oscillations (given by the ciliary beating and the emergent switching timescales given by the ciliary bistability). 

We can account the observation that the activity and length(torque) are constantly changing with height fluctuations. These height fluctuations have important implications for the behavior of the system and include parametric amplification on resonance. Coupled with the saturating nonlinearities of the problem, these dynamics take the form of spikes in the displacement (see figure \ref{fig:fig4}F). 

Parametric oscillators are a simple nonlinear model for a harmonic oscillation which includes a time-dependent parameter (often a modulation of natural frequency, e.g. $\omega_o(t)$). 

A small subset of the phenomena associated with a parametric oscillator include: parametric amplification, parametric mode coupling, parametric down-conversion, and mechanically squeezed thermal fluctuations. 

Quite intriguingly, our above active-elastic resonator takes the form of:
$$
\gamma \dot{x} = -kx + \alpha
$$
and then with $\alpha = ae^{i\phi}$, this means that $\dot \alpha = \dot a \hat \alpha + a\dot{\hat\alpha}$
$$
\dot{\hat \alpha} = -\Gamma \ell \hat \alpha \times \nabla_x E
$$

Notably, assuming $\dot a = 0$ and $\alpha/a << 1$ gives rise to a system of equations reminiscent of the damped harmonic oscillator. If we relax the constant speed assumption, we get something of form (for our 1D problem in the perpendicular frame of reference of the tissue):
$$
\dot \alpha = \dot a \hat\alpha - a\Gamma\ell k x
$$

From our class of single cilia models interacting with a substrate, we know that there are multiple timescales at play in this problem. There are fluctuations in speed which arise from the simple mechanics of ciliary beating (the beat frequency), there are longer-timescale fluctuations (which are fed-back upon by a torque mediated height dynamics and cycle-averaged force dependence upon height).

In the simplest parametric oscillator, the frequency of driving is controlled by an external pump. 

In an effort to build intuition for these dynamics, we begin by studying the low order model where the power injected is modulated by the ciliary beat frequency (which can be height dependent). Subject to $\alpha/a << 1$, we can write this as:
$$
\gamma \dot{x} = -kx + \alpha
$$
$$
\dot \alpha = - a(t)\Gamma\ell k x
$$
$$
a(t) = \epsilon sin(\Omega t)
$$
where $\epsilon$ controls the magnitude of the frequency pump and $\Omega$ controls the pump frequency.

\subsection{Modulating activity alters the effective stiffness of the active elastic resonator}
Periodic activity driving of the active elastic resonator has the effect of modulating the effective stiffness in fashion which is linearly proportional to the driving: $K_{eff}(t) = a(t) \Gamma k$.

\subsection{Parametric amplification}

When the accompanying mode is entrained to the parametric driving, rapid modulation of the effective stiffness can pump energy into the oscillator. Imagine a case where the mass spring system is modulated. If the spring is stiffer as the mass moves toward the center position ($x=0$) than it is when it moves away $x=0$, the spring will add more momentum to the system than it takes away. This means that each swing past the center position will increase in amplitude proportional to the amount of force experienced the mass spends approaching $x=0$. 

This means that there is zero amplification for a periodically driven system at $x=0, p=0$, but the amount of power pumped into the mode increases as the mode amplitude increases. In the presence of dissipation, this parametric amplification can overcome the damping at a critical amplitude pushing the system amplitude up toward the saturating nonlinearity of the system. 

When we plot up these dynamics in response to a small initial displacement, we find that under-damped active-elastic resonators can be amplified up to a saturated limit cycle (see figure \ref{fig:LCs}.)

\subsection{Parametric pumping generates spikes and oscillations in mode amplitude}
We describe the resulting timeseries of periodic active driving of AERs in three classes: i) at low driving amplitude and/or small stimulus, the effective damping overcomes the amplification and the system relaxes to the fixed point at (0,0), ii) at intermediate driving amplitude and intermediate stimulus size, the amplitude of the mode increases rapidly initially due to parametric amplification, but the mode does not entrain to the driving and the mode amplitude ultimately decays back to (0,0), and iii)  when the driving amplitude is high and the amplitude of the stimulus is large, the mode amplitude amplifies up to the saturating non-linearity and stabilizes on a periodic limit cycle. We term the significant growth of the mode amplitude as a mechanical 'spike' and the large amplitude continuous oscillations. 

\subsection{Parametric driving induces a bifurcation unfolding a fixed point into a limit cycle}

The interplay of amplification and nonlinear saturation is a widespread combination in excitable and oscillatory dynamical systems. In our system, the amplification arises from parametric driving and the nonlinear saturation arises from the finite maximum of the active forcing. This combination is sufficient at high driving amplitude to effectively transform the stable fixed point at $x=0, \alpha = 0$ into a stable limit cycle which oscillates around this fixed point. The crossover between these two limiting cases gives rise to a bistable system, where the fixed point at $x=0, \alpha=0$ is stable with a small basin of attraction but the above a threshold of amplitude, the limit cycle attracts the dynamics. This coexistence of behaviors represents an opportunity for tuneable multistable dynamics at the level of single mode amplitudes. 

\subsection{The Mathieu stability diagram}

These dynamics are intimately related to the damped Mathieu equation. These equations in their damped form develop the basis for more general Floquet analysis of periodically driven systems and illustrates a sub-harmonic (or down-conversion) process in which modes which are $1/4$ of the driving frequency are resonate with the parametric driving. 

We can begin to characterize the perturbation response of the parametrically driven active-elastic resonator (when the parametric driving is near to 4$\omega_o$).

Traditional parametric amplifiers increase the energy in the system proportional to the energy already in the system. For this reason, the oscillator with zero displacement and zero momentum will remain stable. However, the Mathieu stability diagram studies the stability of this fixed point by studying the eigenvalues of this stability matrix. When the eigenvalues of this linearization go positive, the point becomes unstable. For the traditional equations, this can be represented as a small 'U' shape (for the damped Mathieu equations) of instability in the parameter space defined by $\epsilon$ and $\Omega$). The frequency $\Omega_\star$ at which the critical $\epsilon_c$ is the lowest lies around 4x the resonate frequency of the oscillator. We define $\Omega_\star$ as the parametric resonance.

To study the dynamics of the parametrically driven active-elastic resonator, we conduct a simple numerical experiment. We begin by choosing a single set of parameters, $\Gamma, k, a$ for our active-elastic resonator. Next, we drive the this resonator with the parametric driving of $a(t) = a(1-\epsilon sin(\Omega t)$. Finally, we mimic a perturbation from equilibrium by setting the initial displacement equal to $x_o$. By monitoring the maximal and sustained evolution of the system response, we can learn about the stability of the whole system to a given perturbation. A timecourse representation of this experiment is visualized in figure 4F.

\subsection{Driving generates a tuneable switching threshold}

The parametrically driven AER exhibits an interesting form of excitable dynamics. The combination of parametric amplification and the saturation of the activity (e.g. a maximum to the amount of force the cell can exert) gives rise to a stable limit cycle. Simultaneously, the parametric driving is not strong enough to overcome the damping to destabilize the point at (x = 0,$\alpha$ = 0). 

This allows us to characterize the response amplitude for a given set of parameters (fixed $\Gamma = 10$, varying $\epsilon, \Omega$ and $x_o$). By plotting up the maximum displacement achieved over the observed timecourse (see figure 4H), we can identify two canonical responses: (1) those that relax down to zero displacement and (2) those which can be excited up to the large amplitude limit cycle. Using this framework, we find that for a fixed drive frequency and amplitude ($\epsilon = \{0, 1\}$ and $\Omega = 4\omega_o$), the excitability behaves like a threshold response (below $T_{excite}(\epsilon)$, it relaxes to x=0 approximately as an underdamped AER where above the threshold it jumps up toward the excited limit cycle). These excitable dynamics hint toward a mechanical mechanism for tuneable sensitivity to external stimuli. 

\section{Generalizing the active elastic resonator to 2D}
While the 1D active-elastic resonator is an illuminating toy model which evokes connections to a wealth of physical topics, the details of the projection for deriving the model plays an important role in the manifestation of the phenomena. In an attempt to look for support of this phenomena outside of the assumptions of the 1D active-elastic resonator model, we develop an intermediate complexity model which avoids the details of the projection into 1D space entirely. 

The primary idea we employ in the development of this intermediate model is moving into the frame of reference of the single cell. Here, the single cell walking in 2D interacts directly with a larger entity which represents the tissue. The cell has four degrees of freedom reflecting the minimal model by construction from the known observations. We assume a minimal construction of the tissue which only keeps track of its current velocity and heading in 2D space. In total the model has 6 degrees of freedom.
$$
\dot{\vec r}_{cell} = -k ({\vec r}_{cell} - {\vec r}_{tissue}) + \vec\alpha_{cell}
$$
$$
\ddot{\vec r}_{tissue} = -\zeta_{tissue}(\dot{\vec r}_{tissue} - \dot{\vec r}_{cell})
$$
$$
\dot{\vec \alpha} = -\Gamma k\vec\ell_{cilia}\times ({\vec r}_{cell} - {\vec r}_{tissue})
$$

where $\zeta_{tissue}$ controls the response time of the tissue to changes to in the cell-scale forcing. We call the inverse of $\zeta_{tissue}$ the behavioral inertia of the tissue where $\dot{\vec r}_{tissue}$ represents the equilibrium position of the cell within the tissue. There can be disagreement between the cell and tissue locomotion which is penalized by a build up of elastic energy. $\ell_{cilia} = \ell \hat \alpha$ in the offset between the basal body and the point on contact with the surface.

\subsection{The emergence of a Goldstone-like mode controlling tissue turning}

In studying this intermediate model, we have introduced an important new conceptual feature: the presence of a massless (energy cost is zero) mode (a.k.a. mechanism). This massless mode arises from the spontaneous symmetry breaking of a continuous degree of freedom. For this reason, it can be classified as a Goldstone mode and is most commonly visualized as the equal energy well around the wine-bottle potential. The physical context of this model suggests that this mode is essential for turning the entire organism.

The timescales of these processes are critical for classifying the qualitative dynamics. The primary two tools for tuning the timescales of the relative processes is altering the inverse of the tissues behavioral inertia $\zeta$ and the amplitude of the cellular active forcing $a$.

When $\zeta << 1$, the behavioral inertia of the tissue is large resulting in a single cell with very little influence over the tissues direction or speed of travel. This results in a cell effectively bouncing around the translating mean position imposed by the tissue. We show that this limit is highly conducive to active elastic oscillations directly analogous to those observed in the toy model.

In the limit where $\zeta >> 1$, the behavioral inertia of the tissue is very small, and the tissue is highly influenced by the dynamics of the cell, we can set the direction of travel of the tissue equal to that of the cell. The qualitative dynamics of system in this limit is characterized by the persistent random walks of an active cell. The most relevant feature of this limit, is that the new Goldstone modes absorb all the fluctuations and no active-elastic oscillations are present. 

At intermediate values of $\zeta \sim \mathcal{O}(1)$, the Goldstone mode leaks power from the active-elastic resonator mode, and as we tune $\zeta$ the model exhibits a crossover between the two limiting behaviors. We study this crossover in figure \ref{fig:fig4}K,M and N, where we define a correlation measurement of underdamped oscillation and turning amplitude and report these summary values for simulations conducted along a grid search of $a$ and $\zeta$. These results support the idea of a smooth crossover between these two limiting behaviors.

\subsection{The coupling between an amplified internal mode and an external mode admits a nice analogy}
We propose a useful analogy to understand the coupling between the oscillating and turning modes of the organism. In this analogy, the oscillating mode which can have a gain greater than 1 as soon as the dissipation is overcome by the parametric amplification is analogous to a optical cavity with a pumped laser medium. With each pass in the resonator, the amplitude of the mode grows. The turning mode is analogous to the free-space output of the laser and is controlled by the output coupler. Here $\zeta$, the behavioral inertia controls this coupling between the amplified resonator and the free space mode similar to the output coupler of the laser. When the output coupler has a very low reflectively, the lasing threshold for the system grows. However, when the reflectively is high, very little light leaks out of the resonator mode. To maximize the power of the laser under fixed pumping power, one needs to find a balance for the output coupler. 

This analogy leads us to calculating the amplitude of the turning mode given a frequency of driving and against $\zeta$. We find that for a fixed parametric driving amplitude, too large of $\zeta$ results in no amplified oscillations, whereas too low results in very little turning.

We can build toward physical interpretation of these results, by positing that the tissue inertia is related to the tissue size: larger tissues will be more difficult to change their course and will thus have a smaller effective $\zeta$. 

\subsection{Prediction: Size dependent behavior of the tissue}
The culmination of the turning output calculations and the size dependence of $\zeta$ is a set of predictions which we find to be qualitatively in line with our anecdotal observations. The first prediction is that small organisms will exhibit very little active-elastic oscillation within the body and will behave much more similarly to a rigid body. The second prediction is that organisms in an intermediate size regime, will exhibit large amplitude oscillations that couple well to rapid changes in direction. The final prediction on the scaling of turning behavior as a function of size, is that the largest organisms will have less turning agility and will exhibit a greater degree of active elastic oscillations. 

To delve deeper into the size dependent behavior of locomoting $T. Adhaerens$, we look beyond the active elastic resonator models to study the full numerics and its link to a collection of coupled active-elastic modes. 

\section{Spatiotemporal tissue dynamics as weakly coupled active elastic resonator modes}
One of the most powerful tools of linear analysis is the concept of modes. Modes of linear systems are unified by the property of superposition and a lack of coupling, that is, if you excite the first mode and the third mode you can predict the displacement time-course from the superposition of these two modes. Next, the modes are disentangled from each other in that power in one mode does not modify another mode. They are independent degrees of freedom of the system under study. While this assumption seems draconian and doomed to failure, for sufficiently small displacements these linear approximations hold very well.

Our goal in this section is to motivate and apply the mode excitation picture to the collective dynamics of an active elastic sheet. Along the way, we will identify places where the simpler formulation of AERs can be applied instead to collective modes instead of single degrees of freedom.

An active elastic sheet is characterized by the combination of activity and elasticity. If we assume that the tissue is governed dominantly by linear elastic contributions in the regime of relevance for our work, we are given a model that takes for the form of:

$$
\frac{\partial}{\partial t} \vec x = -\nabla_{\vec x} E + \vec\alpha
$$
$$
\frac{\partial}{\partial t} \vec \alpha = -L \times \nabla_{\vec x} E
$$

We can write down the energy function for our elastic network as:
$$
E = \sum_{bonds} \alpha \Delta \ell^2
$$

The goal is to extract the stiffness matrix of this material via a direct calculation of the Hessian. Let's start down that path by writing our energy in terms of the degrees of freedom of the problem.
$$
E = \sum_{\delta_{i,j} = 1} \alpha (\sqrt{(x_i - x_j)^2 + (y_i - y_j)^2} - \ell_o)^2
$$

For the purpose of this work, we assume that our neighborhood matrix is fixed to that above. 
\begin{widetext}
Taking the first gradient gives us:
$$
\frac{\partial}{\partial x_i} E = \alpha \sum_{j \in neighbors} \left(\sqrt{(x_i - x_j)^2 + (y_i - y_j)^2} - \ell_o\right)\frac{(x_i-x_j)}{\sqrt{(x_i - x_j)^2 + (y_i - y_j)^2}}
$$

Taking the next partial derivative gives us:
$$
\frac{\partial^2}{\partial x_i \partial x_k} E = \alpha \left(\frac{(x_i-x_k)}{\sqrt{(x_i - x_k)^2 + (y_i - y_k)^2}} - \frac{\left(\sqrt{(x_i - x_k)^2 + (y_i - y_k)^2} - \ell_o\right)}{\sqrt{(x_i - x_k)^2 + (y_i - y_k)^2}}\left( 1 + \frac{3(x_i-x_k)}{(x_i - x_k)^2 + (y_i - y_k)^2} \right) \right)
$$

If $i == k$, then the term looks like:
$$
\frac{\partial^2}{\partial x_i^2} E = \alpha \sum_{j \in neighbors} \left(\frac{(x_i-x_j)}{\sqrt{(x_i - x_j)^2 + (y_i - y_j)^2}} - \frac{\left(\sqrt{(x_i - x_j)^2 + (y_i - y_j)^2} - \ell_o\right)}{\sqrt{(x_i - x_j)^2 + (y_i - y_j)^2}}\left( 1 + \frac{3(x_i-x_j)}{(x_i - x_j)^2 + (y_i - y_j)^2} \right) \right)
$$

We also keep track of the cross-terms. 
$$
\frac{\partial^2}{\partial x_i \partial y_k} E = \alpha \left(\frac{(x_i-x_k)(y_i - y_k)}{(x_i - x_k)^2 + (y_i - y_k)^2} \left(1 - 3\frac{\sqrt{(x_i - x_k)^2 + (y_i - y_k)^2} - \ell_o}{\sqrt{(x_i - x_k)^2 + (y_i - y_k)^2}} \right) \right)
$$

If $i = k$, then we get cross terms of the form:
$$
\frac{\partial^2}{\partial x_i \partial y_i} E = \alpha \sum_{j \in neighbors}\left(\frac{(x_i-x_j)(y_i - y_j)}{(x_i - x_j)^2 + (y_i - y_j)^2} \left(1 - 3\frac{\sqrt{(x_i - x_j)^2 + (y_i - y_j)^2} - \ell_o}{\sqrt{(x_i - x_j)^2 + (y_i - y_j)^2}} \right) \right)
$$
\end{widetext}

This means that the nonzero components goes with the connectivity matrix for both $x$ and $y$ components, with a complex position dependent scaling.

We want to expand around configurations that set $\frac{\partial}{\partial x_i} = 0$ and $\frac{\partial}{\partial y_i} = 0$. 

The simplest solution to this equation (and the only one for the network above with fixed $\ell_o$) is where each length of spring is equal to $\ell_o$. (There are, of course, interesting things to learn as you change the connectivity and relax the $\ell_o = $constant condition (e.g. about the emergence of different types of modes, ZMs, lots of metastable states etc...))

This solution has a fixed relationship with a hexagonal symmetry. 

If $\Delta y = 0$ then $\Delta x = \ell_o$. If $\Delta y = \frac{\ell_o}{\sqrt{3}}$ then $\Delta x = \frac{\ell_o}{2}$. On top of that, we have to get all the negatives and positives correct.

For the condition that we are in this type of equilibrium, the contributions of the second term goes away and our stiffness matrix for x looks like:
$$
\frac{\partial^2}{\partial x_i \partial x_k} E_{\text{all springs at }\ell_o} = \alpha \left(\frac{(x_i-x_k)}{\sqrt{(x_i - x_k)^2 + (y_i - y_k)^2}} \right)
$$
Some further simplification gives us:
$$
\frac{\partial^2}{\partial x_i \partial x_k} E_{\text{all springs at }\ell_o} = \frac{\alpha}{\ell_o} (x_i-x_k)
$$
$$
\frac{\partial^2}{\partial y_i \partial y_k} E_{\text{all springs at }\ell_o} = \frac{\alpha}{\ell_o} (y_i-y_k)
$$

We can study the statistics of the eigenvalues to learn more about the density of states for these types of hexagonal spring networks consistent with the Maxwell lattice with 6 relative constraints for each 6 degrees of freedom. From the rank-nullity theorem of linear algebra, this means that for these elastic networks there should be no zero modes in the absence of states of self stress\cite{huber_topological_2016, Bi2014}.
\subsection{Density of states}
A first essential step in characterizing the collective modes of the active elastic sheet is characterize the distribution of the eigenvalues which tell us about the stiffness against excitation of the standing modes of the network. This density of states for the choice of the cellular triangular network with 6 neighbors shows two strong peaks. The first peak at lower frequency the enriched density of longer wavelength collective modes which are analogous to the acoustic modes of a lattice. The second peak is much higher spatial frequency vibrations of the network and are analogous to the optical modes. It is noteworthy, that the addition of another cell type with a different effective damping would be analogous to the addition of a second mass capable of initiating a band gap between the two peaks in the density of states \cite{huber_topological_2016}.

\subsection{Mode scaling with size}
As we explore the scaling with size of these collective dynamics, a natural question emerges: for a given size tissue, how many modes are softer than a threshold frequency? This question is motivated by the idea that the influence of a mode on the tissue's dynamics is determined by its amplitude. Higher frequency (stiffness) modes will cost more work to excite to the same amplitude than a lower frequency mode. By scaling the size of the network numerically, we can study the extreme value statistics of these eigenvalues using the simple value under threshold technique. In figure \ref{fig:fig5}D, we find that the number of modes under threshold (for a threshold choice much smaller than the acoustic peak) scales anomalously between a slope of 1 and 2 with an exponent we call: $\chi$.  The choice of this threshold we call the \textit{elasto-ciliary} scale because it arises from the competition between the strength of the ciliary activity and the resistance of deformation of the elastic structure. 

We postulate that $\chi$ scales similarly to the 'complexity' of the locomotive behavior of the tissue. While we acknowledge that this prediction is strongly dependent upon the precise measure of complexity, we explore this idea using a simple measure for this work. We begin by defining the correlation length of the spin and amplitude fluctuations. We use $\delta \vec \theta = \vec \theta - \langle \vec \theta \rangle_t$ to extract both the amplitude and the heading following seminal work in flocking\cite{cavagna_silent_2015, mora_are_2011}. Next, we explore how this correlation length grows a function of size for various parameters of the polarized active elastic sheet model with $\epsilon = 0$ for various system parameters including the stiffness of the cell-cell junction network. From this we extract a scaling exponent, $\sigma (k)$, which captures the relationship between the system size and the two point fluctuation correlation length $C_{spin} \propto L^{\sigma_{spin}(k)}$.

We use this scaling of the characteristic correlation lengthscale of the fluctuations with system size to predict complexity by answering the question: how many correlation lengths fit within the organism size? This gives rise to a simple scaling relationship: $N_{domains} \propto \frac{L^2}{L^{2\eta(k)}}$ for numerically accessible organism sizes with $\epsilon = 0$. This simple argument suggests that the number of uncorrelated domains within the tissue will scale with:
\[
N_{domains} \sim L^{\eta} \propto L ^{2(1-\sigma)}
\]

Within the numerical range of study, we found that $\eta$ ranges from 0 to 0.4 and asymptotically approaches a limiting scaling length which is less than 0.5. This back of the envelope argument postulates that the 'complexity' of the tissue behavior scales with size with the exponent, $1 < \eta \leq 2$. This means that larger organisms will incorporate a larger set of these modes to their dynamics and exhibit seemingly more complex internal modes of their locomotive dynamics.

\subsection{Steady state mode power distributions can be modified by driving}
The tissue can tune its effective behavior in an orthogonal fashion to size by exhibiting larger amplitude periodic activity fluctuations. This parametric driving can counteract dissipation through driven amplification of these modes preferentially driving power into the modes which are near resonance. We can study this effect by following a technique which we first saw in work by Ferrante et al \cite{ferrante_elasticity-based_2013}, but has also been presented by Bi et al \cite{Bi2016} and furthered analytically by Henkes et al \cite{henkes_dense_2020} in recent years for the case of persistent random activity. This technique projects the dynamics of the many-body system onto the normal modes of the elastic network resulting in dynamical equations for the time evolution of the mode amplitudes.

The dynamics of a state of cells can be written as:
\[
\frac{\partial}{\partial t} \left|r \right \rangle = |\theta\rangle =  \alpha(|\psi\rangle)\gamma | \phi \rangle + \vec \nabla_r E 
\]

\[
\frac{\partial}{\partial t} \left|\phi \right \rangle =  \Gamma \vec \nabla_r E \times |\phi\rangle
\]

\[
\frac{\partial}{\partial t} \left|\psi \right\rangle = \Omega - J sin(|\psi\rangle)\langle\nabla_r E|\phi\rangle + \xi(t)
\]

where we have defined the ciliary orientation field as $|phi\rangle$, the cell displacement field as $|\theta \rangle$ and the activity phase field as $|\psi\rangle$. The relationship connecting the phase field to the locomotive mechanics is defined as: $\alpha(\psi) = 1 + \epsilon sin(\psi)$ where $\epsilon$ is the non-dimensional quantity that defines the amplitude of periodic driving.

\[
\nabla E(|\psi \rangle) \approx \nabla E|_{|\psi\rangle = |\psi_o\rangle} + \frac{1}{2} D (|\psi\rangle - |\psi_o\rangle)
\]

This means that when we expand around the local minimum the dynamics to leading order are dominated by the Hessian (which is this case is identical to the square of the dynamical matrix) \cite{huber_topological_2016}. If we apply separation of variables, the dynamics become a simple matter of eigenmode with a shared eigenvalue governing the dynamics of that mode. $D |\nu\rangle = \lambda |\nu\rangle $.

These modes define the basis of upcoming calculations:
\[
| \theta \rangle = \sum_{\nu} a_{\nu} |\nu\rangle
\]
\[
| \phi \rangle = \sum_{\nu} b_{\nu} |\nu\rangle
\]

Plugging this mode decomposition into the equations of motion for the individual degrees of freedom gives us equations which govern the evolution of the mode amplitudes. In the simplest limit where $\epsilon \rightarrow 0$ and $\Gamma \rightarrow 0$, Henkes et al\cite{henkes_dense_2020} proposed that the dynamics of the modes are controlled by their mode stiffness: 
\[
\frac{\partial}{\partial t} a_{\nu} = -\lambda_\nu a_\nu + \alpha b_\nu
\]
These dynamics show that the lifetime of the individual mode is controlled by the stiffness of that mode and the dynamics take the form of a overdamped elastic system with a active driving force. The dynamics of the active component can be written to leading order as:
\[
\partial_t b_\nu = -\Gamma \sum_\mu \lambda_\mu a_\mu O_{\nu, \mu}
\]
where $O_{\nu,\mu}$ captures the orthogonal overlap between the modes. By definition the inner product of these modes is zero, and the cross product of these modes represents the coupling between the two modes. Modes which are very similar will have small amplitudes whereas modes that are very different will have strong coupling. 
\[
O_{\nu, \mu} = \sum_{i} \left[(\nu_i \cdot \hat x)(\mu_i \cdot \hat y) - (\nu_i \cdot \hat y)(\mu_i \cdot \hat x)\right]
\]

This overlap coupling of the cross-products gives rise to a new matrix which explicitly couples the active elastic modes mediated by the activity. This coupling network displays a conceptual connection to a linear symmetric synaptic coupling matrix and by studying how power is passed in this network, we can learn more about the native dynamics of a polarized active elastic sheet from the view point of coupled, active-elastic modes.

We can use these modes as the basis set for understanding the dynamics of the actively driven tissue by mapping the observed dynamics on to these modes. This gives us a new N-dimensional representation of the dynamics where we study the mode amplitude space. The advantage of this N dimensional space is that the transients of the very stiff modes are very short as the lifetime is exponentially related to the negative eigenvalue of the mode. This means that on sufficiently long timescales, the transients will decay and the dynamics will effectively live in a lower dimensional space. This simplifies our understanding of the system by encoding effective degrees of freedom along soft directions of the dynamical manifold. 

This observation is born out the numerical dynamics where we study the mode amplitude as a function of time and confirm that we can reconstruct the work of [Henkes] and [Ferrante]. The preferential damping combined with the passing of power between modes, gives rise to a steady state distribution of mode amplitudes which follows a non-universal power law which is sensitive to microscopic parameter choices\cite{alert_active_2021}. 

Adding in parameteric activity driving adds a new power source localizing in the intermediate modes surrounding the parametric resonance. This new power pump modifies the total power spectrum of the modes by shifting net power away from the longest length-scale modes and into the modes near the parametric resonance. With sufficient driving amplitude, the power injected via parametric amplification can exceed the damping and induce length-scale selection via the amplified excitation of intermediate length-scale modes. 

\subsection{Measuring the collective contribution of periodic activity on the response to stimuli}
While the organism can certainly benefit from tuneable modification of its steady-state, the real power of an organism is to respond to its environment. To study the response to stimuli, we revisit the numerical experiments on the tissue response to stimulus and study the role of parametric driving amplitude in the flocking oscillators model, $\epsilon$. How is the response to an identical stimulus altered by periodic activity?

To summarize these numerical results, we report two quantities as a result of the set of simulations: i) the mean polarization, and ii) the change in the tissue direction of travel. The mean polarization illustrates the stability of the longest lengthscale mode which is essential for net animal locomotion. The second serves as a proxy for the sensitivity of the tissue in response to environmental stimuli in the form of the applied force to the same cell. In conducting this numerical experiment, we found that for tissues comprising of $5e3$ cells can increase their turning sensitivity to a stimulus by 450\% while only reducing the net polarization by $\sim 10\%$. This numerical finding supports the conceptual findings of the 2D active elastic resonator in suggesting that periodic activity in active system can parametrically amplify the response of the system to external stimuli without significantly compromising the stability of the collective locomotion.

\subsection{Spatial embodied stimulus response}
The concept of embodiment posits that the sensor and actuator derive their behavior from their morphological context\cite{pfeifer2007self}. This is essential for the division of labor in an organism that has no developmental symmetry breaking. If a cell on the edge can respond to stimuli differently then a cell in the bulk then this 'embodied computation' reduces the need specialized control dependent upon spatial location. This form of self-organized embodied computation is a useful concept for active matter\cite{gilpin_multiscale_2020}.

Here, we study the spatial dependence of the stimulus response, by conducting a full numerical simulation for each cell in the tissue being subjected to an identical external stimulus. This generates a mechanical analogy to the receptive field\cite{dayan_theoretical_2001} where the spatial position of the stimulus can illicit very different responses in the collective mechanics. Continuing with the utility of the polarization and the turning response, we plot the spatial map of the tissue response for each single cell being stimulated. 

One of the key findings is that amplitude of the turning response is strongly dependent and in a non-trivial way on the position of the stimulus. For low $\Gamma = 1$ (see supplementary information), the response adopts a 4-fold symmetry with the highest sensitivity to stimulus observed near the boundaries. However, when $\Gamma = 10$, the propagating waves distorts the shape of this 4 fold symmetry resulting in a region of diminished response close to the boundary below the half way mark up the tissue with the most sensitive region occurring in the top $20\%$ of the tissue (almost twice as strong a response as in the middle of the tissue). This morphological dependence has significant impact on the dynamics of a flocking elastic sheet by biasing the steering of the organism in an information rich force environment. 

\section{Discussion}
Our work furthers the long research program merging the spatio-temporal dynamics of active matter with organism-scale utility by providing important primary data on the collective mechanics underlying the locomotion of a non-neuromuscular animal. This work represents an effort to draw direct line between physical phenomena and the emergent behavior of a whole animal which opens the door to new understanding in how an organism processes a constant stream of environmental stimulus. The joint experimental tractability and conceptual simplicity of the non-neuromsuclar mechanics of \textit{T. Adhaerens}  promises its utility in stretching our understanding toward more general principles in biological information processing in ensembles of cells.

\subsection{Agility from traveling waves and sensitivity}
Along this route, we observed the powerful role of activity fluctuations in generating agile locomotion. We show -- across the spectrum of two toy models and a numerical implementation of the flocking oscillator model -- that the resulting parametric amplification of active-elastic oscillations can generate rapid organism-scale changes in direction in response to stimulus. By harnessing the parametric instability the organism can achieve agile response to mechanical stimuli across a large range of tissue sizes (100s to millions of cells) perhaps enabling even this simple animal to maintain sufficient responsivity to its micro-environment. This can enable effective coupling between the fast contractile mechanics on the top of the organism\cite{armon2018ultrafast} and the locomotive agility conferred by the bottom tissues' amplified ciliary flocking.

We suggest that this mechanism for achieving sensitivity without compromising stability falls into a broader category of phenomena identified by harnessing instability to achieve utility. Echo State machines prescribe initializing the system dynamics on the edge of chaos to maximize performance\cite{tanaka_recent_2019}. Fighter jets use high speed control to stabilize an aerodynamic pitch instability. Are some living systems poised at critical points to leverage the associated diverging susceptibility? The core unifying feature here advances the idea that there is utility to harnessing instability associated with being near a critical point\cite{mora_are_2011}.

Here, we show a mechanism which can be used to increase sensitivity by approaching the flocking transition from a different direction in parameter space. On the scale of the transient dynamics in response to temporally sparse perturbations, we have demonstrated that this direction in phase space amplifies mechanical sensitivity well before the critical point in a fashion that facilitates stable yet sensitive behavior. One of the mechanisms by which this is achieved is that the parametric amplification can serve to counteract the overwhelming damping forces in these tissue systems and establish traveling waves which can propagate across the scale of a sizeable organism.

\subsection{Swarmalators}
This work also provides yet another data point in a growing list of self-organizing biological dynamics which occur on fast timescales\cite{mathijssen_collective_2019, giavazzi_flocking_2018}. One of the values is presenting an experimental living  and behaving dynamical system in which we can begin to access the time and lengthscales of the dynamics and which are to leading order consistent with a broad class of mathematically interesting problems in swarming or flocking oscillators\cite{okeeffe_oscillators_2017, gilpin_multiscale_2020}. While the generalized dynamics presented in this work are microscopically distinct from the models presented elsewhere\cite{okeeffe_oscillators_2017}, this framework may contribute new perspectives on the organismal utility of these emergent spatio-temporal dynamics. 

\subsection{The tissue as a physical reservoir}
Perhaps the most compelling future direction of this work is using these dynamics in an explicitly computational capacity. Here is helps to remember Dikstra's view that computation "is no more about computers than astronomy is about telescopes" \cite{DijkstraQuotes}. Over the last 20 years reservoir computing\cite{jaeger__2001, maass_real-time_2002} has grown into a paradigm for not only achieving cutting-edge performance on forecasting of chaotic dynamics\cite{pathak_model-free_2018} but also outsourcing computation from computers to dynamical systems with the right measure of non-linearity \cite{tanaka_recent_2019}. The intuition behind the power of nonlinear dynamics in an organism arises from a reservoirs' ability to project the incoming dynamics up onto a high dimensional space through which a linear classifier can draw a plane to separate classes. These techniques have recently been extended to study the capacity of flocking 'boids'\cite{lymburn_reservoir_2021}.

With this perspective, we can look toward the frontier of our understanding where the internal state dynamics are tightly coupled to the environment in such a manner that the hybridization of these two dynamical systems generates new dynamics entirely. The goal of the organism is to generate sufficient complexity of dynamics which can be leveraged via a global decoding which maps onto behavior. Perhaps, \textit{T. Adhaerens} is simple enough to shine light on this approach within a behaving animal.

\section{Conclusion}

In this work, we have reported the spatio-temporal dynamics of ciliary flocking mediated by sub-second ciliary reorientations in the bottom layer of the animal, \textit{T. Adhaerens}. We present a modeling by construction framework to propose a many-body flocking oscillator model. Through direct measurement of the tissue dynamics, we identify tissue oscillations and traveling waves which we reconcile with the model through an analytically tractable toy model which maps the single cell dynamics onto a weakly nonlinear active-elastic resonator. In addition to the observed longer timescale tissue oscillations, we identify the relevant periodic forcing which arises from ciliary walking which leads us to propose a mechanism for parametrically amplifying the active elastic resonator mode through periodic fluctuations in the active forcing near 4 times the native frequency of the resonator. This amplification phenomenon is confirmed in both a 2D model and the full numerical simulation. The many modes of the active-elastic tissue mean that fine tuning is not necessarily as long as the tissue is larger than the characteristic scale of the amplified mode. This work presents alternative microscale inspired mechanisms for achieving length scale selection and under-damped traveling waves in polarized active matter. We expect these results will be of use to communities ranging from self-organizing dynamics of sensitive active matter to the makers of distributed robotic systems.

\section{Acknowledgements}
We thank all members of the PrakashLab for scientific discussions and comments. In particular, we thank Pranav Vyas, Shahaf Armon, Grace Zhong, Hazel Soto-Montoya and Laurel Kroo in the lab for their contributions to a vibrant research community. M.S.B. was supported by the National Science Foundation Graduate Research Fellowship (DGE-1147470) and the Stanford University BioX Fellows Program. This work was supported by HHMI Faculty Fellows Award (M.P), BioHub Investigator Fellowship (M.P), Pew Fellowship (M.P), Schmidt Futures Fellowship, NSF Career Award (M.P), NSF CCC (DBI-1548297) and Moore Foundation. 

\bibliography{mybib.bib}

\begin{thebibliography}{84}%
\makeatletter
\providecommand \@ifxundefined [1]{%
 \@ifx{#1\undefined}
}%
\providecommand \@ifnum [1]{%
 \ifnum #1\expandafter \@firstoftwo
 \else \expandafter \@secondoftwo
 \fi
}%
\providecommand \@ifx [1]{%
 \ifx #1\expandafter \@firstoftwo
 \else \expandafter \@secondoftwo
 \fi
}%
\providecommand \natexlab [1]{#1}%
\providecommand \enquote  [1]{``#1''}%
\providecommand \bibnamefont  [1]{#1}%
\providecommand \bibfnamefont [1]{#1}%
\providecommand \citenamefont [1]{#1}%
\providecommand \href@noop [0]{\@secondoftwo}%
\providecommand \href [0]{\begingroup \@sanitize@url \@href}%
\providecommand \@href[1]{\@@startlink{#1}\@@href}%
\providecommand \@@href[1]{\endgroup#1\@@endlink}%
\providecommand \@sanitize@url [0]{\catcode `\\12\catcode `\$12\catcode
  `\&12\catcode `\#12\catcode `\^12\catcode `\_12\catcode `\%12\relax}%
\providecommand \@@startlink[1]{}%
\providecommand \@@endlink[0]{}%
\providecommand \url  [0]{\begingroup\@sanitize@url \@url }%
\providecommand \@url [1]{\endgroup\@href {#1}{\urlprefix }}%
\providecommand \urlprefix  [0]{URL }%
\providecommand \Eprint [0]{\href }%
\providecommand \doibase [0]{https://doi.org/}%
\providecommand \selectlanguage [0]{\@gobble}%
\providecommand \bibinfo  [0]{\@secondoftwo}%
\providecommand \bibfield  [0]{\@secondoftwo}%
\providecommand \translation [1]{[#1]}%
\providecommand \BibitemOpen [0]{}%
\providecommand \bibitemStop [0]{}%
\providecommand \bibitemNoStop [0]{.\EOS\space}%
\providecommand \EOS [0]{\spacefactor3000\relax}%
\providecommand \BibitemShut  [1]{\csname bibitem#1\endcsname}%
\let\auto@bib@innerbib\@empty
\bibitem [{\citenamefont {Bull}\ \emph {et~al.}(2021)\citenamefont {Bull},
  \citenamefont {Kroo},\ and\ \citenamefont {Prakash}}]{bullpart1}%
  \BibitemOpen
  \bibfield  {author} {\bibinfo {author} {\bibfnamefont {M.~S.}\ \bibnamefont
  {Bull}}, \bibinfo {author} {\bibfnamefont {L.}~\bibnamefont {Kroo}},\ and\
  \bibinfo {author} {\bibfnamefont {M.}~\bibnamefont {Prakash}},\ }\bibfield
  {title} {\bibinfo {title} {Excitable mechanics embodied in a walking
  cilium},\ }\href@noop {} {\bibfield  {journal} {\bibinfo  {journal} {arXiv}\
  } (\bibinfo {year} {2021})}\BibitemShut {NoStop}%
\bibitem [{\citenamefont {Bull}\ and\ \citenamefont
  {Prakash}(2021)}]{bullpart3}%
  \BibitemOpen
  \bibfield  {author} {\bibinfo {author} {\bibfnamefont {M.~S.}\ \bibnamefont
  {Bull}}\ and\ \bibinfo {author} {\bibfnamefont {M.}~\bibnamefont {Prakash}},\
  }\bibfield  {title} {\bibinfo {title} {Mobile defects born from an energy
  cascade shape the locomotive behavior of a headless animal},\ }\href@noop {}
  {\bibfield  {journal} {\bibinfo  {journal} {arXiv}\ } (\bibinfo {year}
  {2021})}\BibitemShut {NoStop}%
\bibitem [{\citenamefont {Keijzer}(2015)}]{keijzer_moving_2015}%
  \BibitemOpen
  \bibfield  {author} {\bibinfo {author} {\bibfnamefont {F.}~\bibnamefont
  {Keijzer}},\ }\bibfield  {title} {\bibinfo {title} {Moving and sensing
  without input and output: early nervous systems and the origins of the animal
  sensorimotor organization},\ }\href
  {https://doi.org/10.1007/s10539-015-9483-1} {\bibfield  {journal} {\bibinfo
  {journal} {Biology \& Philosophy}\ }\textbf {\bibinfo {volume} {30}},\
  \bibinfo {pages} {311} (\bibinfo {year} {2015})}\BibitemShut {NoStop}%
\bibitem [{\citenamefont {Moroz}(2009)}]{moroz_independent_2009}%
  \BibitemOpen
  \bibfield  {author} {\bibinfo {author} {\bibfnamefont {L.~L.}\ \bibnamefont
  {Moroz}},\ }\bibfield  {title} {\bibinfo {title} {On the {Independent}
  {Origins} of {Complex} {Brains} and {Neurons}},\ }\href
  {https://doi.org/10.1159/000258665} {\bibfield  {journal} {\bibinfo
  {journal} {Brain, Behavior and Evolution}\ }\textbf {\bibinfo {volume}
  {74}},\ \bibinfo {pages} {177} (\bibinfo {year} {2009})},\ \bibinfo {note}
  {publisher: Karger Publishers}\BibitemShut {NoStop}%
\bibitem [{\citenamefont {Jékely}(2021)}]{jekely_chemical_2021}%
  \BibitemOpen
  \bibfield  {author} {\bibinfo {author} {\bibfnamefont {G.}~\bibnamefont
  {Jékely}},\ }\bibfield  {title} {\bibinfo {title} {The chemical brain
  hypothesis for the origin of nervous systems},\ }\href
  {https://doi.org/10.1098/rstb.2019.0761} {\bibfield  {journal} {\bibinfo
  {journal} {Philosophical Transactions of the Royal Society B: Biological
  Sciences}\ }\textbf {\bibinfo {volume} {376}},\ \bibinfo {pages} {20190761}
  (\bibinfo {year} {2021})},\ \bibinfo {note} {publisher: Royal
  Society}\BibitemShut {NoStop}%
\bibitem [{\citenamefont {J{\'e}kely}(2010)}]{jekely2010origin}%
  \BibitemOpen
  \bibfield  {author} {\bibinfo {author} {\bibfnamefont {G.}~\bibnamefont
  {J{\'e}kely}},\ }\bibfield  {title} {\bibinfo {title} {Origin and early
  evolution of neural circuits for the control of ciliary locomotion},\
  }\href@noop {} {\bibfield  {journal} {\bibinfo  {journal} {Proceedings of the
  Royal Society B: Biological Sciences}\ }\textbf {\bibinfo {volume} {278}},\
  \bibinfo {pages} {914} (\bibinfo {year} {2010})}\BibitemShut {NoStop}%
\bibitem [{\citenamefont {Wan}\ and\ \citenamefont
  {Jékely}(2021)}]{wan_origins_2021}%
  \BibitemOpen
  \bibfield  {author} {\bibinfo {author} {\bibfnamefont {K.~Y.}\ \bibnamefont
  {Wan}}\ and\ \bibinfo {author} {\bibfnamefont {G.}~\bibnamefont {Jékely}},\
  }\bibfield  {title} {\bibinfo {title} {Origins of eukaryotic excitability},\
  }\href {https://doi.org/10.1098/rstb.2019.0758} {\bibfield  {journal}
  {\bibinfo  {journal} {Philosophical Transactions of the Royal Society B:
  Biological Sciences}\ }\textbf {\bibinfo {volume} {376}},\ \bibinfo {pages}
  {20190758} (\bibinfo {year} {2021})},\ \bibinfo {note} {publisher: Royal
  Society}\BibitemShut {NoStop}%
\bibitem [{\citenamefont {Sponberg}(2017)}]{sponberg_emergent_2017}%
  \BibitemOpen
  \bibfield  {author} {\bibinfo {author} {\bibfnamefont {S.}~\bibnamefont
  {Sponberg}},\ }\bibfield  {title} {\bibinfo {title} {The emergent physics of
  animal locomotion},\ }\href {https://doi.org/10.1063/PT.3.3691} {\bibfield
  {journal} {\bibinfo  {journal} {Physics Today}\ }\textbf {\bibinfo {volume}
  {70}},\ \bibinfo {pages} {34} (\bibinfo {year} {2017})}\BibitemShut {NoStop}%
\bibitem [{\citenamefont {More}\ and\ \citenamefont
  {Donelan}()}]{more_scaling_nodate}%
  \BibitemOpen
  \bibfield  {author} {\bibinfo {author} {\bibfnamefont {H.~L.}\ \bibnamefont
  {More}}\ and\ \bibinfo {author} {\bibfnamefont {J.~M.}\ \bibnamefont
  {Donelan}},\ }\bibfield  {title} {\bibinfo {title} {Scaling of sensorimotor
  delays in terrestrial mammals},\ }\href
  {https://doi.org/10.1098/rspb.2018.0613} {\bibfield  {journal} {\bibinfo
  {journal} {Proceedings of the Royal Society B: Biological Sciences}\ }\textbf
  {\bibinfo {volume} {285}},\ \bibinfo {pages} {20180613}},\ \bibinfo {note}
  {publisher: Royal Society}\BibitemShut {NoStop}%
\bibitem [{\citenamefont {Pratt}\ \emph {et~al.}(2017)\citenamefont {Pratt},
  \citenamefont {Deora}, \citenamefont {Mohren},\ and\ \citenamefont
  {Daniel}}]{pratt_neural_2017}%
  \BibitemOpen
  \bibfield  {author} {\bibinfo {author} {\bibfnamefont {B.}~\bibnamefont
  {Pratt}}, \bibinfo {author} {\bibfnamefont {T.}~\bibnamefont {Deora}},
  \bibinfo {author} {\bibfnamefont {T.}~\bibnamefont {Mohren}},\ and\ \bibinfo
  {author} {\bibfnamefont {T.}~\bibnamefont {Daniel}},\ }\bibfield  {title}
  {\bibinfo {title} {Neural evidence supports a dual sensory-motor role for
  insect wings},\ }\href {https://doi.org/10.1098/rspb.2017.0969} {\bibfield
  {journal} {\bibinfo  {journal} {Proceedings of the Royal Society B:
  Biological Sciences}\ }\textbf {\bibinfo {volume} {284}},\ \bibinfo {pages}
  {20170969} (\bibinfo {year} {2017})},\ \bibinfo {note} {publisher: Royal
  Society}\BibitemShut {NoStop}%
\bibitem [{\citenamefont {Grosberg}\ and\ \citenamefont
  {Strathmann}(2007)}]{grosberg_evolution_2007}%
  \BibitemOpen
  \bibfield  {author} {\bibinfo {author} {\bibfnamefont {R.~K.}\ \bibnamefont
  {Grosberg}}\ and\ \bibinfo {author} {\bibfnamefont {R.~R.}\ \bibnamefont
  {Strathmann}},\ }\bibfield  {title} {\bibinfo {title} {The {Evolution} of
  {Multicellularity}: {A} {Minor} {Major} {Transition}?},\ }\href
  {https://doi.org/10.1146/annurev.ecolsys.36.102403.114735} {\bibfield
  {journal} {\bibinfo  {journal} {Annual Review of Ecology, Evolution, and
  Systematics}\ }\textbf {\bibinfo {volume} {38}},\ \bibinfo {pages} {621}
  (\bibinfo {year} {2007})},\ \bibinfo {note} {\_eprint:
  https://doi.org/10.1146/annurev.ecolsys.36.102403.114735}\BibitemShut
  {NoStop}%
\bibitem [{\citenamefont {Pentz}\ \emph {et~al.}(2020)\citenamefont {Pentz},
  \citenamefont {Márquez-Zacarías}, \citenamefont {Bozdag}, \citenamefont
  {Burnetti}, \citenamefont {Yunker}, \citenamefont {Libby},\ and\
  \citenamefont {Ratcliff}}]{pentz_ecological_2020}%
  \BibitemOpen
  \bibfield  {author} {\bibinfo {author} {\bibfnamefont {J.~T.}\ \bibnamefont
  {Pentz}}, \bibinfo {author} {\bibfnamefont {P.}~\bibnamefont
  {Márquez-Zacarías}}, \bibinfo {author} {\bibfnamefont {G.~O.}\ \bibnamefont
  {Bozdag}}, \bibinfo {author} {\bibfnamefont {A.}~\bibnamefont {Burnetti}},
  \bibinfo {author} {\bibfnamefont {P.~J.}\ \bibnamefont {Yunker}}, \bibinfo
  {author} {\bibfnamefont {E.}~\bibnamefont {Libby}},\ and\ \bibinfo {author}
  {\bibfnamefont {W.~C.}\ \bibnamefont {Ratcliff}},\ }\bibfield  {title}
  {\bibinfo {title} {Ecological {Advantages} and {Evolutionary} {Limitations}
  of {Aggregative} {Multicellular} {Development}},\ }\href
  {https://doi.org/10.1016/j.cub.2020.08.006} {\bibfield  {journal} {\bibinfo
  {journal} {Current biology: CB}\ }\textbf {\bibinfo {volume} {30}},\ \bibinfo
  {pages} {4155} (\bibinfo {year} {2020})}\BibitemShut {NoStop}%
\bibitem [{\citenamefont {Marchetti}\ \emph {et~al.}(2013)\citenamefont
  {Marchetti}, \citenamefont {Joanny}, \citenamefont {Ramaswamy}, \citenamefont
  {Liverpool}, \citenamefont {Prost}, \citenamefont {Rao},\ and\ \citenamefont
  {Simha}}]{marchetti_hydrodynamics_2013}%
  \BibitemOpen
  \bibfield  {author} {\bibinfo {author} {\bibfnamefont {M.~C.}\ \bibnamefont
  {Marchetti}}, \bibinfo {author} {\bibfnamefont {J.~F.}\ \bibnamefont
  {Joanny}}, \bibinfo {author} {\bibfnamefont {S.}~\bibnamefont {Ramaswamy}},
  \bibinfo {author} {\bibfnamefont {T.~B.}\ \bibnamefont {Liverpool}}, \bibinfo
  {author} {\bibfnamefont {J.}~\bibnamefont {Prost}}, \bibinfo {author}
  {\bibfnamefont {M.}~\bibnamefont {Rao}},\ and\ \bibinfo {author}
  {\bibfnamefont {R.~A.}\ \bibnamefont {Simha}},\ }\bibfield  {title} {\bibinfo
  {title} {Hydrodynamics of soft active matter},\ }\href
  {https://doi.org/10.1103/RevModPhys.85.1143} {\bibfield  {journal} {\bibinfo
  {journal} {Reviews of Modern Physics}\ }\textbf {\bibinfo {volume} {85}},\
  \bibinfo {pages} {1143} (\bibinfo {year} {2013})},\ \bibinfo {note}
  {publisher: American Physical Society}\BibitemShut {NoStop}%
\bibitem [{\citenamefont {Gompper}\ \emph {et~al.}(2020)\citenamefont
  {Gompper}, \citenamefont {Winkler}, \citenamefont {Speck}, \citenamefont
  {Solon}, \citenamefont {Nardini}, \citenamefont {Peruani}, \citenamefont
  {Löwen}, \citenamefont {Golestanian}, \citenamefont {Kaupp}, \citenamefont
  {Alvarez}, \citenamefont {Kiørboe}, \citenamefont {Lauga}, \citenamefont
  {Poon}, \citenamefont {DeSimone}, \citenamefont {Muiños-Landin},
  \citenamefont {Fischer}, \citenamefont {Söker}, \citenamefont {Cichos},
  \citenamefont {Kapral}, \citenamefont {Gaspard}, \citenamefont {Ripoll},
  \citenamefont {Sagues}, \citenamefont {Doostmohammadi}, \citenamefont
  {Yeomans}, \citenamefont {Aranson}, \citenamefont {Bechinger}, \citenamefont
  {Stark}, \citenamefont {Hemelrijk}, \citenamefont {Nedelec}, \citenamefont
  {Sarkar}, \citenamefont {Aryaksama}, \citenamefont {Lacroix}, \citenamefont
  {Duclos}, \citenamefont {Yashunsky}, \citenamefont {Silberzan}, \citenamefont
  {Arroyo},\ and\ \citenamefont {Kale}}]{gompper_2020_2020}%
  \BibitemOpen
  \bibfield  {author} {\bibinfo {author} {\bibfnamefont {G.}~\bibnamefont
  {Gompper}}, \bibinfo {author} {\bibfnamefont {R.~G.}\ \bibnamefont
  {Winkler}}, \bibinfo {author} {\bibfnamefont {T.}~\bibnamefont {Speck}},
  \bibinfo {author} {\bibfnamefont {A.}~\bibnamefont {Solon}}, \bibinfo
  {author} {\bibfnamefont {C.}~\bibnamefont {Nardini}}, \bibinfo {author}
  {\bibfnamefont {F.}~\bibnamefont {Peruani}}, \bibinfo {author} {\bibfnamefont
  {H.}~\bibnamefont {Löwen}}, \bibinfo {author} {\bibfnamefont
  {R.}~\bibnamefont {Golestanian}}, \bibinfo {author} {\bibfnamefont {U.~B.}\
  \bibnamefont {Kaupp}}, \bibinfo {author} {\bibfnamefont {L.}~\bibnamefont
  {Alvarez}}, \bibinfo {author} {\bibfnamefont {T.}~\bibnamefont {Kiørboe}},
  \bibinfo {author} {\bibfnamefont {E.}~\bibnamefont {Lauga}}, \bibinfo
  {author} {\bibfnamefont {W.~C.~K.}\ \bibnamefont {Poon}}, \bibinfo {author}
  {\bibfnamefont {A.}~\bibnamefont {DeSimone}}, \bibinfo {author}
  {\bibfnamefont {S.}~\bibnamefont {Muiños-Landin}}, \bibinfo {author}
  {\bibfnamefont {A.}~\bibnamefont {Fischer}}, \bibinfo {author} {\bibfnamefont
  {N.~A.}\ \bibnamefont {Söker}}, \bibinfo {author} {\bibfnamefont
  {F.}~\bibnamefont {Cichos}}, \bibinfo {author} {\bibfnamefont
  {R.}~\bibnamefont {Kapral}}, \bibinfo {author} {\bibfnamefont
  {P.}~\bibnamefont {Gaspard}}, \bibinfo {author} {\bibfnamefont
  {M.}~\bibnamefont {Ripoll}}, \bibinfo {author} {\bibfnamefont
  {F.}~\bibnamefont {Sagues}}, \bibinfo {author} {\bibfnamefont
  {A.}~\bibnamefont {Doostmohammadi}}, \bibinfo {author} {\bibfnamefont
  {J.~M.}\ \bibnamefont {Yeomans}}, \bibinfo {author} {\bibfnamefont {I.~S.}\
  \bibnamefont {Aranson}}, \bibinfo {author} {\bibfnamefont {C.}~\bibnamefont
  {Bechinger}}, \bibinfo {author} {\bibfnamefont {H.}~\bibnamefont {Stark}},
  \bibinfo {author} {\bibfnamefont {C.~K.}\ \bibnamefont {Hemelrijk}}, \bibinfo
  {author} {\bibfnamefont {F.~J.}\ \bibnamefont {Nedelec}}, \bibinfo {author}
  {\bibfnamefont {T.}~\bibnamefont {Sarkar}}, \bibinfo {author} {\bibfnamefont
  {T.}~\bibnamefont {Aryaksama}}, \bibinfo {author} {\bibfnamefont
  {M.}~\bibnamefont {Lacroix}}, \bibinfo {author} {\bibfnamefont
  {G.}~\bibnamefont {Duclos}}, \bibinfo {author} {\bibfnamefont
  {V.}~\bibnamefont {Yashunsky}}, \bibinfo {author} {\bibfnamefont
  {P.}~\bibnamefont {Silberzan}}, \bibinfo {author} {\bibfnamefont
  {M.}~\bibnamefont {Arroyo}},\ and\ \bibinfo {author} {\bibfnamefont
  {S.}~\bibnamefont {Kale}},\ }\bibfield  {title} {\bibinfo {title} {The 2020
  motile active matter roadmap},\ }\href
  {https://doi.org/10.1088/1361-648X/ab6348} {\bibfield  {journal} {\bibinfo
  {journal} {Journal of Physics: Condensed Matter}\ }\textbf {\bibinfo {volume}
  {32}},\ \bibinfo {pages} {193001} (\bibinfo {year} {2020})},\ \bibinfo {note}
  {publisher: IOP Publishing}\BibitemShut {NoStop}%
\bibitem [{\citenamefont {Shaebani}\ \emph {et~al.}(2020)\citenamefont
  {Shaebani}, \citenamefont {Wysocki}, \citenamefont {Winkler}, \citenamefont
  {Gompper},\ and\ \citenamefont {Rieger}}]{shaebani_computational_2020}%
  \BibitemOpen
  \bibfield  {author} {\bibinfo {author} {\bibfnamefont {M.~R.}\ \bibnamefont
  {Shaebani}}, \bibinfo {author} {\bibfnamefont {A.}~\bibnamefont {Wysocki}},
  \bibinfo {author} {\bibfnamefont {R.~G.}\ \bibnamefont {Winkler}}, \bibinfo
  {author} {\bibfnamefont {G.}~\bibnamefont {Gompper}},\ and\ \bibinfo {author}
  {\bibfnamefont {H.}~\bibnamefont {Rieger}},\ }\bibfield  {title} {\bibinfo
  {title} {Computational models for active matter},\ }\href
  {https://doi.org/10.1038/s42254-020-0152-1} {\bibfield  {journal} {\bibinfo
  {journal} {Nature Reviews Physics}\ }\textbf {\bibinfo {volume} {2}},\
  \bibinfo {pages} {181} (\bibinfo {year} {2020})},\ \bibinfo {note} {number: 4
  Publisher: Nature Publishing Group}\BibitemShut {NoStop}%
\bibitem [{\citenamefont {Alert}\ \emph {et~al.}(2021)\citenamefont {Alert},
  \citenamefont {Casademunt},\ and\ \citenamefont
  {Joanny}}]{alert_active_2021}%
  \BibitemOpen
  \bibfield  {author} {\bibinfo {author} {\bibfnamefont {R.}~\bibnamefont
  {Alert}}, \bibinfo {author} {\bibfnamefont {J.}~\bibnamefont {Casademunt}},\
  and\ \bibinfo {author} {\bibfnamefont {J.-F.}\ \bibnamefont {Joanny}},\
  }\bibfield  {title} {\bibinfo {title} {Active {Turbulence}},\ }\href
  {http://arxiv.org/abs/2104.02122} {\bibfield  {journal} {\bibinfo  {journal}
  {arXiv:2104.02122 [cond-mat, physics:nlin, physics:physics]}\ } (\bibinfo
  {year} {2021})},\ \bibinfo {note} {arXiv: 2104.02122}\BibitemShut {NoStop}%
\bibitem [{\citenamefont {Giavazzi}\ \emph {et~al.}(2018)\citenamefont
  {Giavazzi}, \citenamefont {Paoluzzi}, \citenamefont {Macchi}, \citenamefont
  {Bi}, \citenamefont {Scita}, \citenamefont {Manning}, \citenamefont
  {Cerbino},\ and\ \citenamefont {Marchetti}}]{giavazzi_flocking_2018}%
  \BibitemOpen
  \bibfield  {author} {\bibinfo {author} {\bibfnamefont {F.}~\bibnamefont
  {Giavazzi}}, \bibinfo {author} {\bibfnamefont {M.}~\bibnamefont {Paoluzzi}},
  \bibinfo {author} {\bibfnamefont {M.}~\bibnamefont {Macchi}}, \bibinfo
  {author} {\bibfnamefont {D.}~\bibnamefont {Bi}}, \bibinfo {author}
  {\bibfnamefont {G.}~\bibnamefont {Scita}}, \bibinfo {author} {\bibfnamefont
  {M.~L.}\ \bibnamefont {Manning}}, \bibinfo {author} {\bibfnamefont
  {R.}~\bibnamefont {Cerbino}},\ and\ \bibinfo {author} {\bibfnamefont {M.~C.}\
  \bibnamefont {Marchetti}},\ }\bibfield  {title} {\bibinfo {title} {Flocking
  transitions in confluent tissues},\ }\href
  {https://doi.org/10.1039/C8SM00126J} {\bibfield  {journal} {\bibinfo
  {journal} {Soft Matter}\ }\textbf {\bibinfo {volume} {14}},\ \bibinfo {pages}
  {3471} (\bibinfo {year} {2018})},\ \bibinfo {note} {publisher: The Royal
  Society of Chemistry}\BibitemShut {NoStop}%
\bibitem [{\citenamefont {Szabó}\ \emph {et~al.}(2006)\citenamefont {Szabó},
  \citenamefont {Szöllösi}, \citenamefont {Gönci}, \citenamefont {Jurányi},
  \citenamefont {Selmeczi},\ and\ \citenamefont {Vicsek}}]{szabo_phase_2006}%
  \BibitemOpen
  \bibfield  {author} {\bibinfo {author} {\bibfnamefont {B.}~\bibnamefont
  {Szabó}}, \bibinfo {author} {\bibfnamefont {G.~J.}\ \bibnamefont
  {Szöllösi}}, \bibinfo {author} {\bibfnamefont {B.}~\bibnamefont {Gönci}},
  \bibinfo {author} {\bibfnamefont {Z.}~\bibnamefont {Jurányi}}, \bibinfo
  {author} {\bibfnamefont {D.}~\bibnamefont {Selmeczi}},\ and\ \bibinfo
  {author} {\bibfnamefont {T.}~\bibnamefont {Vicsek}},\ }\bibfield  {title}
  {\bibinfo {title} {Phase transition in the collective migration of tissue
  cells: {Experiment} and model},\ }\href
  {https://doi.org/10.1103/PhysRevE.74.061908} {\bibfield  {journal} {\bibinfo
  {journal} {Physical Review E}\ }\textbf {\bibinfo {volume} {74}},\ \bibinfo
  {pages} {061908} (\bibinfo {year} {2006})},\ \bibinfo {note} {publisher:
  American Physical Society}\BibitemShut {NoStop}%
\bibitem [{\citenamefont {Ferrante}\ \emph {et~al.}(2013)\citenamefont
  {Ferrante}, \citenamefont {Turgut}, \citenamefont {Dorigo},\ and\
  \citenamefont {Huepe}}]{ferrante_elasticity-based_2013}%
  \BibitemOpen
  \bibfield  {author} {\bibinfo {author} {\bibfnamefont {E.}~\bibnamefont
  {Ferrante}}, \bibinfo {author} {\bibfnamefont {A.~E.}\ \bibnamefont
  {Turgut}}, \bibinfo {author} {\bibfnamefont {M.}~\bibnamefont {Dorigo}},\
  and\ \bibinfo {author} {\bibfnamefont {C.}~\bibnamefont {Huepe}},\ }\bibfield
   {title} {\bibinfo {title} {Elasticity-{Based} {Mechanism} for the
  {Collective} {Motion} of {Self}-{Propelled} {Particles} with {Springlike}
  {Interactions}: {A} {Model} {System} for {Natural} and {Artificial}
  {Swarms}},\ }\href {https://doi.org/10.1103/PhysRevLett.111.268302}
  {\bibfield  {journal} {\bibinfo  {journal} {Physical Review Letters}\
  }\textbf {\bibinfo {volume} {111}},\ \bibinfo {pages} {268302} (\bibinfo
  {year} {2013})},\ \bibinfo {note} {publisher: American Physical
  Society}\BibitemShut {NoStop}%
\bibitem [{\citenamefont {Henkes}\ \emph {et~al.}(2020)\citenamefont {Henkes},
  \citenamefont {Kostanjevec}, \citenamefont {Collinson}, \citenamefont
  {Sknepnek},\ and\ \citenamefont {Bertin}}]{henkes_dense_2020}%
  \BibitemOpen
  \bibfield  {author} {\bibinfo {author} {\bibfnamefont {S.}~\bibnamefont
  {Henkes}}, \bibinfo {author} {\bibfnamefont {K.}~\bibnamefont {Kostanjevec}},
  \bibinfo {author} {\bibfnamefont {J.~M.}\ \bibnamefont {Collinson}}, \bibinfo
  {author} {\bibfnamefont {R.}~\bibnamefont {Sknepnek}},\ and\ \bibinfo
  {author} {\bibfnamefont {E.}~\bibnamefont {Bertin}},\ }\bibfield  {title}
  {\bibinfo {title} {Dense active matter model of motion patterns in confluent
  cell monolayers},\ }\href {https://doi.org/10.1038/s41467-020-15164-5}
  {\bibfield  {journal} {\bibinfo  {journal} {Nature Communications}\ }\textbf
  {\bibinfo {volume} {11}},\ \bibinfo {pages} {1405} (\bibinfo {year}
  {2020})},\ \bibinfo {note} {number: 1 Publisher: Nature Publishing
  Group}\BibitemShut {NoStop}%
\bibitem [{\citenamefont {van~der Vaart}\ \emph {et~al.}(2020)\citenamefont
  {van~der Vaart}, \citenamefont {Sinhuber}, \citenamefont {Reynolds},\ and\
  \citenamefont {Ouellette}}]{van_der_vaart_environmental_2020}%
  \BibitemOpen
  \bibfield  {author} {\bibinfo {author} {\bibfnamefont {K.}~\bibnamefont
  {van~der Vaart}}, \bibinfo {author} {\bibfnamefont {M.}~\bibnamefont
  {Sinhuber}}, \bibinfo {author} {\bibfnamefont {A.~M.}\ \bibnamefont
  {Reynolds}},\ and\ \bibinfo {author} {\bibfnamefont {N.~T.}\ \bibnamefont
  {Ouellette}},\ }\bibfield  {title} {\bibinfo {title} {Environmental
  perturbations induce correlations in midge swarms},\ }\href
  {https://doi.org/10.1098/rsif.2020.0018} {\bibfield  {journal} {\bibinfo
  {journal} {Journal of The Royal Society Interface}\ }\textbf {\bibinfo
  {volume} {17}},\ \bibinfo {pages} {20200018} (\bibinfo {year} {2020})},\
  \bibinfo {note} {publisher: Royal Society}\BibitemShut {NoStop}%
\bibitem [{\citenamefont {Attanasi}\ \emph {et~al.}(2014)\citenamefont
  {Attanasi}, \citenamefont {Cavagna}, \citenamefont {Del~Castello},
  \citenamefont {Giardina}, \citenamefont {Grigera}, \citenamefont {Jelić},
  \citenamefont {Melillo}, \citenamefont {Parisi}, \citenamefont {Pohl},
  \citenamefont {Shen},\ and\ \citenamefont
  {Viale}}]{attanasi_information_2014}%
  \BibitemOpen
  \bibfield  {author} {\bibinfo {author} {\bibfnamefont {A.}~\bibnamefont
  {Attanasi}}, \bibinfo {author} {\bibfnamefont {A.}~\bibnamefont {Cavagna}},
  \bibinfo {author} {\bibfnamefont {L.}~\bibnamefont {Del~Castello}}, \bibinfo
  {author} {\bibfnamefont {I.}~\bibnamefont {Giardina}}, \bibinfo {author}
  {\bibfnamefont {T.~S.}\ \bibnamefont {Grigera}}, \bibinfo {author}
  {\bibfnamefont {A.}~\bibnamefont {Jelić}}, \bibinfo {author} {\bibfnamefont
  {S.}~\bibnamefont {Melillo}}, \bibinfo {author} {\bibfnamefont
  {L.}~\bibnamefont {Parisi}}, \bibinfo {author} {\bibfnamefont
  {O.}~\bibnamefont {Pohl}}, \bibinfo {author} {\bibfnamefont {E.}~\bibnamefont
  {Shen}},\ and\ \bibinfo {author} {\bibfnamefont {M.}~\bibnamefont {Viale}},\
  }\bibfield  {title} {\bibinfo {title} {Information transfer and behavioural
  inertia in starling flocks},\ }\href {https://doi.org/10.1038/nphys3035}
  {\bibfield  {journal} {\bibinfo  {journal} {Nature Physics}\ }\textbf
  {\bibinfo {volume} {10}},\ \bibinfo {pages} {691} (\bibinfo {year} {2014})},\
  \bibinfo {note} {number: 9 Publisher: Nature Publishing Group}\BibitemShut
  {NoStop}%
\bibitem [{\citenamefont {Tunstrøm}\ \emph {et~al.}(2013)\citenamefont
  {Tunstrøm}, \citenamefont {Katz}, \citenamefont {Ioannou}, \citenamefont
  {Huepe}, \citenamefont {Lutz},\ and\ \citenamefont
  {Couzin}}]{tunstrom_collective_2013}%
  \BibitemOpen
  \bibfield  {author} {\bibinfo {author} {\bibfnamefont {K.}~\bibnamefont
  {Tunstrøm}}, \bibinfo {author} {\bibfnamefont {Y.}~\bibnamefont {Katz}},
  \bibinfo {author} {\bibfnamefont {C.~C.}\ \bibnamefont {Ioannou}}, \bibinfo
  {author} {\bibfnamefont {C.}~\bibnamefont {Huepe}}, \bibinfo {author}
  {\bibfnamefont {M.~J.}\ \bibnamefont {Lutz}},\ and\ \bibinfo {author}
  {\bibfnamefont {I.~D.}\ \bibnamefont {Couzin}},\ }\bibfield  {title}
  {\bibinfo {title} {Collective {States}, {Multistability} and {Transitional}
  {Behavior} in {Schooling} {Fish}},\ }\href
  {https://doi.org/10.1371/journal.pcbi.1002915} {\bibfield  {journal}
  {\bibinfo  {journal} {PLOS Computational Biology}\ }\textbf {\bibinfo
  {volume} {9}},\ \bibinfo {pages} {e1002915} (\bibinfo {year} {2013})},\
  \bibinfo {note} {publisher: Public Library of Science}\BibitemShut {NoStop}%
\bibitem [{\citenamefont {Katz}\ \emph {et~al.}(2011)\citenamefont {Katz},
  \citenamefont {Tunstrøm}, \citenamefont {Ioannou}, \citenamefont {Huepe},\
  and\ \citenamefont {Couzin}}]{katz_inferring_2011}%
  \BibitemOpen
  \bibfield  {author} {\bibinfo {author} {\bibfnamefont {Y.}~\bibnamefont
  {Katz}}, \bibinfo {author} {\bibfnamefont {K.}~\bibnamefont {Tunstrøm}},
  \bibinfo {author} {\bibfnamefont {C.~C.}\ \bibnamefont {Ioannou}}, \bibinfo
  {author} {\bibfnamefont {C.}~\bibnamefont {Huepe}},\ and\ \bibinfo {author}
  {\bibfnamefont {I.~D.}\ \bibnamefont {Couzin}},\ }\bibfield  {title}
  {\bibinfo {title} {Inferring the structure and dynamics of interactions in
  schooling fish},\ }\href@noop {} {\bibfield  {journal} {\bibinfo  {journal}
  {Proceedings of the National Academy of Sciences}\ }\textbf {\bibinfo
  {volume} {108}},\ \bibinfo {pages} {18720} (\bibinfo {year}
  {2011})}\BibitemShut {NoStop}%
\bibitem [{\citenamefont {Blanch-Mercader}\ and\ \citenamefont
  {Casademunt}(2017)}]{blanch-mercader_hydrodynamic_2017}%
  \BibitemOpen
  \bibfield  {author} {\bibinfo {author} {\bibfnamefont {C.}~\bibnamefont
  {Blanch-Mercader}}\ and\ \bibinfo {author} {\bibfnamefont {J.}~\bibnamefont
  {Casademunt}},\ }\bibfield  {title} {\bibinfo {title} {Hydrodynamic
  instabilities, waves and turbulence in spreading epithelia},\ }\href
  {https://doi.org/10.1039/C7SM01128H} {\bibfield  {journal} {\bibinfo
  {journal} {Soft Matter}\ }\textbf {\bibinfo {volume} {13}},\ \bibinfo {pages}
  {6913} (\bibinfo {year} {2017})},\ \bibinfo {note} {publisher: The Royal
  Society of Chemistry}\BibitemShut {NoStop}%
\bibitem [{\citenamefont {Gilpin}\ \emph {et~al.}(2020)\citenamefont {Gilpin},
  \citenamefont {Bull},\ and\ \citenamefont
  {Prakash}}]{gilpin_multiscale_2020}%
  \BibitemOpen
  \bibfield  {author} {\bibinfo {author} {\bibfnamefont {W.}~\bibnamefont
  {Gilpin}}, \bibinfo {author} {\bibfnamefont {M.~S.}\ \bibnamefont {Bull}},\
  and\ \bibinfo {author} {\bibfnamefont {M.}~\bibnamefont {Prakash}},\
  }\bibfield  {title} {\bibinfo {title} {The multiscale physics of cilia and
  flagella},\ }\href {https://doi.org/10.1038/s42254-019-0129-0} {\bibfield
  {journal} {\bibinfo  {journal} {Nature Reviews Physics}\ }\textbf {\bibinfo
  {volume} {2}},\ \bibinfo {pages} {74} (\bibinfo {year} {2020})},\ \bibinfo
  {note} {number: 2 Publisher: Nature Publishing Group}\BibitemShut {NoStop}%
\bibitem [{\citenamefont {Mongera}\ \emph {et~al.}(2018)\citenamefont
  {Mongera}, \citenamefont {Rowghanian}, \citenamefont {Gustafson},
  \citenamefont {Shelton}, \citenamefont {Kealhofer}, \citenamefont {Carn},
  \citenamefont {Serwane}, \citenamefont {Lucio}, \citenamefont {Giammona},\
  and\ \citenamefont {Camp{\`{a}}s}}]{Mongera2018AElongation}%
  \BibitemOpen
  \bibfield  {author} {\bibinfo {author} {\bibfnamefont {A.}~\bibnamefont
  {Mongera}}, \bibinfo {author} {\bibfnamefont {P.}~\bibnamefont {Rowghanian}},
  \bibinfo {author} {\bibfnamefont {H.~J.}\ \bibnamefont {Gustafson}}, \bibinfo
  {author} {\bibfnamefont {E.}~\bibnamefont {Shelton}}, \bibinfo {author}
  {\bibfnamefont {D.~A.}\ \bibnamefont {Kealhofer}}, \bibinfo {author}
  {\bibfnamefont {E.~K.}\ \bibnamefont {Carn}}, \bibinfo {author}
  {\bibfnamefont {F.}~\bibnamefont {Serwane}}, \bibinfo {author} {\bibfnamefont
  {A.~A.}\ \bibnamefont {Lucio}}, \bibinfo {author} {\bibfnamefont
  {J.}~\bibnamefont {Giammona}},\ and\ \bibinfo {author} {\bibfnamefont
  {O.}~\bibnamefont {Camp{\`{a}}s}},\ }\bibfield  {title} {\bibinfo {title} {{A
  fluid-to-solid jamming transition underlies vertebrate body axis
  elongation}},\ }\bibfield  {journal} {\bibinfo  {journal} {Nature}\ }\href
  {https://doi.org/10.1038/s41586-018-0479-2} {10.1038/s41586-018-0479-2}
  (\bibinfo {year} {2018})\BibitemShut {NoStop}%
\bibitem [{\citenamefont {Gilpin}\ \emph
  {et~al.}(2017{\natexlab{a}})\citenamefont {Gilpin}, \citenamefont {Prakash},\
  and\ \citenamefont {Prakash}}]{Gilpin2017}%
  \BibitemOpen
  \bibfield  {author} {\bibinfo {author} {\bibfnamefont {W.}~\bibnamefont
  {Gilpin}}, \bibinfo {author} {\bibfnamefont {V.~N.}\ \bibnamefont
  {Prakash}},\ and\ \bibinfo {author} {\bibfnamefont {M.}~\bibnamefont
  {Prakash}},\ }\bibfield  {title} {\bibinfo {title} {{Flowtrace: simple
  visualization of coherent structures in biological fluid flows}},\ }\href
  {https://doi.org/10.1242/jeb.162511} {\bibfield  {journal} {\bibinfo
  {journal} {The Journal of Experimental Biology}\ ,\ \bibinfo {pages}
  {jeb.162511}} (\bibinfo {year} {2017}{\natexlab{a}})}\BibitemShut {NoStop}%
\bibitem [{\citenamefont {Geyer}\ \emph {et~al.}(2018)\citenamefont {Geyer},
  \citenamefont {Morin},\ and\ \citenamefont {Bartolo}}]{geyer_sounds_2018}%
  \BibitemOpen
  \bibfield  {author} {\bibinfo {author} {\bibfnamefont {D.}~\bibnamefont
  {Geyer}}, \bibinfo {author} {\bibfnamefont {A.}~\bibnamefont {Morin}},\ and\
  \bibinfo {author} {\bibfnamefont {D.}~\bibnamefont {Bartolo}},\ }\bibfield
  {title} {\bibinfo {title} {Sounds and hydrodynamics of polar active fluids},\
  }\href {https://doi.org/10.1038/s41563-018-0123-4} {\bibfield  {journal}
  {\bibinfo  {journal} {Nature Materials}\ }\textbf {\bibinfo {volume} {17}},\
  \bibinfo {pages} {789} (\bibinfo {year} {2018})},\ \bibinfo {note} {number: 9
  Publisher: Nature Publishing Group}\BibitemShut {NoStop}%
\bibitem [{\citenamefont {Tu}\ \emph {et~al.}(1998)\citenamefont {Tu},
  \citenamefont {Toner},\ and\ \citenamefont {Ulm}}]{tu_sound_1998}%
  \BibitemOpen
  \bibfield  {author} {\bibinfo {author} {\bibfnamefont {Y.}~\bibnamefont
  {Tu}}, \bibinfo {author} {\bibfnamefont {J.}~\bibnamefont {Toner}},\ and\
  \bibinfo {author} {\bibfnamefont {M.}~\bibnamefont {Ulm}},\ }\bibfield
  {title} {\bibinfo {title} {Sound {Waves} and the {Absence} of {Galilean}
  {Invariance} in {Flocks}},\ }\href
  {https://doi.org/10.1103/PhysRevLett.80.4819} {\bibfield  {journal} {\bibinfo
   {journal} {Physical Review Letters}\ }\textbf {\bibinfo {volume} {80}},\
  \bibinfo {pages} {4819} (\bibinfo {year} {1998})},\ \bibinfo {note}
  {publisher: American Physical Society}\BibitemShut {NoStop}%
\bibitem [{\citenamefont {Pajic-Lijakovic}\ and\ \citenamefont
  {Milivojevic}(2020)}]{pajic-lijakovic_mechanical_2020}%
  \BibitemOpen
  \bibfield  {author} {\bibinfo {author} {\bibfnamefont {I.}~\bibnamefont
  {Pajic-Lijakovic}}\ and\ \bibinfo {author} {\bibfnamefont {M.}~\bibnamefont
  {Milivojevic}},\ }\bibfield  {title} {\bibinfo {title} {Mechanical
  {Oscillations} in {2D} {Collective} {Cell} {Migration}: {The} {Elastic}
  {Turbulence}},\ }\bibfield  {journal} {\bibinfo  {journal} {Frontiers in
  Physics}\ }\textbf {\bibinfo {volume} {8}},\ \href
  {https://doi.org/10.3389/fphy.2020.585681} {10.3389/fphy.2020.585681}
  (\bibinfo {year} {2020}),\ \bibinfo {note} {publisher: Frontiers}\BibitemShut
  {NoStop}%
\bibitem [{\citenamefont {Mathijssen}\ \emph {et~al.}(2019)\citenamefont
  {Mathijssen}, \citenamefont {Culver}, \citenamefont {Bhamla},\ and\
  \citenamefont {Prakash}}]{mathijssen_collective_2019}%
  \BibitemOpen
  \bibfield  {author} {\bibinfo {author} {\bibfnamefont {A.~J. T.~M.}\
  \bibnamefont {Mathijssen}}, \bibinfo {author} {\bibfnamefont
  {J.}~\bibnamefont {Culver}}, \bibinfo {author} {\bibfnamefont {M.~S.}\
  \bibnamefont {Bhamla}},\ and\ \bibinfo {author} {\bibfnamefont
  {M.}~\bibnamefont {Prakash}},\ }\bibfield  {title} {\bibinfo {title}
  {Collective intercellular communication through ultra-fast hydrodynamic
  trigger waves},\ }\href {https://doi.org/10.1038/s41586-019-1387-9}
  {\bibfield  {journal} {\bibinfo  {journal} {Nature}\ }\textbf {\bibinfo
  {volume} {571}},\ \bibinfo {pages} {560} (\bibinfo {year} {2019})},\ \bibinfo
  {note} {number: 7766 Publisher: Nature Publishing Group}\BibitemShut
  {NoStop}%
\bibitem [{\citenamefont {Armon}\ \emph {et~al.}(2020)\citenamefont {Armon},
  \citenamefont {Bull}, \citenamefont {Moriel}, \citenamefont {Aharoni},\ and\
  \citenamefont {Prakash}}]{armon_epithelial_2020}%
  \BibitemOpen
  \bibfield  {author} {\bibinfo {author} {\bibfnamefont {S.}~\bibnamefont
  {Armon}}, \bibinfo {author} {\bibfnamefont {M.~S.}\ \bibnamefont {Bull}},
  \bibinfo {author} {\bibfnamefont {A.}~\bibnamefont {Moriel}}, \bibinfo
  {author} {\bibfnamefont {H.}~\bibnamefont {Aharoni}},\ and\ \bibinfo {author}
  {\bibfnamefont {M.}~\bibnamefont {Prakash}},\ }\bibfield  {title} {\bibinfo
  {title} {Epithelial {Tissues} as {Active} {Solids}: {From} {Nonlinear}
  {Contraction} {Pulses} to {Rupture} {Resistance}},\ }\href
  {https://doi.org/10.1101/2020.06.15.153163} {\bibfield  {journal} {\bibinfo
  {journal} {bioRxiv}\ ,\ \bibinfo {pages} {2020.06.15.153163}} (\bibinfo
  {year} {2020})},\ \bibinfo {note} {publisher: Cold Spring Harbor Laboratory
  Section: New Results}\BibitemShut {NoStop}%
\bibitem [{\citenamefont {Boocock}\ \emph {et~al.}(2021)\citenamefont
  {Boocock}, \citenamefont {Hino}, \citenamefont {Ruzickova}, \citenamefont
  {Hirashima},\ and\ \citenamefont {Hannezo}}]{boocock_theory_2021}%
  \BibitemOpen
  \bibfield  {author} {\bibinfo {author} {\bibfnamefont {D.}~\bibnamefont
  {Boocock}}, \bibinfo {author} {\bibfnamefont {N.}~\bibnamefont {Hino}},
  \bibinfo {author} {\bibfnamefont {N.}~\bibnamefont {Ruzickova}}, \bibinfo
  {author} {\bibfnamefont {T.}~\bibnamefont {Hirashima}},\ and\ \bibinfo
  {author} {\bibfnamefont {E.}~\bibnamefont {Hannezo}},\ }\bibfield  {title}
  {\bibinfo {title} {Theory of mechanochemical patterning and optimal migration
  in cell monolayers},\ }\href {https://doi.org/10.1038/s41567-020-01037-7}
  {\bibfield  {journal} {\bibinfo  {journal} {Nature Physics}\ }\textbf
  {\bibinfo {volume} {17}},\ \bibinfo {pages} {267} (\bibinfo {year} {2021})},\
  \bibinfo {note} {number: 2 Publisher: Nature Publishing Group}\BibitemShut
  {NoStop}%
\bibitem [{\citenamefont {Banerjee}\ \emph {et~al.}(2015)\citenamefont
  {Banerjee}, \citenamefont {Utuje},\ and\ \citenamefont
  {Marchetti}}]{banerjee_propagating_2015}%
  \BibitemOpen
  \bibfield  {author} {\bibinfo {author} {\bibfnamefont {S.}~\bibnamefont
  {Banerjee}}, \bibinfo {author} {\bibfnamefont {K.~J.}\ \bibnamefont
  {Utuje}},\ and\ \bibinfo {author} {\bibfnamefont {M.~C.}\ \bibnamefont
  {Marchetti}},\ }\bibfield  {title} {\bibinfo {title} {Propagating {Stress}
  {Waves} {During} {Epithelial} {Expansion}},\ }\href
  {https://doi.org/10.1103/PhysRevLett.114.228101} {\bibfield  {journal}
  {\bibinfo  {journal} {Physical Review Letters}\ }\textbf {\bibinfo {volume}
  {114}},\ \bibinfo {pages} {228101} (\bibinfo {year} {2015})},\ \bibinfo
  {note} {publisher: American Physical Society}\BibitemShut {NoStop}%
\bibitem [{\citenamefont {Peyret}\ \emph {et~al.}(2019)\citenamefont {Peyret},
  \citenamefont {Mueller}, \citenamefont {d’Alessandro}, \citenamefont
  {Begnaud}, \citenamefont {Marcq}, \citenamefont {Mège}, \citenamefont
  {Yeomans}, \citenamefont {Doostmohammadi},\ and\ \citenamefont
  {Ladoux}}]{peyret_sustained_2019}%
  \BibitemOpen
  \bibfield  {author} {\bibinfo {author} {\bibfnamefont {G.}~\bibnamefont
  {Peyret}}, \bibinfo {author} {\bibfnamefont {R.}~\bibnamefont {Mueller}},
  \bibinfo {author} {\bibfnamefont {J.}~\bibnamefont {d’Alessandro}},
  \bibinfo {author} {\bibfnamefont {S.}~\bibnamefont {Begnaud}}, \bibinfo
  {author} {\bibfnamefont {P.}~\bibnamefont {Marcq}}, \bibinfo {author}
  {\bibfnamefont {R.-M.}\ \bibnamefont {Mège}}, \bibinfo {author}
  {\bibfnamefont {J.~M.}\ \bibnamefont {Yeomans}}, \bibinfo {author}
  {\bibfnamefont {A.}~\bibnamefont {Doostmohammadi}},\ and\ \bibinfo {author}
  {\bibfnamefont {B.}~\bibnamefont {Ladoux}},\ }\bibfield  {title} {\bibinfo
  {title} {Sustained {Oscillations} of {Epithelial} {Cell} {Sheets}},\ }\href
  {https://doi.org/10.1016/j.bpj.2019.06.013} {\bibfield  {journal} {\bibinfo
  {journal} {Biophysical Journal}\ }\textbf {\bibinfo {volume} {117}},\
  \bibinfo {pages} {464} (\bibinfo {year} {2019})}\BibitemShut {NoStop}%
\bibitem [{\citenamefont {Bi}\ \emph {et~al.}(2015)\citenamefont {Bi},
  \citenamefont {Lopez}, \citenamefont {Schwarz},\ and\ \citenamefont
  {Manning}}]{Bi2015a}%
  \BibitemOpen
  \bibfield  {author} {\bibinfo {author} {\bibfnamefont {D.}~\bibnamefont
  {Bi}}, \bibinfo {author} {\bibfnamefont {J.~H.}\ \bibnamefont {Lopez}},
  \bibinfo {author} {\bibfnamefont {J.~M.}\ \bibnamefont {Schwarz}},\ and\
  \bibinfo {author} {\bibfnamefont {M.~L.}\ \bibnamefont {Manning}},\
  }\bibfield  {title} {\bibinfo {title} {{A density-independent rigidity
  transition in biological tissues}},\ }\href
  {https://doi.org/10.1038/nphys3471} {\bibfield  {journal} {\bibinfo
  {journal} {Nature Physics}\ }\textbf {\bibinfo {volume} {11}},\ \bibinfo
  {pages} {1074} (\bibinfo {year} {2015})}\BibitemShut {NoStop}%
\bibitem [{\citenamefont {Serra-Picamal}\ \emph {et~al.}(2012)\citenamefont
  {Serra-Picamal}, \citenamefont {Conte}, \citenamefont {Vincent},
  \citenamefont {Anon}, \citenamefont {Tambe}, \citenamefont {Bazellieres},
  \citenamefont {Butler}, \citenamefont {Fredberg},\ and\ \citenamefont
  {Trepat}}]{serra-picamal_mechanical_2012}%
  \BibitemOpen
  \bibfield  {author} {\bibinfo {author} {\bibfnamefont {X.}~\bibnamefont
  {Serra-Picamal}}, \bibinfo {author} {\bibfnamefont {V.}~\bibnamefont
  {Conte}}, \bibinfo {author} {\bibfnamefont {R.}~\bibnamefont {Vincent}},
  \bibinfo {author} {\bibfnamefont {E.}~\bibnamefont {Anon}}, \bibinfo {author}
  {\bibfnamefont {D.~T.}\ \bibnamefont {Tambe}}, \bibinfo {author}
  {\bibfnamefont {E.}~\bibnamefont {Bazellieres}}, \bibinfo {author}
  {\bibfnamefont {J.~P.}\ \bibnamefont {Butler}}, \bibinfo {author}
  {\bibfnamefont {J.~J.}\ \bibnamefont {Fredberg}},\ and\ \bibinfo {author}
  {\bibfnamefont {X.}~\bibnamefont {Trepat}},\ }\bibfield  {title} {\bibinfo
  {title} {Mechanical waves during tissue expansion},\ }\href
  {https://doi.org/10.1038/nphys2355} {\bibfield  {journal} {\bibinfo
  {journal} {Nature Physics}\ }\textbf {\bibinfo {volume} {8}},\ \bibinfo
  {pages} {628} (\bibinfo {year} {2012})},\ \bibinfo {note} {number: 8
  Publisher: Nature Publishing Group}\BibitemShut {NoStop}%
\bibitem [{\citenamefont {Camalet}\ and\ \citenamefont
  {J{\"u}licher}(2000)}]{camalet2000generic}%
  \BibitemOpen
  \bibfield  {author} {\bibinfo {author} {\bibfnamefont {S.}~\bibnamefont
  {Camalet}}\ and\ \bibinfo {author} {\bibfnamefont {F.}~\bibnamefont
  {J{\"u}licher}},\ }\bibfield  {title} {\bibinfo {title} {Generic aspects of
  axonemal beating},\ }\href@noop {} {\bibfield  {journal} {\bibinfo  {journal}
  {New Journal of Physics}\ }\textbf {\bibinfo {volume} {2}},\ \bibinfo {pages}
  {24} (\bibinfo {year} {2000})}\BibitemShut {NoStop}%
\bibitem [{\citenamefont {Ma}\ \emph {et~al.}(2014)\citenamefont {Ma},
  \citenamefont {Klindt}, \citenamefont {Riedel-Kruse}, \citenamefont
  {J{\"u}licher},\ and\ \citenamefont {Friedrich}}]{ma2014active}%
  \BibitemOpen
  \bibfield  {author} {\bibinfo {author} {\bibfnamefont {R.}~\bibnamefont
  {Ma}}, \bibinfo {author} {\bibfnamefont {G.~S.}\ \bibnamefont {Klindt}},
  \bibinfo {author} {\bibfnamefont {I.~H.}\ \bibnamefont {Riedel-Kruse}},
  \bibinfo {author} {\bibfnamefont {F.}~\bibnamefont {J{\"u}licher}},\ and\
  \bibinfo {author} {\bibfnamefont {B.~M.}\ \bibnamefont {Friedrich}},\
  }\bibfield  {title} {\bibinfo {title} {Active phase and amplitude
  fluctuations of flagellar beating},\ }\href@noop {} {\bibfield  {journal}
  {\bibinfo  {journal} {Physical review letters}\ }\textbf {\bibinfo {volume}
  {113}},\ \bibinfo {pages} {048101} (\bibinfo {year} {2014})}\BibitemShut
  {NoStop}%
\bibitem [{\citenamefont {Rojas}\ \emph {et~al.}(2014)\citenamefont {Rojas},
  \citenamefont {Theriot},\ and\ \citenamefont {Huang}}]{rojas_response_2014}%
  \BibitemOpen
  \bibfield  {author} {\bibinfo {author} {\bibfnamefont {E.}~\bibnamefont
  {Rojas}}, \bibinfo {author} {\bibfnamefont {J.~A.}\ \bibnamefont {Theriot}},\
  and\ \bibinfo {author} {\bibfnamefont {K.~C.}\ \bibnamefont {Huang}},\
  }\bibfield  {title} {\bibinfo {title} {Response of {Escherichia} coli growth
  rate to osmotic shock},\ }\bibfield  {journal} {\bibinfo  {journal}
  {Proceedings of the National Academy of Sciences}\ }\href
  {https://doi.org/10.1073/pnas.1402591111} {10.1073/pnas.1402591111} (\bibinfo
  {year} {2014}),\ \bibinfo {note} {publisher: National Academy of Sciences
  Section: Biological Sciences}\BibitemShut {NoStop}%
\bibitem [{\citenamefont {Russell}\ \emph {et~al.}(2017)\citenamefont
  {Russell}, \citenamefont {Theriot}, \citenamefont {Sood}, \citenamefont
  {Marshall}, \citenamefont {Landweber}, \citenamefont {Fritz-Laylin},
  \citenamefont {Polka}, \citenamefont {Oliferenko}, \citenamefont {Gerbich},
  \citenamefont {Gladfelter}, \citenamefont {Umen}, \citenamefont {Bezanilla},
  \citenamefont {Lancaster}, \citenamefont {He}, \citenamefont {Gibson},
  \citenamefont {Goldstein}, \citenamefont {Tanaka}, \citenamefont {Hu},\ and\
  \citenamefont {Brunet}}]{russell_non-model_2017}%
  \BibitemOpen
  \bibfield  {author} {\bibinfo {author} {\bibfnamefont {J.~J.}\ \bibnamefont
  {Russell}}, \bibinfo {author} {\bibfnamefont {J.~A.}\ \bibnamefont
  {Theriot}}, \bibinfo {author} {\bibfnamefont {P.}~\bibnamefont {Sood}},
  \bibinfo {author} {\bibfnamefont {W.~F.}\ \bibnamefont {Marshall}}, \bibinfo
  {author} {\bibfnamefont {L.~F.}\ \bibnamefont {Landweber}}, \bibinfo {author}
  {\bibfnamefont {L.}~\bibnamefont {Fritz-Laylin}}, \bibinfo {author}
  {\bibfnamefont {J.~K.}\ \bibnamefont {Polka}}, \bibinfo {author}
  {\bibfnamefont {S.}~\bibnamefont {Oliferenko}}, \bibinfo {author}
  {\bibfnamefont {T.}~\bibnamefont {Gerbich}}, \bibinfo {author} {\bibfnamefont
  {A.}~\bibnamefont {Gladfelter}}, \bibinfo {author} {\bibfnamefont
  {J.}~\bibnamefont {Umen}}, \bibinfo {author} {\bibfnamefont {M.}~\bibnamefont
  {Bezanilla}}, \bibinfo {author} {\bibfnamefont {M.~A.}\ \bibnamefont
  {Lancaster}}, \bibinfo {author} {\bibfnamefont {S.}~\bibnamefont {He}},
  \bibinfo {author} {\bibfnamefont {M.~C.}\ \bibnamefont {Gibson}}, \bibinfo
  {author} {\bibfnamefont {B.}~\bibnamefont {Goldstein}}, \bibinfo {author}
  {\bibfnamefont {E.~M.}\ \bibnamefont {Tanaka}}, \bibinfo {author}
  {\bibfnamefont {C.-K.}\ \bibnamefont {Hu}},\ and\ \bibinfo {author}
  {\bibfnamefont {A.}~\bibnamefont {Brunet}},\ }\bibfield  {title} {\bibinfo
  {title} {Non-model model organisms},\ }\href
  {https://doi.org/10.1186/s12915-017-0391-5} {\bibfield  {journal} {\bibinfo
  {journal} {BMC Biology}\ }\textbf {\bibinfo {volume} {15}},\ \bibinfo {pages}
  {55} (\bibinfo {year} {2017})}\BibitemShut {NoStop}%
\bibitem [{\citenamefont {Smith}\ \emph
  {et~al.}(2014{\natexlab{a}})\citenamefont {Smith}, \citenamefont
  {Varoqueaux}, \citenamefont {Kittelmann}, \citenamefont {Azzam},
  \citenamefont {Cooper}, \citenamefont {Winters}, \citenamefont {Eitel},
  \citenamefont {Fasshauer},\ and\ \citenamefont {Reese}}]{Smith2014}%
  \BibitemOpen
  \bibfield  {author} {\bibinfo {author} {\bibfnamefont {C.~L.}\ \bibnamefont
  {Smith}}, \bibinfo {author} {\bibfnamefont {F.}~\bibnamefont {Varoqueaux}},
  \bibinfo {author} {\bibfnamefont {M.}~\bibnamefont {Kittelmann}}, \bibinfo
  {author} {\bibfnamefont {R.~N.}\ \bibnamefont {Azzam}}, \bibinfo {author}
  {\bibfnamefont {B.}~\bibnamefont {Cooper}}, \bibinfo {author} {\bibfnamefont
  {C.~A.}\ \bibnamefont {Winters}}, \bibinfo {author} {\bibfnamefont
  {M.}~\bibnamefont {Eitel}}, \bibinfo {author} {\bibfnamefont
  {D.}~\bibnamefont {Fasshauer}},\ and\ \bibinfo {author} {\bibfnamefont
  {T.~S.}\ \bibnamefont {Reese}},\ }\bibfield  {title} {\bibinfo {title}
  {{Novel cell types, neurosecretory cells, and body plan of the
  early-diverging metazoan Trichoplax adhaerens}},\ }\href
  {https://doi.org/10.1016/j.cub.2014.05.046} {\bibfield  {journal} {\bibinfo
  {journal} {Current Biology}\ }\textbf {\bibinfo {volume} {24}},\ \bibinfo
  {pages} {1565} (\bibinfo {year} {2014}{\natexlab{a}})},\ \Eprint
  {https://arxiv.org/abs/NIHMS150003} {arXiv:NIHMS150003} \BibitemShut
  {NoStop}%
\bibitem [{\citenamefont {DuBuc}\ \emph {et~al.}(2019)\citenamefont {DuBuc},
  \citenamefont {Ryan},\ and\ \citenamefont
  {Martindale}}]{dubuc_dorsalventral_2019}%
  \BibitemOpen
  \bibfield  {author} {\bibinfo {author} {\bibfnamefont {T.~Q.}\ \bibnamefont
  {DuBuc}}, \bibinfo {author} {\bibfnamefont {J.~F.}\ \bibnamefont {Ryan}},\
  and\ \bibinfo {author} {\bibfnamefont {M.~Q.}\ \bibnamefont {Martindale}},\
  }\bibfield  {title} {\bibinfo {title} {“{Dorsal}–{Ventral}” {Genes}
  {Are} {Part} of an {Ancient} {Axial} {Patterning} {System}: {Evidence} from
  {Trichoplax} adhaerens ({Placozoa})},\ }\href
  {https://doi.org/10.1093/molbev/msz025} {\bibfield  {journal} {\bibinfo
  {journal} {Molecular Biology and Evolution}\ }\textbf {\bibinfo {volume}
  {36}},\ \bibinfo {pages} {966} (\bibinfo {year} {2019})}\BibitemShut
  {NoStop}%
\bibitem [{\citenamefont {Smith}\ \emph {et~al.}(2015)\citenamefont {Smith},
  \citenamefont {Pivovarova},\ and\ \citenamefont
  {Reese}}]{smith2015coordinated}%
  \BibitemOpen
  \bibfield  {author} {\bibinfo {author} {\bibfnamefont {C.~L.}\ \bibnamefont
  {Smith}}, \bibinfo {author} {\bibfnamefont {N.}~\bibnamefont {Pivovarova}},\
  and\ \bibinfo {author} {\bibfnamefont {T.~S.}\ \bibnamefont {Reese}},\
  }\bibfield  {title} {\bibinfo {title} {Coordinated feeding behavior in
  trichoplax, an animal without synapses},\ }\href@noop {} {\bibfield
  {journal} {\bibinfo  {journal} {PLoS One}\ }\textbf {\bibinfo {volume}
  {10}},\ \bibinfo {pages} {e0136098} (\bibinfo {year} {2015})}\BibitemShut
  {NoStop}%
\bibitem [{\citenamefont {Smith}\ and\ \citenamefont
  {Reese}(2016)}]{Smith2016AdherensAdhaerens.}%
  \BibitemOpen
  \bibfield  {author} {\bibinfo {author} {\bibfnamefont {C.~L.}\ \bibnamefont
  {Smith}}\ and\ \bibinfo {author} {\bibfnamefont {T.~S.}\ \bibnamefont
  {Reese}},\ }\bibfield  {title} {\bibinfo {title} {{Adherens Junctions
  Modulate Diffusion between Epithelial Cells in Trichoplax adhaerens.}},\
  }\href {https://doi.org/10.1086/691069} {\bibfield  {journal} {\bibinfo
  {journal} {The Biological bulletin}\ }\textbf {\bibinfo {volume} {231}},\
  \bibinfo {pages} {216} (\bibinfo {year} {2016})}\BibitemShut {NoStop}%
\bibitem [{\citenamefont {Senatore}\ \emph {et~al.}(2017)\citenamefont
  {Senatore}, \citenamefont {Reese},\ and\ \citenamefont
  {Smith}}]{Senatore2017NeuropeptidergicSynapses.}%
  \BibitemOpen
  \bibfield  {author} {\bibinfo {author} {\bibfnamefont {A.}~\bibnamefont
  {Senatore}}, \bibinfo {author} {\bibfnamefont {T.~S.}\ \bibnamefont
  {Reese}},\ and\ \bibinfo {author} {\bibfnamefont {C.~L.}\ \bibnamefont
  {Smith}},\ }\bibfield  {title} {\bibinfo {title} {{Neuropeptidergic
  integration of behavior in Trichoplax adhaerens, an animal without
  synapses.}},\ }\href {https://doi.org/10.1242/jeb.162396} {\bibfield
  {journal} {\bibinfo  {journal} {The Journal of experimental biology}\
  }\textbf {\bibinfo {volume} {220}},\ \bibinfo {pages} {3381} (\bibinfo {year}
  {2017})}\BibitemShut {NoStop}%
\bibitem [{\citenamefont {Ueda}\ \emph {et~al.}(1999)\citenamefont {Ueda},
  \citenamefont {Koya},\ and\ \citenamefont
  {Maruyama}}]{Ueda1999DynamicAdhaerence}%
  \BibitemOpen
  \bibfield  {author} {\bibinfo {author} {\bibfnamefont {T.}~\bibnamefont
  {Ueda}}, \bibinfo {author} {\bibfnamefont {S.}~\bibnamefont {Koya}},\ and\
  \bibinfo {author} {\bibfnamefont {Y.~K.}\ \bibnamefont {Maruyama}},\
  }\bibfield  {title} {\bibinfo {title} {{Dynamic patterns in the locomotion
  and feeding behaviors by the placozoan Trichoplax adhaerence}},\ }\href
  {https://doi.org/10.1016/S0303-2647(99)00066-0} {\bibfield  {journal}
  {\bibinfo  {journal} {Biosystems}\ }\textbf {\bibinfo {volume} {54}},\
  \bibinfo {pages} {65} (\bibinfo {year} {1999})}\BibitemShut {NoStop}%
\bibitem [{\citenamefont {Romanova}\ \emph {et~al.}(2020)\citenamefont
  {Romanova}, \citenamefont {Smirnov}, \citenamefont {Nikitin}, \citenamefont
  {Kohn}, \citenamefont {Borman}, \citenamefont {Malyshev}, \citenamefont
  {Balaban},\ and\ \citenamefont {Moroz}}]{romanova_sodium_2020}%
  \BibitemOpen
  \bibfield  {author} {\bibinfo {author} {\bibfnamefont {D.~Y.}\ \bibnamefont
  {Romanova}}, \bibinfo {author} {\bibfnamefont {I.~V.}\ \bibnamefont
  {Smirnov}}, \bibinfo {author} {\bibfnamefont {M.~A.}\ \bibnamefont
  {Nikitin}}, \bibinfo {author} {\bibfnamefont {A.~B.}\ \bibnamefont {Kohn}},
  \bibinfo {author} {\bibfnamefont {A.~I.}\ \bibnamefont {Borman}}, \bibinfo
  {author} {\bibfnamefont {A.~Y.}\ \bibnamefont {Malyshev}}, \bibinfo {author}
  {\bibfnamefont {P.~M.}\ \bibnamefont {Balaban}},\ and\ \bibinfo {author}
  {\bibfnamefont {L.~L.}\ \bibnamefont {Moroz}},\ }\bibfield  {title} {\bibinfo
  {title} {Sodium action potentials in placozoa: {Insights} into behavioral
  integration and evolution of nerveless animals},\ }\href
  {https://doi.org/10.1016/j.bbrc.2020.08.020} {\bibfield  {journal} {\bibinfo
  {journal} {Biochemical and Biophysical Research Communications}\ }\textbf
  {\bibinfo {volume} {532}},\ \bibinfo {pages} {120} (\bibinfo {year}
  {2020})}\BibitemShut {NoStop}%
\bibitem [{\citenamefont {Smith}\ \emph {et~al.}(2019)\citenamefont {Smith},
  \citenamefont {Reese}, \citenamefont {Govezensky},\ and\ \citenamefont
  {Barrio}}]{smith_coherent_2019}%
  \BibitemOpen
  \bibfield  {author} {\bibinfo {author} {\bibfnamefont {C.~L.}\ \bibnamefont
  {Smith}}, \bibinfo {author} {\bibfnamefont {T.~S.}\ \bibnamefont {Reese}},
  \bibinfo {author} {\bibfnamefont {T.}~\bibnamefont {Govezensky}},\ and\
  \bibinfo {author} {\bibfnamefont {R.~A.}\ \bibnamefont {Barrio}},\ }\bibfield
   {title} {\bibinfo {title} {Coherent directed movement toward food modeled in
  {Trichoplax}, a ciliated animal lacking a nervous system},\ }\href
  {https://doi.org/10.1073/pnas.1815655116} {\bibfield  {journal} {\bibinfo
  {journal} {Proceedings of the National Academy of Sciences}\ }\textbf
  {\bibinfo {volume} {116}},\ \bibinfo {pages} {8901} (\bibinfo {year}
  {2019})},\ \bibinfo {note} {publisher: National Academy of Sciences Section:
  PNAS Plus}\BibitemShut {NoStop}%
\bibitem [{\citenamefont {Armon}\ \emph {et~al.}(2018)\citenamefont {Armon},
  \citenamefont {Bull}, \citenamefont {Aranda-Diaz},\ and\ \citenamefont
  {Prakash}}]{armon2018ultrafast}%
  \BibitemOpen
  \bibfield  {author} {\bibinfo {author} {\bibfnamefont {S.}~\bibnamefont
  {Armon}}, \bibinfo {author} {\bibfnamefont {M.~S.}\ \bibnamefont {Bull}},
  \bibinfo {author} {\bibfnamefont {A.}~\bibnamefont {Aranda-Diaz}},\ and\
  \bibinfo {author} {\bibfnamefont {M.}~\bibnamefont {Prakash}},\ }\bibfield
  {title} {\bibinfo {title} {Ultrafast epithelial contractions provide insights
  into contraction speed limits and tissue integrity},\ }\href@noop {}
  {\bibfield  {journal} {\bibinfo  {journal} {Proceedings of the National
  Academy of Sciences}\ }\textbf {\bibinfo {volume} {115}},\ \bibinfo {pages}
  {E10333} (\bibinfo {year} {2018})}\BibitemShut {NoStop}%
\bibitem [{\citenamefont {Prakash}\ \emph {et~al.}(2021)\citenamefont
  {Prakash}, \citenamefont {Bull},\ and\ \citenamefont
  {Prakash}}]{prakash_motility-induced_2021}%
  \BibitemOpen
  \bibfield  {author} {\bibinfo {author} {\bibfnamefont {V.~N.}\ \bibnamefont
  {Prakash}}, \bibinfo {author} {\bibfnamefont {M.~S.}\ \bibnamefont {Bull}},\
  and\ \bibinfo {author} {\bibfnamefont {M.}~\bibnamefont {Prakash}},\
  }\bibfield  {title} {\bibinfo {title} {Motility-induced fracture reveals a
  ductile-to-brittle crossover in a simple animal’s epithelia},\ }\href
  {https://doi.org/10.1038/s41567-020-01134-7} {\bibfield  {journal} {\bibinfo
  {journal} {Nature Physics}\ }\textbf {\bibinfo {volume} {17}},\ \bibinfo
  {pages} {504} (\bibinfo {year} {2021})},\ \bibinfo {note} {number: 4
  Publisher: Nature Publishing Group}\BibitemShut {NoStop}%
\bibitem [{\citenamefont {Nguyen}\ \emph {et~al.}(2016)\citenamefont {Nguyen},
  \citenamefont {Shipley}, \citenamefont {Linder}, \citenamefont {Plummer},
  \citenamefont {Liu}, \citenamefont {Setru}, \citenamefont {Shaevitz},\ and\
  \citenamefont {Leifer}}]{nguyen_whole-brain_2016}%
  \BibitemOpen
  \bibfield  {author} {\bibinfo {author} {\bibfnamefont {J.~P.}\ \bibnamefont
  {Nguyen}}, \bibinfo {author} {\bibfnamefont {F.~B.}\ \bibnamefont {Shipley}},
  \bibinfo {author} {\bibfnamefont {A.~N.}\ \bibnamefont {Linder}}, \bibinfo
  {author} {\bibfnamefont {G.~S.}\ \bibnamefont {Plummer}}, \bibinfo {author}
  {\bibfnamefont {M.}~\bibnamefont {Liu}}, \bibinfo {author} {\bibfnamefont
  {S.~U.}\ \bibnamefont {Setru}}, \bibinfo {author} {\bibfnamefont {J.~W.}\
  \bibnamefont {Shaevitz}},\ and\ \bibinfo {author} {\bibfnamefont {A.~M.}\
  \bibnamefont {Leifer}},\ }\bibfield  {title} {\bibinfo {title} {Whole-brain
  calcium imaging with cellular resolution in freely behaving {Caenorhabditis}
  elegans},\ }\href@noop {} {\bibfield  {journal} {\bibinfo  {journal}
  {Proceedings of the National Academy of Sciences}\ }\textbf {\bibinfo
  {volume} {113}},\ \bibinfo {pages} {E1074} (\bibinfo {year}
  {2016})}\BibitemShut {NoStop}%
\bibitem [{\citenamefont {Kovesi}(1997)}]{kovesi_symmetry_1997}%
  \BibitemOpen
  \bibfield  {author} {\bibinfo {author} {\bibfnamefont {P.}~\bibnamefont
  {Kovesi}},\ }\bibfield  {title} {\bibinfo {title} {Symmetry and {Asymmetry}
  from {Local} {Phase}},\ }in\ \href@noop {} {\emph {\bibinfo {booktitle}
  {Tenth {Australian} {Joint} {Converence} on {Artificial} {Intelligence}}}}\
  (\bibinfo {year} {1997})\ pp.\ \bibinfo {pages} {2--4}\BibitemShut {NoStop}%
\bibitem [{\citenamefont {Thielicke}\ and\ \citenamefont
  {Stamhuis}(2014)}]{Thielicke2014}%
  \BibitemOpen
  \bibfield  {author} {\bibinfo {author} {\bibfnamefont {W.}~\bibnamefont
  {Thielicke}}\ and\ \bibinfo {author} {\bibfnamefont {E.~J.}\ \bibnamefont
  {Stamhuis}},\ }\bibfield  {title} {\bibinfo {title} {{PIVlab – Towards
  User-friendly, Affordable and Accurate Digital Particle Image Velocimetry in
  MATLAB}},\ }\bibfield  {journal} {\bibinfo  {journal} {Journal of Open
  Research Software}\ }\textbf {\bibinfo {volume} {2}},\ \href
  {https://doi.org/10.5334/jors.bl} {10.5334/jors.bl} (\bibinfo {year}
  {2014})\BibitemShut {NoStop}%
\bibitem [{\citenamefont {Tamm}\ and\ \citenamefont
  {Horridge}(1970)}]{tamm_relation_1970}%
  \BibitemOpen
  \bibfield  {author} {\bibinfo {author} {\bibfnamefont {S.~L.}\ \bibnamefont
  {Tamm}}\ and\ \bibinfo {author} {\bibfnamefont {G.~A.}\ \bibnamefont
  {Horridge}},\ }\bibfield  {title} {\bibinfo {title} {The relation between the
  orientation of the central fibrils and the direction of beat in cilia of
  {Opalina}},\ }\href {https://doi.org/10.1098/rspb.1970.0020} {\bibfield
  {journal} {\bibinfo  {journal} {Proceedings of the Royal Society of London.
  Series B. Biological Sciences}\ }\textbf {\bibinfo {volume} {175}},\ \bibinfo
  {pages} {219} (\bibinfo {year} {1970})},\ \bibinfo {note} {publisher: Royal
  Society}\BibitemShut {NoStop}%
\bibitem [{\citenamefont {Słomka}\ and\ \citenamefont
  {Dunkel}(2017)}]{slomka_geometry-dependent_2017}%
  \BibitemOpen
  \bibfield  {author} {\bibinfo {author} {\bibfnamefont {J.}~\bibnamefont
  {Słomka}}\ and\ \bibinfo {author} {\bibfnamefont {J.}~\bibnamefont
  {Dunkel}},\ }\bibfield  {title} {\bibinfo {title} {Geometry-dependent
  viscosity reduction in sheared active fluids},\ }\href
  {https://doi.org/10.1103/PhysRevFluids.2.043102} {\bibfield  {journal}
  {\bibinfo  {journal} {Physical Review Fluids}\ }\textbf {\bibinfo {volume}
  {2}},\ \bibinfo {pages} {043102} (\bibinfo {year} {2017})},\ \bibinfo {note}
  {publisher: American Physical Society}\BibitemShut {NoStop}%
\bibitem [{\citenamefont {Cavagna}\ \emph {et~al.}(2015)\citenamefont
  {Cavagna}, \citenamefont {Giardina}, \citenamefont {Grigera}, \citenamefont
  {Jelic}, \citenamefont {Levine}, \citenamefont {Ramaswamy},\ and\
  \citenamefont {Viale}}]{cavagna_silent_2015}%
  \BibitemOpen
  \bibfield  {author} {\bibinfo {author} {\bibfnamefont {A.}~\bibnamefont
  {Cavagna}}, \bibinfo {author} {\bibfnamefont {I.}~\bibnamefont {Giardina}},
  \bibinfo {author} {\bibfnamefont {T.~S.}\ \bibnamefont {Grigera}}, \bibinfo
  {author} {\bibfnamefont {A.}~\bibnamefont {Jelic}}, \bibinfo {author}
  {\bibfnamefont {D.}~\bibnamefont {Levine}}, \bibinfo {author} {\bibfnamefont
  {S.}~\bibnamefont {Ramaswamy}},\ and\ \bibinfo {author} {\bibfnamefont
  {M.}~\bibnamefont {Viale}},\ }\bibfield  {title} {\bibinfo {title} {Silent
  {Flocks}: {Constraints} on {Signal} {Propagation} {Across} {Biological}
  {Groups}},\ }\href {https://doi.org/10.1103/PhysRevLett.114.218101}
  {\bibfield  {journal} {\bibinfo  {journal} {Physical Review Letters}\
  }\textbf {\bibinfo {volume} {114}},\ \bibinfo {pages} {218101} (\bibinfo
  {year} {2015})},\ \bibinfo {note} {publisher: American Physical
  Society}\BibitemShut {NoStop}%
\bibitem [{\citenamefont {Cestnik}\ and\ \citenamefont
  {Rosenblum}(2018)}]{cestnik_inferring_2018}%
  \BibitemOpen
  \bibfield  {author} {\bibinfo {author} {\bibfnamefont {R.}~\bibnamefont
  {Cestnik}}\ and\ \bibinfo {author} {\bibfnamefont {M.}~\bibnamefont
  {Rosenblum}},\ }\bibfield  {title} {\bibinfo {title} {Inferring the phase
  response curve from observation of a continuously perturbed oscillator},\
  }\href {https://doi.org/10.1038/s41598-018-32069-y} {\bibfield  {journal}
  {\bibinfo  {journal} {Scientific Reports}\ }\textbf {\bibinfo {volume} {8}},\
  \bibinfo {pages} {13606} (\bibinfo {year} {2018})},\ \bibinfo {note} {number:
  1 Publisher: Nature Publishing Group}\BibitemShut {NoStop}%
\bibitem [{\citenamefont {Dayan}\ and\ \citenamefont
  {Abbott}(2001)}]{dayan_theoretical_2001}%
  \BibitemOpen
  \bibfield  {author} {\bibinfo {author} {\bibfnamefont {P.}~\bibnamefont
  {Dayan}}\ and\ \bibinfo {author} {\bibfnamefont {L.~F.}\ \bibnamefont
  {Abbott}},\ }\href@noop {} {\emph {\bibinfo {title} {Theoretical
  {Neuroscience}: {Computational} and {Mathematical} {Modeling} of {Neural}
  {Systems}}}}\ (\bibinfo  {publisher} {Massachusetts Institute of Technology
  Press},\ \bibinfo {year} {2001})\ \bibinfo {note} {google-Books-ID:
  5GSKQgAACAAJ}\BibitemShut {NoStop}%
\bibitem [{\citenamefont {O’Keeffe}\ \emph {et~al.}(2017)\citenamefont
  {O’Keeffe}, \citenamefont {Hong},\ and\ \citenamefont
  {Strogatz}}]{okeeffe_oscillators_2017}%
  \BibitemOpen
  \bibfield  {author} {\bibinfo {author} {\bibfnamefont {K.~P.}\ \bibnamefont
  {O’Keeffe}}, \bibinfo {author} {\bibfnamefont {H.}~\bibnamefont {Hong}},\
  and\ \bibinfo {author} {\bibfnamefont {S.~H.}\ \bibnamefont {Strogatz}},\
  }\bibfield  {title} {\bibinfo {title} {Oscillators that sync and swarm},\
  }\href {https://doi.org/10.1038/s41467-017-01190-3} {\bibfield  {journal}
  {\bibinfo  {journal} {Nature Communications}\ }\textbf {\bibinfo {volume}
  {8}},\ \bibinfo {pages} {1504} (\bibinfo {year} {2017})},\ \bibinfo {note}
  {number: 1 Publisher: Nature Publishing Group}\BibitemShut {NoStop}%
\bibitem [{\citenamefont {Cobb}(2020)}]{cobb_idea_2020}%
  \BibitemOpen
  \bibfield  {author} {\bibinfo {author} {\bibfnamefont {M.}~\bibnamefont
  {Cobb}},\ }\href@noop {} {\emph {\bibinfo {title} {The {Idea} of the {Brain}:
  {The} {Past} and {Future} of {Neuroscience}}}}\ (\bibinfo  {publisher} {Basic
  Books},\ \bibinfo {year} {2020})\ \bibinfo {note} {google-Books-ID:
  RAiqDwAAQBAJ}\BibitemShut {NoStop}%
\bibitem [{\citenamefont {Gilpin}\ \emph
  {et~al.}(2017{\natexlab{b}})\citenamefont {Gilpin}, \citenamefont {Prakash},\
  and\ \citenamefont {Prakash}}]{gilpin2017flowtrace}%
  \BibitemOpen
  \bibfield  {author} {\bibinfo {author} {\bibfnamefont {W.}~\bibnamefont
  {Gilpin}}, \bibinfo {author} {\bibfnamefont {V.~N.}\ \bibnamefont
  {Prakash}},\ and\ \bibinfo {author} {\bibfnamefont {M.}~\bibnamefont
  {Prakash}},\ }\bibfield  {title} {\bibinfo {title} {Flowtrace: simple
  visualization of coherent structures in biological fluid flows},\ }\href@noop
  {} {\bibfield  {journal} {\bibinfo  {journal} {Journal of Experimental
  Biology}\ }\textbf {\bibinfo {volume} {220}},\ \bibinfo {pages} {3411}
  (\bibinfo {year} {2017}{\natexlab{b}})}\BibitemShut {NoStop}%
\bibitem [{\citenamefont {Yang}\ and\ \citenamefont
  {Marchetti}(2015)}]{yang_hydrodynamics_2015}%
  \BibitemOpen
  \bibfield  {author} {\bibinfo {author} {\bibfnamefont {X.}~\bibnamefont
  {Yang}}\ and\ \bibinfo {author} {\bibfnamefont {M.}~\bibnamefont
  {Marchetti}},\ }\bibfield  {title} {\bibinfo {title} {Hydrodynamics of
  {Turning} {Flocks}},\ }\href {https://doi.org/10.1103/PhysRevLett.115.258101}
  {\bibfield  {journal} {\bibinfo  {journal} {Physical Review Letters}\
  }\textbf {\bibinfo {volume} {115}},\ \bibinfo {pages} {258101} (\bibinfo
  {year} {2015})},\ \bibinfo {note} {publisher: American Physical
  Society}\BibitemShut {NoStop}%
\bibitem [{\citenamefont {Copenhagen}\ \emph {et~al.}(2018)\citenamefont
  {Copenhagen}, \citenamefont {Malet-Engra}, \citenamefont {Yu}, \citenamefont
  {Scita}, \citenamefont {Gov},\ and\ \citenamefont
  {Gopinathan}}]{copenhagen_frustration-induced_2018}%
  \BibitemOpen
  \bibfield  {author} {\bibinfo {author} {\bibfnamefont {K.}~\bibnamefont
  {Copenhagen}}, \bibinfo {author} {\bibfnamefont {G.}~\bibnamefont
  {Malet-Engra}}, \bibinfo {author} {\bibfnamefont {W.}~\bibnamefont {Yu}},
  \bibinfo {author} {\bibfnamefont {G.}~\bibnamefont {Scita}}, \bibinfo
  {author} {\bibfnamefont {N.}~\bibnamefont {Gov}},\ and\ \bibinfo {author}
  {\bibfnamefont {A.}~\bibnamefont {Gopinathan}},\ }\bibfield  {title}
  {\bibinfo {title} {Frustration-induced phases in migrating cell clusters},\
  }\href {https://doi.org/10.1126/sciadv.aar8483} {\bibfield  {journal}
  {\bibinfo  {journal} {Science Advances}\ }\textbf {\bibinfo {volume} {4}},\
  \bibinfo {pages} {eaar8483} (\bibinfo {year} {2018})},\ \bibinfo {note}
  {publisher: American Association for the Advancement of Science Section:
  Research Article}\BibitemShut {NoStop}%
\bibitem [{\citenamefont {Geyer}\ \emph {et~al.}(2016)\citenamefont {Geyer},
  \citenamefont {Sartori}, \citenamefont {Friedrich}, \citenamefont
  {Jülicher},\ and\ \citenamefont {Howard}}]{geyer_independent_2016}%
  \BibitemOpen
  \bibfield  {author} {\bibinfo {author} {\bibfnamefont {V.}~\bibnamefont
  {Geyer}}, \bibinfo {author} {\bibfnamefont {P.}~\bibnamefont {Sartori}},
  \bibinfo {author} {\bibfnamefont {B.}~\bibnamefont {Friedrich}}, \bibinfo
  {author} {\bibfnamefont {F.}~\bibnamefont {Jülicher}},\ and\ \bibinfo
  {author} {\bibfnamefont {J.}~\bibnamefont {Howard}},\ }\bibfield  {title}
  {\bibinfo {title} {Independent {Control} of the {Static} and {Dynamic}
  {Components} of the {Chlamydomonas} {Flagellar} {Beat}},\ }\href
  {https://doi.org/10.1016/j.cub.2016.02.053} {\bibfield  {journal} {\bibinfo
  {journal} {Current Biology}\ }\textbf {\bibinfo {volume} {26}},\ \bibinfo
  {pages} {1098} (\bibinfo {year} {2016})}\BibitemShut {NoStop}%
\bibitem [{\citenamefont {Huber}(2016)}]{huber_topological_2016}%
  \BibitemOpen
  \bibfield  {author} {\bibinfo {author} {\bibfnamefont {S.~D.}\ \bibnamefont
  {Huber}},\ }\bibfield  {title} {\bibinfo {title} {Topological mechanics},\
  }\href {https://doi.org/10.1038/nphys3801} {\bibfield  {journal} {\bibinfo
  {journal} {Nature Physics}\ }\textbf {\bibinfo {volume} {12}},\ \bibinfo
  {pages} {621} (\bibinfo {year} {2016})},\ \bibinfo {note} {number: 7
  Publisher: Nature Publishing Group}\BibitemShut {NoStop}%
\bibitem [{\citenamefont {Shankar}\ \emph {et~al.}(2020)\citenamefont
  {Shankar}, \citenamefont {Souslov}, \citenamefont {Bowick}, \citenamefont
  {Marchetti},\ and\ \citenamefont {Vitelli}}]{shankar_topological_2020}%
  \BibitemOpen
  \bibfield  {author} {\bibinfo {author} {\bibfnamefont {S.}~\bibnamefont
  {Shankar}}, \bibinfo {author} {\bibfnamefont {A.}~\bibnamefont {Souslov}},
  \bibinfo {author} {\bibfnamefont {M.~J.}\ \bibnamefont {Bowick}}, \bibinfo
  {author} {\bibfnamefont {M.~C.}\ \bibnamefont {Marchetti}},\ and\ \bibinfo
  {author} {\bibfnamefont {V.}~\bibnamefont {Vitelli}},\ }\bibfield  {title}
  {\bibinfo {title} {Topological active matter},\ }\href
  {http://arxiv.org/abs/2010.00364} {\bibfield  {journal} {\bibinfo  {journal}
  {arXiv:2010.00364 [cond-mat]}\ } (\bibinfo {year} {2020})},\ \bibinfo {note}
  {arXiv: 2010.00364}\BibitemShut {NoStop}%
\bibitem [{\citenamefont {Kim}\ \emph {et~al.}(2021)\citenamefont {Kim},
  \citenamefont {Pochitaloff}, \citenamefont {Stooke-Vaughan},\ and\
  \citenamefont {Campàs}}]{kim_embryonic_2021}%
  \BibitemOpen
  \bibfield  {author} {\bibinfo {author} {\bibfnamefont {S.}~\bibnamefont
  {Kim}}, \bibinfo {author} {\bibfnamefont {M.}~\bibnamefont {Pochitaloff}},
  \bibinfo {author} {\bibfnamefont {G.~A.}\ \bibnamefont {Stooke-Vaughan}},\
  and\ \bibinfo {author} {\bibfnamefont {O.}~\bibnamefont {Campàs}},\
  }\bibfield  {title} {\bibinfo {title} {Embryonic tissues as active foams},\
  }\href {https://doi.org/10.1038/s41567-021-01215-1} {\bibfield  {journal}
  {\bibinfo  {journal} {Nature Physics}\ ,\ \bibinfo {pages} {1}} (\bibinfo
  {year} {2021})},\ \bibinfo {note} {publisher: Nature Publishing
  Group}\BibitemShut {NoStop}%
\bibitem [{\citenamefont {Boyd}(2020)}]{boyd_nonlinear_2020}%
  \BibitemOpen
  \bibfield  {author} {\bibinfo {author} {\bibfnamefont {R.~W.}\ \bibnamefont
  {Boyd}},\ }\href@noop {} {\emph {\bibinfo {title} {Nonlinear {Optics}}}}\
  (\bibinfo  {publisher} {Academic Press},\ \bibinfo {year} {2020})\ \bibinfo
  {note} {google-Books-ID: 54vZDwAAQBAJ}\BibitemShut {NoStop}%
\bibitem [{\citenamefont {Hill}\ \emph {et~al.}(2010)\citenamefont {Hill},
  \citenamefont {Swaminathan}, \citenamefont {Estes}, \citenamefont {Cribb},
  \citenamefont {O'Brien}, \citenamefont {Davis},\ and\ \citenamefont
  {Superfine}}]{hill2010force}%
  \BibitemOpen
  \bibfield  {author} {\bibinfo {author} {\bibfnamefont {D.~B.}\ \bibnamefont
  {Hill}}, \bibinfo {author} {\bibfnamefont {V.}~\bibnamefont {Swaminathan}},
  \bibinfo {author} {\bibfnamefont {A.}~\bibnamefont {Estes}}, \bibinfo
  {author} {\bibfnamefont {J.}~\bibnamefont {Cribb}}, \bibinfo {author}
  {\bibfnamefont {E.~T.}\ \bibnamefont {O'Brien}}, \bibinfo {author}
  {\bibfnamefont {C.~W.}\ \bibnamefont {Davis}},\ and\ \bibinfo {author}
  {\bibfnamefont {R.}~\bibnamefont {Superfine}},\ }\bibfield  {title} {\bibinfo
  {title} {Force generation and dynamics of individual cilia under external
  loading},\ }\href@noop {} {\bibfield  {journal} {\bibinfo  {journal}
  {Biophysical journal}\ }\textbf {\bibinfo {volume} {98}},\ \bibinfo {pages}
  {57} (\bibinfo {year} {2010})}\BibitemShut {NoStop}%
\bibitem [{\citenamefont {Klindt}\ \emph {et~al.}(2016)\citenamefont {Klindt},
  \citenamefont {Ruloff}, \citenamefont {Wagner},\ and\ \citenamefont
  {Friedrich}}]{klindt2016load}%
  \BibitemOpen
  \bibfield  {author} {\bibinfo {author} {\bibfnamefont {G.~S.}\ \bibnamefont
  {Klindt}}, \bibinfo {author} {\bibfnamefont {C.}~\bibnamefont {Ruloff}},
  \bibinfo {author} {\bibfnamefont {C.}~\bibnamefont {Wagner}},\ and\ \bibinfo
  {author} {\bibfnamefont {B.~M.}\ \bibnamefont {Friedrich}},\ }\bibfield
  {title} {\bibinfo {title} {Load response of the flagellar beat},\ }\href@noop
  {} {\bibfield  {journal} {\bibinfo  {journal} {Physical review letters}\
  }\textbf {\bibinfo {volume} {117}},\ \bibinfo {pages} {258101} (\bibinfo
  {year} {2016})}\BibitemShut {NoStop}%
\bibitem [{\citenamefont {Smith}\ \emph
  {et~al.}(2014{\natexlab{b}})\citenamefont {Smith}, \citenamefont
  {Varoqueaux}, \citenamefont {Kittelmann}, \citenamefont {Azzam},
  \citenamefont {Cooper}, \citenamefont {Winters}, \citenamefont {Eitel},
  \citenamefont {Fasshauer},\ and\ \citenamefont {Reese}}]{Smith2014a}%
  \BibitemOpen
  \bibfield  {author} {\bibinfo {author} {\bibfnamefont {C.}~\bibnamefont
  {Smith}}, \bibinfo {author} {\bibfnamefont {F.}~\bibnamefont {Varoqueaux}},
  \bibinfo {author} {\bibfnamefont {M.}~\bibnamefont {Kittelmann}}, \bibinfo
  {author} {\bibfnamefont {R.}~\bibnamefont {Azzam}}, \bibinfo {author}
  {\bibfnamefont {B.}~\bibnamefont {Cooper}}, \bibinfo {author} {\bibfnamefont
  {C.}~\bibnamefont {Winters}}, \bibinfo {author} {\bibfnamefont
  {M.}~\bibnamefont {Eitel}}, \bibinfo {author} {\bibfnamefont
  {D.}~\bibnamefont {Fasshauer}},\ and\ \bibinfo {author} {\bibfnamefont
  {T.}~\bibnamefont {Reese}},\ }\bibfield  {title} {\bibinfo {title} {{Novel
  Cell Types, Neurosecretory Cells, and Body Plan of the Early-Diverging
  Metazoan Trichoplax adhaerens}},\ }\bibfield  {journal} {\bibinfo  {journal}
  {Current Biology}\ }\href {https://doi.org/10.1016/j.cub.2014.05.046}
  {10.1016/j.cub.2014.05.046} (\bibinfo {year}
  {2014}{\natexlab{b}})\BibitemShut {NoStop}%
\bibitem [{\citenamefont {Grell}\ and\ \citenamefont
  {Benwitz}(1974)}]{Grell1974ElektronenmikroskopischePlacozoa}%
  \BibitemOpen
  \bibfield  {author} {\bibinfo {author} {\bibfnamefont {K.~G.}\ \bibnamefont
  {Grell}}\ and\ \bibinfo {author} {\bibfnamefont {G.}~\bibnamefont
  {Benwitz}},\ }\bibfield  {title} {\bibinfo {title} {{Elektronenmikroskopische
  beobachtungen ??ber das wachstum der eizelle und die bildung der
  "befruchtungsmembran" von Trichoplax adhaerens F. E. Schulze (Placozoa)}},\
  }\href {https://doi.org/10.1007/BF00277511} {\bibfield  {journal} {\bibinfo
  {journal} {Zeitschrift f??r Morphologie der Tiere}\ }\textbf {\bibinfo
  {volume} {79}},\ \bibinfo {pages} {295} (\bibinfo {year} {1974})}\BibitemShut
  {NoStop}%
\bibitem [{\citenamefont {Toyjanova}\ \emph {et~al.}(2014)\citenamefont
  {Toyjanova}, \citenamefont {Hannen}, \citenamefont {Bar-Kochba},
  \citenamefont {Darling}, \citenamefont {Henann},\ and\ \citenamefont
  {Franck}}]{toyjanova_3d_2014}%
  \BibitemOpen
  \bibfield  {author} {\bibinfo {author} {\bibfnamefont {J.}~\bibnamefont
  {Toyjanova}}, \bibinfo {author} {\bibfnamefont {E.}~\bibnamefont {Hannen}},
  \bibinfo {author} {\bibfnamefont {E.}~\bibnamefont {Bar-Kochba}}, \bibinfo
  {author} {\bibfnamefont {E.~M.}\ \bibnamefont {Darling}}, \bibinfo {author}
  {\bibfnamefont {D.~L.}\ \bibnamefont {Henann}},\ and\ \bibinfo {author}
  {\bibfnamefont {C.}~\bibnamefont {Franck}},\ }\bibfield  {title} {\bibinfo
  {title} {{3D} {Viscoelastic} {Traction} {Force} {Microscopy}},\ }\href
  {https://doi.org/10.1039/c4sm01271b} {\bibfield  {journal} {\bibinfo
  {journal} {Soft matter}\ }\textbf {\bibinfo {volume} {10}},\ \bibinfo {pages}
  {8095} (\bibinfo {year} {2014})}\BibitemShut {NoStop}%
\bibitem [{\citenamefont {Mora}\ and\ \citenamefont
  {Bialek}(2011)}]{mora_are_2011}%
  \BibitemOpen
  \bibfield  {author} {\bibinfo {author} {\bibfnamefont {T.}~\bibnamefont
  {Mora}}\ and\ \bibinfo {author} {\bibfnamefont {W.}~\bibnamefont {Bialek}},\
  }\bibfield  {title} {\bibinfo {title} {Are {Biological} {Systems} {Poised} at
  {Criticality}?},\ }\href {https://doi.org/10.1007/s10955-011-0229-4}
  {\bibfield  {journal} {\bibinfo  {journal} {Journal of Statistical Physics}\
  }\textbf {\bibinfo {volume} {144}},\ \bibinfo {pages} {268} (\bibinfo {year}
  {2011})}\BibitemShut {NoStop}%
\bibitem [{\citenamefont {Bi}\ \emph {et~al.}(2016)\citenamefont {Bi},
  \citenamefont {Yang}, \citenamefont {Marchetti},\ and\ \citenamefont
  {Manning}}]{Bi2016}%
  \BibitemOpen
  \bibfield  {author} {\bibinfo {author} {\bibfnamefont {D.}~\bibnamefont
  {Bi}}, \bibinfo {author} {\bibfnamefont {X.}~\bibnamefont {Yang}}, \bibinfo
  {author} {\bibfnamefont {M.~C.}\ \bibnamefont {Marchetti}},\ and\ \bibinfo
  {author} {\bibfnamefont {M.~L.}\ \bibnamefont {Manning}},\ }\bibfield
  {title} {\bibinfo {title} {{Motility-driven glass and jamming transitions in
  biological tissues}},\ }\href {https://doi.org/10.1103/PhysRevX.6.021011}
  {\bibfield  {journal} {\bibinfo  {journal} {Physical Review X}\ }\textbf
  {\bibinfo {volume} {6}},\ \bibinfo {pages} {1} (\bibinfo {year}
  {2016})}\BibitemShut {NoStop}%
\bibitem [{\citenamefont {Pfeifer}\ \emph {et~al.}(2007)\citenamefont
  {Pfeifer}, \citenamefont {Lungarella},\ and\ \citenamefont
  {Iida}}]{pfeifer2007self}%
  \BibitemOpen
  \bibfield  {author} {\bibinfo {author} {\bibfnamefont {R.}~\bibnamefont
  {Pfeifer}}, \bibinfo {author} {\bibfnamefont {M.}~\bibnamefont
  {Lungarella}},\ and\ \bibinfo {author} {\bibfnamefont {F.}~\bibnamefont
  {Iida}},\ }\bibfield  {title} {\bibinfo {title} {Self-organization,
  embodiment, and biologically inspired robotics},\ }\href@noop {} {\bibfield
  {journal} {\bibinfo  {journal} {Science}\ }\textbf {\bibinfo {volume}
  {318}},\ \bibinfo {pages} {1088} (\bibinfo {year} {2007})}\BibitemShut
  {NoStop}%
\bibitem [{\citenamefont {Tanaka}\ \emph {et~al.}(2019)\citenamefont {Tanaka},
  \citenamefont {Yamane}, \citenamefont {Héroux}, \citenamefont {Nakane},
  \citenamefont {Kanazawa}, \citenamefont {Takeda}, \citenamefont {Numata},
  \citenamefont {Nakano},\ and\ \citenamefont {Hirose}}]{tanaka_recent_2019}%
  \BibitemOpen
  \bibfield  {author} {\bibinfo {author} {\bibfnamefont {G.}~\bibnamefont
  {Tanaka}}, \bibinfo {author} {\bibfnamefont {T.}~\bibnamefont {Yamane}},
  \bibinfo {author} {\bibfnamefont {J.~B.}\ \bibnamefont {Héroux}}, \bibinfo
  {author} {\bibfnamefont {R.}~\bibnamefont {Nakane}}, \bibinfo {author}
  {\bibfnamefont {N.}~\bibnamefont {Kanazawa}}, \bibinfo {author}
  {\bibfnamefont {S.}~\bibnamefont {Takeda}}, \bibinfo {author} {\bibfnamefont
  {H.}~\bibnamefont {Numata}}, \bibinfo {author} {\bibfnamefont
  {D.}~\bibnamefont {Nakano}},\ and\ \bibinfo {author} {\bibfnamefont
  {A.}~\bibnamefont {Hirose}},\ }\bibfield  {title} {\bibinfo {title} {Recent
  advances in physical reservoir computing: {A} review},\ }\href
  {https://doi.org/10.1016/j.neunet.2019.03.005} {\bibfield  {journal}
  {\bibinfo  {journal} {Neural Networks}\ }\textbf {\bibinfo {volume} {115}},\
  \bibinfo {pages} {100} (\bibinfo {year} {2019})}\BibitemShut {NoStop}%
\bibitem [{Dij()}]{DijkstraQuotes}%
  \BibitemOpen
  \href@noop {} {\bibinfo {title} {Dijkstra quotes}}\BibitemShut {NoStop}%
\bibitem [{\citenamefont {Jaeger}(2001)}]{jaeger__2001}%
  \BibitemOpen
  \bibfield  {author} {\bibinfo {author} {\bibfnamefont {H.}~\bibnamefont
  {Jaeger}},\ }\bibfield  {title} {\bibinfo {title} {The" echo state" approach
  to analysing and training recurrent neural networks-with an erratum note'},\
  }\href@noop {} {\bibfield  {journal} {\bibinfo  {journal} {Bonn, Germany:
  German National Research Center for Information Technology GMD Technical
  Report}\ }\textbf {\bibinfo {volume} {148}} (\bibinfo {year}
  {2001})}\BibitemShut {NoStop}%
\bibitem [{\citenamefont {Maass}\ \emph {et~al.}(2002)\citenamefont {Maass},
  \citenamefont {Natschläger},\ and\ \citenamefont
  {Markram}}]{maass_real-time_2002}%
  \BibitemOpen
  \bibfield  {author} {\bibinfo {author} {\bibfnamefont {W.}~\bibnamefont
  {Maass}}, \bibinfo {author} {\bibfnamefont {T.}~\bibnamefont
  {Natschläger}},\ and\ \bibinfo {author} {\bibfnamefont {H.}~\bibnamefont
  {Markram}},\ }\bibfield  {title} {\bibinfo {title} {Real-time computing
  without stable states: a new framework for neural computation based on
  perturbations},\ }\href {https://doi.org/10.1162/089976602760407955}
  {\bibfield  {journal} {\bibinfo  {journal} {Neural Computation}\ }\textbf
  {\bibinfo {volume} {14}},\ \bibinfo {pages} {2531} (\bibinfo {year}
  {2002})}\BibitemShut {NoStop}%
\bibitem [{\citenamefont {Pathak}\ \emph {et~al.}(2018)\citenamefont {Pathak},
  \citenamefont {Hunt}, \citenamefont {Girvan}, \citenamefont {Lu},\ and\
  \citenamefont {Ott}}]{pathak_model-free_2018}%
  \BibitemOpen
  \bibfield  {author} {\bibinfo {author} {\bibfnamefont {J.}~\bibnamefont
  {Pathak}}, \bibinfo {author} {\bibfnamefont {B.}~\bibnamefont {Hunt}},
  \bibinfo {author} {\bibfnamefont {M.}~\bibnamefont {Girvan}}, \bibinfo
  {author} {\bibfnamefont {Z.}~\bibnamefont {Lu}},\ and\ \bibinfo {author}
  {\bibfnamefont {E.}~\bibnamefont {Ott}},\ }\bibfield  {title} {\bibinfo
  {title} {Model-{Free} {Prediction} of {Large} {Spatiotemporally} {Chaotic}
  {Systems} from {Data}: {A} {Reservoir} {Computing} {Approach}},\ }\href
  {https://doi.org/10.1103/PhysRevLett.120.024102} {\bibfield  {journal}
  {\bibinfo  {journal} {Physical Review Letters}\ }\textbf {\bibinfo {volume}
  {120}},\ \bibinfo {pages} {024102} (\bibinfo {year} {2018})},\ \bibinfo
  {note} {publisher: American Physical Society}\BibitemShut {NoStop}%
\bibitem [{\citenamefont {Lymburn}\ \emph {et~al.}(2021)\citenamefont
  {Lymburn}, \citenamefont {Algar}, \citenamefont {Small},\ and\ \citenamefont
  {Jüngling}}]{lymburn_reservoir_2021}%
  \BibitemOpen
  \bibfield  {author} {\bibinfo {author} {\bibfnamefont {T.}~\bibnamefont
  {Lymburn}}, \bibinfo {author} {\bibfnamefont {S.~D.}\ \bibnamefont {Algar}},
  \bibinfo {author} {\bibfnamefont {M.}~\bibnamefont {Small}},\ and\ \bibinfo
  {author} {\bibfnamefont {T.}~\bibnamefont {Jüngling}},\ }\bibfield  {title}
  {\bibinfo {title} {Reservoir computing with swarms},\ }\href
  {https://doi.org/10.1063/5.0039745} {\bibfield  {journal} {\bibinfo
  {journal} {Chaos: An Interdisciplinary Journal of Nonlinear Science}\
  }\textbf {\bibinfo {volume} {31}},\ \bibinfo {pages} {033121} (\bibinfo
  {year} {2021})},\ \bibinfo {note} {publisher: American Institute of
  Physics}\BibitemShut {NoStop}%
\end{thebibliography}%

\section{Figures}
\onecolumngrid
\newpage

\begin{figure} [t!]
  \caption{FIG 1. (Previous page.) \textbf{Ciliary flocking underlies collective dynamics in \textit{T. adhaerens}.} A) The \textit{T. adhaerens} is an emerging model organism for studying the minimal ingredients of tissue mechanics. Experiments are conducted across numerous imaging modalities, length scales and boundary conditions to reveal agile collective locomotion from hundreds of cells to millions. B) The organism can be viewed as two mechanically distinct tissues stitched together around the edge to form a flat three-layered architecture with the top and bottom tissues separated by network of fiber cells. C) The bottom tissue contacts the surface mediated by single-cilia which are $\sim 20 \mu$m. D.i) High resolution (100 Hz) trans-illumination microscopy using a 60x oil Objective (Nikon TIRF, NA=1.43) captures rapid organism wide ciliary dynamics characterized by an oscillatory stepping force and dynamic orientation patterns, such as the +1 defect shown here.  D.ii) A digital crop of the center of the defect shows a circular orientation pattern with decreasing local order towards the defect center. D.iii) Far from the +1 defect core, the orientation field of the cilia is largely uniform. E) Coloring the orientation of each cilia, $\phi$ from $-\pi $ to $\pi$ shows the dynamics of the ciliary orientation. F) Computationally, we simultaneously extract the tissue velocity field to show its close agreement with the local ciliary orientation field. G) Using both segmented ciliary orientation $\hat\phi_i (t)$ and H) the local tissue velocity $\hat\theta_i (t)$, I) we discover a torque mediated coupling between the rate of change of the ciliary orientation and the direction of the tissue displacement by way of the locally evaluated difference.  J) Conceptually, this gives the intuition that while these two fields are related the transient dynamics of how they approach agreement is highly relevant for the dynamics of this phenomena which we call ‘ciliary flocking’.  H) The resulting collective dynamics across millions of cells generates agility at time scales of seconds, all without a neuromuscular system.}
\end{figure}

\newpage
\begin{figure}
\includegraphics[width = \textwidth]{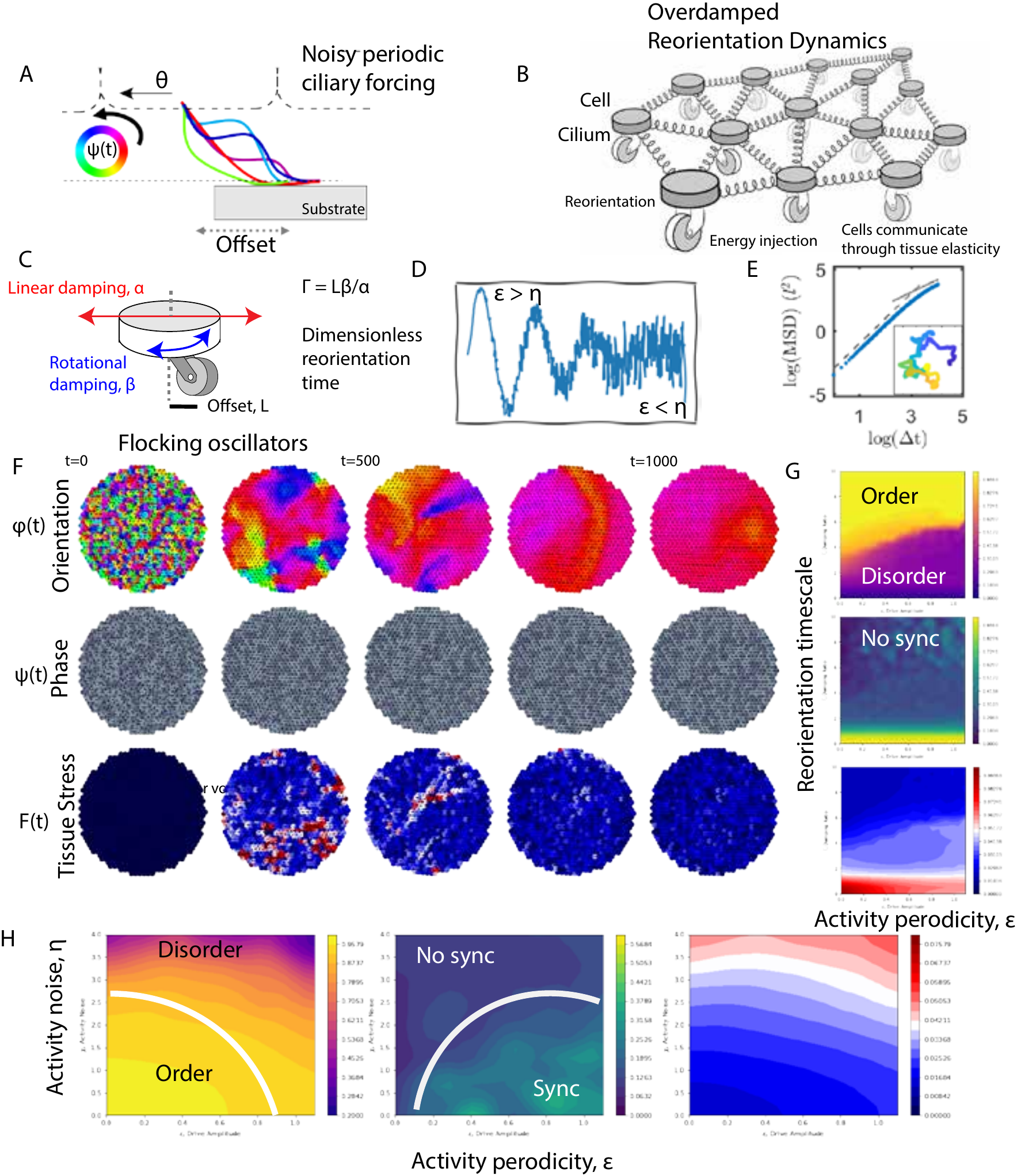}
\caption{\textbf{FIG 2}: (Caption next page.)}
  \label{fig:fig2}
\end{figure}

\begin{figure} [t!]
\caption{FIG 2. (previous page) \textbf{Flocking oscillator model for collective dynamics in the \textit{T. adhaerens.}} We combine periodic active forcing and overdamped reorientation dynamics to propose a flocking oscillator model as a physical analogy to the collective dynamics observed in this simple animal. A) Ciliary walking can be minimally modeled as a noisy periodic force. B) For simplicity, we model the yield-stress nature of the tissue as a network of fixed topology springs (see supplementary information). 
The full model captures periodic active power injection, $a(t) = a_o + \epsilon sin(\Omega t) + \eta$, and drives reorientation through the applied torque applied on the cilium via the collective tissue forces. C) The single cilium has two damping imposed time scales, the time to reorient under a given force (transformed to a torque via a characteristic length-scale) and the time to displace in response to the same force. We can non-dimensionalize this timescale ratio as, $\Gamma = \frac{L\beta}{\alpha}$. D) The dynamics of the periodic activity is characterized by the two parameters which control the amplitude of the periodic signal and the noise. This relationship can be summarized by a signal-to-noise ratio. E) Allowing the simulation to run for large number of time steps captures a ballistic to diffusive roll-off the the center of mass motion of the tissue which is strongly dependent upon the noise.  F) The collective dynamics of flocking oscillators can be represented with the cell-resolution dynamics of three fields i) the ciliary orientation field $\phi(r,t)$, the oscillator phase field $\psi(r,t)$ and the force acting on each cell $|F(r,t)|$. When initialized from a random initial condition, the resulting steady state shows the coarsening of defects toward a global agreement in orientation, uniform density of phase angles and reduced internal tissue forces. G) The steady state of these dynamics can be summarized by plotting three summary statistics across a 2D phase space characterized by the reorientation timescale and the activity periodicity. The bulk of the choices of parameter result in a ordered state with low-mean-synchronization.  H) The activity periodicity versus activity noise space captures two distinct crossovers between polarized and disordered, and synchronized and unsynchronized. The shape of these distinct crossovers shows that all $2^2$ combinations are possible: i) unpolarized and unsynchronized at high noise, ii) High polarization and low synchronization at low $\epsilon$ and low noise, iii) high polarization and high synchronization at intermediate $\epsilon$ and low noise, and iv) synchronized disorder at high $\epsilon$. }
\end{figure}

\newpage
\begin{figure}
\includegraphics[width = \textwidth]{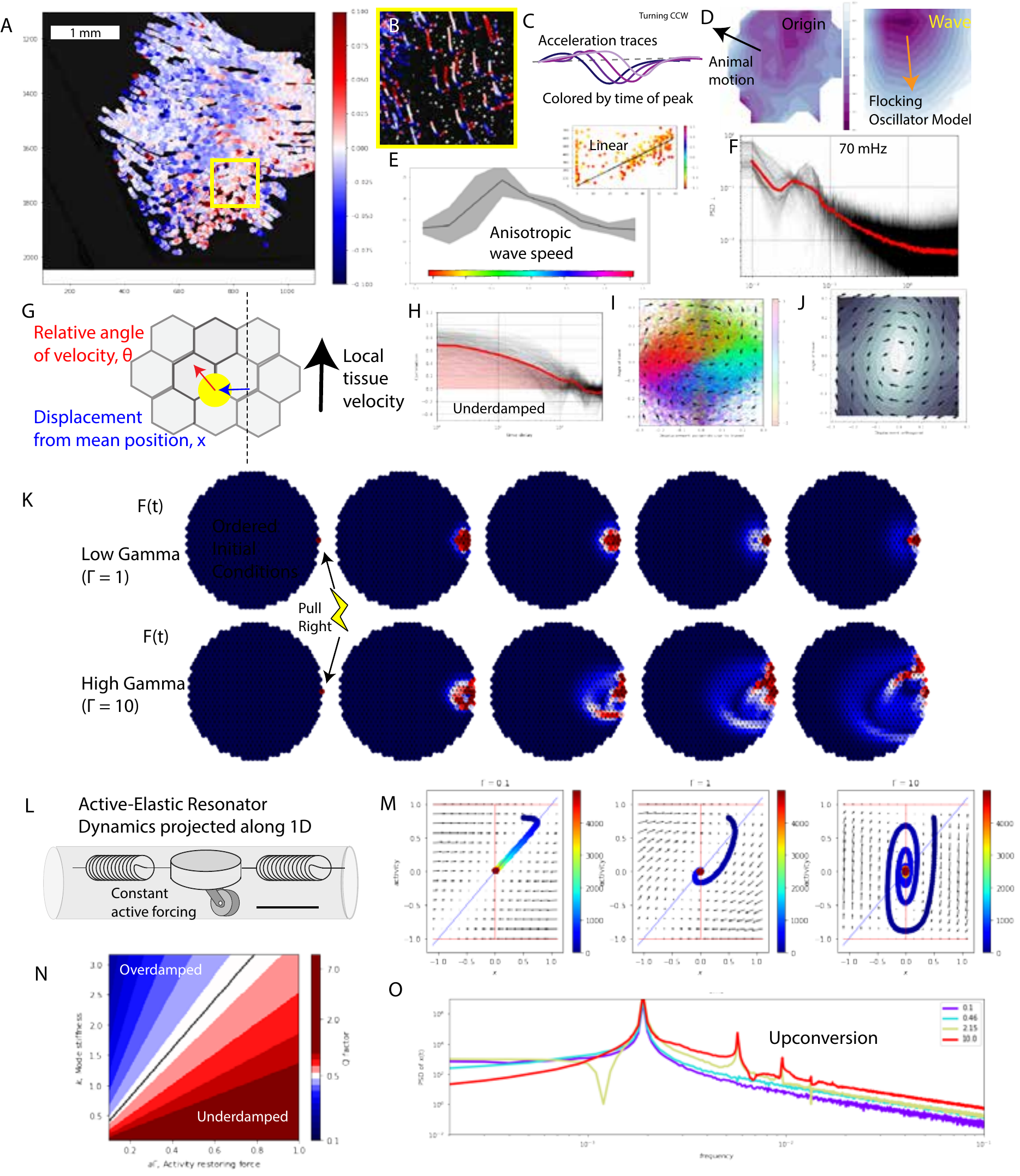}
\caption{\textbf{FIG 3}: (Caption next page.)}
  \label{fig:fig3}
\end{figure}

\begin{figure} [t!]
\caption{FIG 3 (previous page) \textbf{The interplay of reorientation and overdamped density waves generates underdamped active-elastic oscillations and traveling waves.} A) By tracking chemically attached fluorescent micro-spheres on top layer epithelium (see supplementary information), 
we track the tissue displacement through rapid directional changes over several hours (with 0.1 Hz resolution). Each time point is colored by its clockwise (blue) or counterclockwise(red) acceleration. B) Individual micro-sphere trajectories (computationally thinned to aide visualization) show rapid accelerations ($\sim 10^{-1} \mu m/s^2$). C) Trajectory acceleration dynamics: color indicates time at which it reaches its local peak acceleration perpendicular to its instantaneous direction of travel (in this case, turning CW). D) Spatial position of acceleration peaks reveals a traveling wave with an anisotropic wave speed. 
We compare these experimental results with the analogous dynamics in the flocking oscillator model. Waves manifest as smooth gradients from dark purple to light blue. E) 
Distance from the putative origin of the wave versus the time of the peak shows a linear speed wave at $\sim 100 \mu m/s$. Fitting the wave speed for the initial disturbance for different bins of wave speed direction $angle(\vec K)$ reveals a strong dependence of the wave speed on direction with a peak speed orthogonal to the direction of travel $angle(\vec K) = 0$ around $\sim$2x faster than the measured speeds $50^o +$ bins. F) Spectral content of fluctuations in tissue speed orthogonal to the direction of travel show a peak around 70 mHz (see supplementary information). 
G) Projection of the relative velocity and the displacement from the mean position in a two dimensional space.  H) Further support of underdamped-like dynamics shows up in the two point correlation of the orthogonal component of the displacement where a number of the correlations oscillate around zero indicating underdamped like dynamics. I) For each point, we color the angle of its next step. Taking an ensemble average of all the next steps in each box gives us an average flow in the relative velocity versus mean position phase space. The circulation of these trajectories indicate an underdamped like response. J) The density plot shows that the highest probability is to be close to (0,0) in this space lending support that (0,0) is a fixed point and the transients are similar qualitatively to an underdamped relaxation. K) In our flocking oscillator model, we studied the response of the tissue to small, short time stimuli applied on the right side of the tissue for different choices of the non-dimensional parameter $\Gamma$. When $\Gamma = 1$, the response to the stimulus is highly local and is characterized by an overdamped diffusive wave through the media. However, at sufficiently high $\Gamma = 10$, an identical stimulus initiates an effectively underdamped traveling wave which passes all the way across the tissue.  ...(continued on next page) ...}
\end{figure}
\begin{figure}[t!]
\caption{FIG 3 ...(continued from previous page) ...
L) Conceptual understanding of the emergence of this underdamped phenomena by deriving a 1D toy model from the 2D dynamical system of the reorienting active force (here with fixed amplitude, $\epsilon = 0$ and $\eta = 0$). M) Studying this toy model numerically reveals the same dependence of the underdamped transients on the non-dimensional ratio of timescales, $\Gamma$, where large $\Gamma = 10$ exhibits an effectively underdamped like response to stimulus. N) This two degree of freedom model maps in the low amplitude of excitation limit onto the simple harmonic oscillator where activity term plays the role of momentum. The effective Q factor of the oscillation is determined by the activity restoring force arising from $a\Gamma$ and the stiffness, $k$, which acts like a damping. For high stiffness, the system is overdamped, where at low stiffness and high $a\Gamma$, the dynamics become underdamped (Q$\geq$0.5). O) The response to driving of this system with the nonlinear contributions from the activity results in a second-harmonic-generation-like upconversion with power being passed up the harmonics of the drive frequency.
}
\end{figure}

\newpage
\begin{figure}
\includegraphics[width = \textwidth]{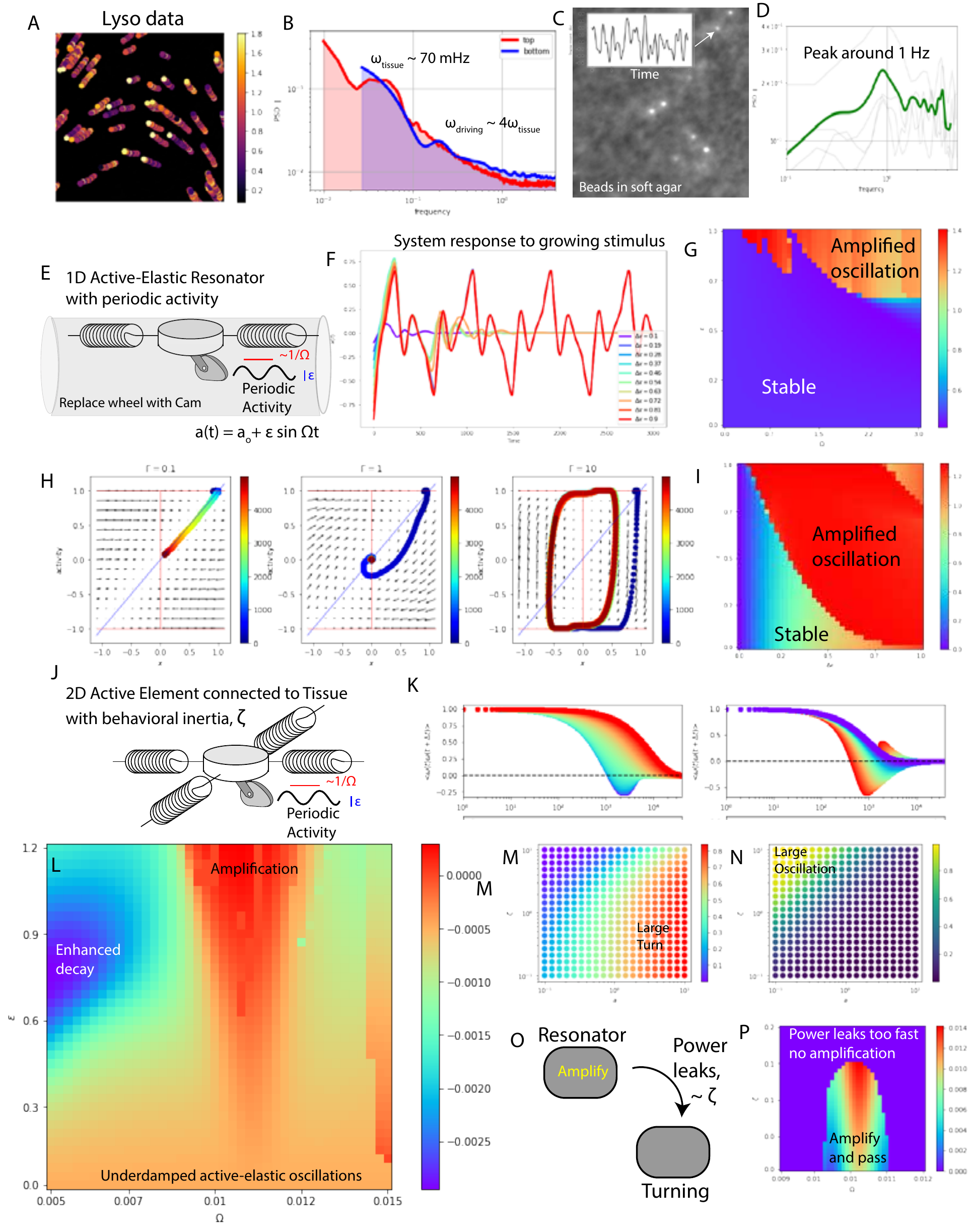}
\caption{\textbf{FIG 4}: (Caption next page.)}
  \label{fig:fig4}
\end{figure}

\begin{figure} [t!]
\caption{FIG 4. (previous page) \textbf{Periodic fluctuations in the active force result in a parametric amplification of the active-elastic resonator mode.} A) By staining the lipophilic cells with lysotracker, we can track the displacement dynamics of the bottom tissue with high spatio-temporal resolution. B) By projecting the fluctuations in the displacement speed onto the direction of travel, we gain access to a proxy for the periodic fluctuations in forcing. We find that the peak in the power spectrum of these local fluctuations is 4x times the peak in the power spectrum of orthogonal fluctuations (active-elastic modes) consistent with a parametric type resonance pumping energy from high frequency modulation of a system parameter into lower frequency modes through a parametric instability. C) We look for further evidence of a fluctuating driving force by imaging the fluctuation of beads suspended in a soft agar surface (using protocols from Traction force microscopy). D) These spectra reveal a peak consistent with the previously measured\cite{bullpart1} 1 Hz frequency for ciliary steps. Coupled together, these measurements support the idea that high frequency periodic activity fluctuations may contribute to the nonlinear driving of lower frequency tissue modes. E) We study the role of periodic activity fluctuations in the previously derived active-elastic resonator model by the physical analogy of replacing a round wheel on our motorized castor for a oblong cam. F) We can study the response of this system to a stimulus by plotting the displacement timeseries for different magnitudes of stimulus under a fixed driving amplitude. For small driving amplitudes, these dynamics are qualitatively identical to underdamped dynamics but above a critical stimulus size, the dynamics approach a periodic beat of mechanical ‘spikes’ which are consistent with the dynamics approaching a steady limit cycle instead of a fixed point at (0,0). G) We can study the system response in the Mathieu diagram space where we compare the maximum response of the parametrically driven active-elastic resonator for different choices of drive amplitude $\epsilon$ and frequency $\Omega$. This reveals a tongue of instability extending down to $\epsilon \sim 0.5$ for a choice of driving frequency of 4 times the natural frequency of the active-elastic resonator. H) We can repeat our study for the dependence on $\Gamma$ and show that even for large driving, small $\Gamma$ exhibits no amplification. Thus sufficiently large $\Gamma$ is necessary for both underdamped like dynamics and parametric amplification.  I) By studying the maximum response of the system to different size stimuli $\delta x$ with different driving amplitudes. $\epsilon$, we find that the threshold stimulus for exciting the system up to the saturating limit cycle is dependent upon the amplitude of driving. Thus by tuning the drive amplitude, an active-elastic resonator can adjust its sensitivity to external mechanical stimuli. J) In an effort to corroborate the lessons of our toy model apply, we present a two-dimensional model where we couple a single active castor wheel to a slow adjusting tissue with a behavioral inertia which is related to the inverse of our coupling tuning parameter $\zeta$.  K) A common method to extract signatures of underdamped like dynamics from a timeseries is to study the minimum value of a two-point correlation indicating the time at which the system is maximally anti-correlated.  We illustrate the sensitivity of damping to both changing the coupling parameter, $\zeta$ and the force magnitude of the active cell $a_o$. Increasing $zeta$ results in more overdamped like dynamics and increasing $a_o$ results in more underdamped dynamics.  ...(continued on next page) ...}
\end{figure}
\begin{figure}[t!]
\caption{FIG 4 ...(continued from previous page) ...  L) Recreating the 2d analog of the forcing amplitude, $\epsilon$ versus frequency, $\Omega$ gives rise to another Mathieu tongue of amplification around 4 times the resonator frequency. M) Turning agility is essential for an organism attempting to escape a dangerous stimulus, so we studied how much the 2D tissue coupled system turns in response to a stimulus for different values of $\zeta$ and $a$. At large $\zeta$ the tissue is more responsive to fluctuations N) but is also much more overdamped. O) Conceptually, we can imagine two modes, one which is amplified by parametric resonance and the other which is responsible for turning (AER and a Goldstone like mode).  The coupling parameter $\zeta$ controls how much power is passed from the amplifier to the turning mode. P) This results in an interesting system (analogous to a laser) where leaked power is the output of the dynamics, but if you leak too much power too quickly, it becomes harder to achieve the amplification threshold (analogous to the lasing threshold) and for small stimulus you get no amplification. This careful balance suggests that an organism below a certain size (e.g. $\zeta$ large) will exhibit no amplification and will only couple into Goldstone like modes, whereas larger organisms may exhibit amplification facilitated turning dynamics.  }
\end{figure}

\newpage
\begin{figure}
\includegraphics[width = \textwidth]{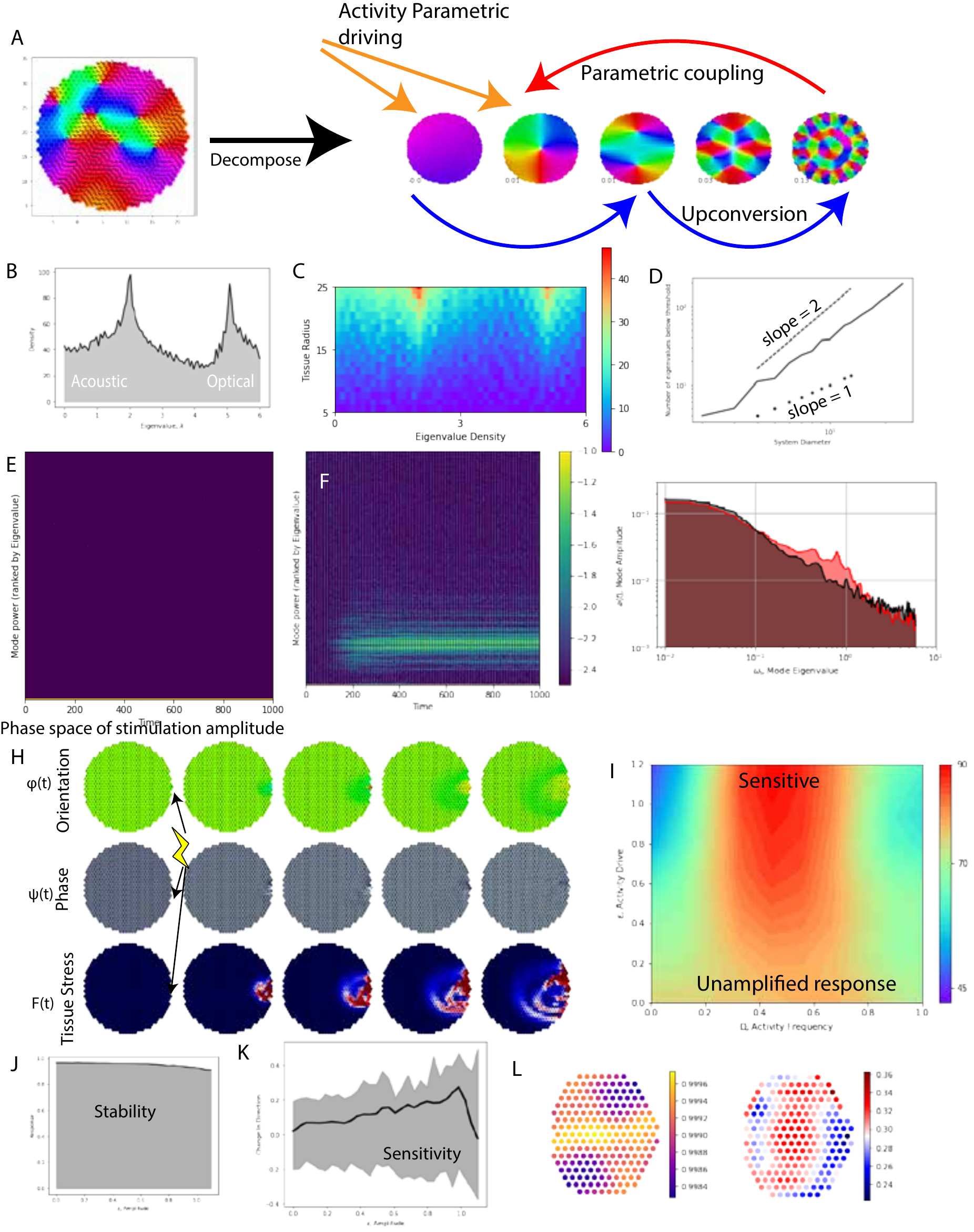}
\caption{\textbf{FIG 5}: (Caption next page.)}
  \label{fig:fig5}
\end{figure}

\begin{figure} [t!]
\caption{FIG 5 (previous page) \textbf{The stable yet sensitive dynamics of flocking oscillators can be decomposed into a weakly coupled spectrum of parametrically driven active-elastic resonator modes.} A) The rich spatio-temporal dynamics of flocking oscillators, can be projected onto the active-elastic modes of the tissue. These mode amplitudes then pass power between each other using parametric down conversion and three-wave mixing like upconversion. B) Studying the density of states of this spectrum of these modes reveals two enriched peaks corresponding to lower frequency acoustic-like modes and higher frequency optical modes. C) Since, these tissues exist at many sizes, it is natural to ask how this density of states scales with tissue size showing not only a marked increase in the size of the acoustic peak but also more dense packing of the low frequency modes. D) Studying how the number of modes below a threshold scales with size reveals an anomalous scaling between 1 and 2 (but approaching 2) which suggests that as the characteristic lengthscale of the tissue increases the number of modes which are ‘easily’ excited to large amplitudes grows nearly with the square. Larger tissues have more accessible modes and thus more complex spatio-temporal evolution. E) Next, we can ask how the power in each of these modes evolves in time without (E) and with (F) parametric activity driving. Notice that when activity driving amplitude $\epsilon >0$, more power is passed out of the perfectly polarized mode and into a band of intermediate modes. This is clear by comparing the power-spectra of the $\epsilon = 0$ (black) case and the $\epsilon = 1$ case (red). H) We can recreate the numerical experiments of the flocking oscillators in response to stimulus from figure 3 and study the system response as a function of the driving amplitude, $\epsilon$ versus frequency space $\Omega$. I) We find that there are a number of Mathieu tongues corresponding to collections of modes on resonance. On parametric resonance, the tissue is more sensitive to stimulus as measured by mean tissue force. To study the trade off between J) Stability of the polarization state and K) sensitivity of the tissue to stimulus as measured through turning amplitude, we find that for a $\sim$10\% reduction of polarization, the tissue can become 450 $\%$ more agile (as measured by ability to turn) in response to the same stimulus by increasing $\epsilon$ from 0 toward 1. L) Importantly, the response of the tissue is not agnostic to the location of the stimulus. We study this form of embodiment by summarizing N=500 simulations (one for each cell in the tissue) by i) how much the polarization changes for the stimulus applied to the ith cell and ii) how much the organism changed its center of mass heading in response to a stimulus of the ith cell. Intriguingly, there is a strong (50\% of the overall signal) sensitivity to the location of the stimulus which has important implications for how a tissue of flocking oscillators responds to stimulus.}
\end{figure}

\end{document}